\documentclass[twoside,11pt]{article}

\usepackage[accepted]{melba}

\usepackage{amsmath,amsfonts}
\usepackage{xspace} 
\usepackage{subfig}
\usepackage{array}
\makeatletter
\newcommand{\clearsubcaptcounter}{\setcounter{sub\@captype}{0}}
\makeatother
\usepackage{etoolbox}
\usepackage{verbatim} 
\usepackage{makecell}
\usepackage{lipsum}
\usepackage{multicol, blindtext}

\newcommand*\rot{\rotatebox{90}}
\usepackage[inline]{enumitem}
\usepackage{svg}
\usepackage{graphicx}
\usepackage{dirtytalk}

\usepackage{tikz}
\usepackage{booktabs}
\usetikzlibrary{calc}
\usepackage{colortbl}

\newcommand{\tikzmark}[1]{\tikz[overlay,remember picture] \node (#1) {};}
\newcommand{\DrawBox}[3][]{%
    \tikz[overlay,remember picture]{
    \draw[black,#1]
      ($(#2)+(-0.1em,4.1ex)$) rectangle
      ($(#3)+(0.75em,-0.75ex)$);}
}

\melbaid{2023:007}  
\melbaauthors{Khun Jush, Biele, Dueppenbecker and Maier}
\volume{2}
\firstpageno{202}  
\melbayear{2023}  
\datesubmitted{12/2021}  
\datepublished{05/2023}  

\ShortHeadings{Impacts of Training Data Diversity on Stability and Robustness}{Khun Jush, Biele, Dueppenbecker and Maier}

\title{Deep Learning for Ultrasound Speed-of-Sound Reconstruction: Impacts of Training Data Diversity on Stability and Robustness}
 
\author{\firstname Farnaz \surname Khun Jush \orcid{0000-0002-4860-1775} \email farnaz.khun.jush@fau.de \\  
	\addr Technology Excellence, Siemens Healthcare GmbH, Erlangen, Germany \\
	\addr Pattern Recognition Lab, Friedrich-Alexander-University, Erlangen, Germany 
	\AND
	\name Markus Biele \email  markus.biele@siemens-healthineers.com \\  
	\addr Technology Excellence, Siemens Healthcare GmbH, Erlangen, Germany
	\AND
	\name Peter M. Dueppenbecker \email peter.dueppenbecker@siemens-healthineers.com \\  
	\addr Technology Excellence, Siemens Healthcare GmbH, Erlangen, Germany
    \AND
	\name Andreas Maier \orcid{0000-0002-9550-5284} \email andreas.maier@fau.de \\
	\addr Pattern Recognition Lab, Friedrich-Alexander-University, Erlangen, Germany 
}

\begin{document}

\maketitle

\begin{abstract}%
Ultrasound b-mode imaging is a qualitative approach and diagnostic quality strongly depends on operators' training and experience.
Quantitative approaches can provide information about tissue properties; therefore, can be used for identifying various tissue types, e.g., speed-of-sound in the tissue can be used as a biomarker for tissue malignancy, especially in breast imaging.
Recent studies showed the possibility of speed-of-sound reconstruction using deep neural networks that are fully trained on simulated data. 
However, because of the ever-present domain shift between simulated and measured data, the stability and performance of these models in real setups are still under debate. 
In prior works, for training data generation, tissue structures were modeled as simplified geometrical structures which does not reflect the complexity of the real tissues.
In this study, we proposed a new simulation setup for training data generation based on Tomosynthesis images. We combined our approach with the simplified geometrical model and investigated the impacts of training data diversity on the stability and robustness of an existing network architecture. 
We studied the sensitivity of the trained network to different simulation parameters, e.g., echogenicity, number of scatterers, noise, and geometry. 
We showed that the network trained with the joint set of data is more stable on out-of-domain simulated data as well as measured phantom data.

\end{abstract}

\begin{keywords}
Speed-of-Sound, Medical Ultrasound, Deep Neural Networks
\end{keywords}

\section{Introduction}

Ultrasound b-mode imaging is a qualitative approach and is highly dependent on the operators' experience. 
In b-mode imaging, different tissue types can result in very similar image impressions, for example, in breast cancer screening b-mode images have low specificity for differentiating malignant and benign lesions \citep{carra2014sonography,chung2015ultrasonography}. 
This complicates the diagnosis.
Quantitative information, e.g., elasticity, density, speed-of-sound (SoS), and attenuation can increase ultrasound specificity and ease interpretation. 
Studies show that SoS values carry diagnostic information that can be used for tissue characterization \citep{hachiya1994relationship,khodr2015determinants, sak2017using,sanabria2018breast}.

Application-specific systems, e.g., ultrasound tomography systems (USCT) have emerged to measure SoS and attenuation in tissues \citep{duric2007detection}.
There are several methods of reconstruction for USCT, e.g., approximate models such as ray-tracing methods \citep{jirik2012sound,opielinski2018multimodal,javaherian2020refraction,li2009vivo,li2010refraction} that are often fast but have low spatial resolutions; or the full-wave inversion (FWI) methods that directly solve the wave equation and allow the whole measured data to be used in the inversion (time-series or all frequencies), thus, resulting in high-resolution images \citep{wiskin2007full,roy2010sound,agudo20183d,li2014toward,wang2015waveform,perez2017time,lucka2021high}. 
The drawback of FWI methods is that they have a huge memory footprint and are computationally expensive \citep{lucka2021high}. 
Additionally, from the hardware point of view, USCT typically needs double-sided access to the tissue with multiple transducers  \citep{malik2018quantitative,duric2007detection,gemmeke20073d} which makes the setups bulky and costly.

SoS values can be estimated using conventional pulse-echo setups \citep{qu2012average,benjamin2018surgery,sanabria2018breast}. 
In practice, SoS in tissue can be estimated in various ways. One method is based on the idea that the assumed SoS used in b-mode image reconstruction closely matches the actual SoS if the b-mode  image quality is maximized \citep{gyongy2015variation,benjamin2018surgery}.
\cite{stahli2020improved} proposed computed ultrasound tomography in echo mode (CUTE). This method is based on spatial phase shifts of beamformed echoes. 
In \cite{huang2004experimental,sanabria2018speed} a passive reflector is used and a limited angle ill-posed inverse problem is formulated. 
\cite{stahli2020bayesian} proposed a Bayesian framework to solve the inverse problem which includes an a priori model based on b-mode image segmentation.


Model-based SoS reconstruction is a non-trivial task and extracting SoS information by analytical or optimization methods requires prior knowledge, carefully chosen regularization, and optimization methods \citep{vishnevskiy2019deep}. 
Furthermore, there are challenges in implementing these methods for real-time imaging, only a few methods have the potential to be used in real-time imaging systems (e.g., CUTE \citep{stahli2020improved}) but often these methods are computationally expensive with varying run times.

Deep-learning-based methods move the computational burden to the training phase and therefore have a great potential for real-time applications.
In recent years, these techniques are vastly being used in the medical imaging field \citep{suzuki2017overview,lee2017deep,mainak2019state,maier2019gentle,lundervold2019overview,wang2018image}. 
As such, in the field of medical ultrasound, deep learning techniques are gaining rapid attention, especially for segmentation and image formation (a.k.a. beamforming), an overview can be found in \cite{van2019deep, mischi2020deep}. 


Studies on deep-learning-based quantitative ultrasound are sparse compared to image formation approaches \citep{vishnevskiy2019deep,hoerig2018data,feigin2019deep, feigin2020detecting,jush2020dnn,jush2021data, bernhardt2020training,gao2019learning,mohammadi2021ultrasound,heller2021deep,oh2021neural}.
Deep-learning-based SoS reconstruction for pulse-echo ultrasound was first addressed in \cite{feigin2019deep}. 
Inspired by \cite{feigin2019deep}, in \cite{jush2020dnn} we investigated the possibility of deep-learning-based SoS reconstruction for automated breast ultrasound using a single PW acquisition.
These studies rely on simulated training data. 

\section{Thesis and Aim}

One of the main obstacles to developing deep learning techniques is the collection of training datasets alongside their corresponding labels or Ground Truth (GT) maps. 
This process is often time-consuming and costly. 
The problem further expands when the labeled data or GT is difficult to measure. 
Particularly, for SoS reconstruction there is no gold-standard reference method capable of creating SoS GT from ultrasound reflection data.
Plus, there are only a few phantoms available with known heterogeneous SoS. Therefore, using simulated data for the training phase is a common approach in deep-learning-based algorithms \citep{feigin2019deep,feigin2020detecting,jush2020dnn, jush2021data,vishnevskiy2019deep,oh2021neural,heller2021deep}.
These studies employed a simplified setup for tissue modeling in which inclusions are assumed to be simple geometric structures, e.g., ellipsoids, randomly located in a homogeneous background.

Although the previously proposed tissue models enabled deep-learning-based SoS reconstruction, the simplified approach does not reflect the complexity of real tissues, thus, resulting in stability and reproducibility challenges when tested on measured data, especially in the presence of noise.
We encountered such challenges in \cite{jush2020dnn} that motivated the current study. 
Creating large diverse training data is a crucial task. 
Deep-learning-based models only deal with generalization within the distributions of the training data. 
Otherwise, the model performs poorly or fails in real-world setups.
Therefore, the broader the distribution of the training data the more the likelihood of success of the deep neural networks in real setups.
One possibility which is extensively studied is to train the models with pools of different datasets or multiple datasets each providing a different aspect of the underlying model \citep{goyal2020inductive}. 

This study aims to investigate the stability and pitfalls of an existing deep-learning-based method for SoS reconstruction and improve its performance by introducing diversity to tissue modeling. 
Additionally, since tissue modeling is a non-trivial task, investigations of the sensitivity of the networks are helpful to create a stable setup for data generation which is missing from previous studies. 
Therefore, we include experiments with digital phantoms of various properties and provide the reader with insights regarding the sensitivity of the investigated network encountering out-of-domain simulation characteristics, e.g., varying echogenicity, geometry, speckle density, and noise. 

The contributions of this study are as follows: 

\begin{enumerate}[label=(\roman*)]
 \item We proposed a new simulation setup based on Tomosynthesis images to expand the diversity of the tissue modeling that integrates tissue-like structures for ultrasound training data generation.
    
    \item We augmented the dataset created from our proposed setup to a baseline simplified tissue model proposed by \cite{feigin2019deep, feigin2020detecting} and showed that without any modifications the existing architecture (an encoder-decoder network using a single plane-wave acquisition) is capable of reconstructing the complicated SoS maps as well as simple structures. 
    
    \item We analyzed the stability of the network trained with augmented data in comparison with the baseline setup (simplified tissue model) in the presence of out-of-domain simulated data to find the crucial parameters to which the network shows high sensitivity. 
    
    \item We compared the stability and robustness of the network trained with the simplified dataset versus the augmented dataset on measured phantom data and showed that the network trained with a joint set of data is more stable when tested on real data.  
\end{enumerate}

\section{Materials and Methods}

We used the k-Wave toolbox (Version 1.3) \citep{treeby2010k} for simulating the training data. 
K-Wave is a MATLAB toolbox that allows simulation of linear and non-linear wave propagation, arbitrary heterogeneous parameters for mediums, and acoustic absorption \citep{treeby2010k}.
Studies demonstrated that acoustic non-linearity and absorption are well modeled in k-wave, and when there is a good representation of medium properties and geometry, the models are accurate \citep{wang2012modelling,treeby2012modeling,martin2019experimental,agrawal2021modeling}.

We generated a training dataset using two simulation setups, one based on prior works proposed by \cite{feigin2019deep} and one is a new simulation setup that we introduce in this study. 
We used a single PW and 2D medium of size $3.8 \times 7.6$~$cm$ (grid size: $1536 \times 3072$) where the transducer is placed in the central section of the mediums (for more details on transducer setup, the geometry of the medium and acoustic properties see Appendix~A.1 and A.2). 

\subsection{Simulation Setup}

\subsubsection{Prior Work: Ellipsoids Setup} 
\begin{figure}[!t]
\centering
\subfloat[]{\raisebox{+0.03\totalheight}{\includegraphics[scale=0.2]{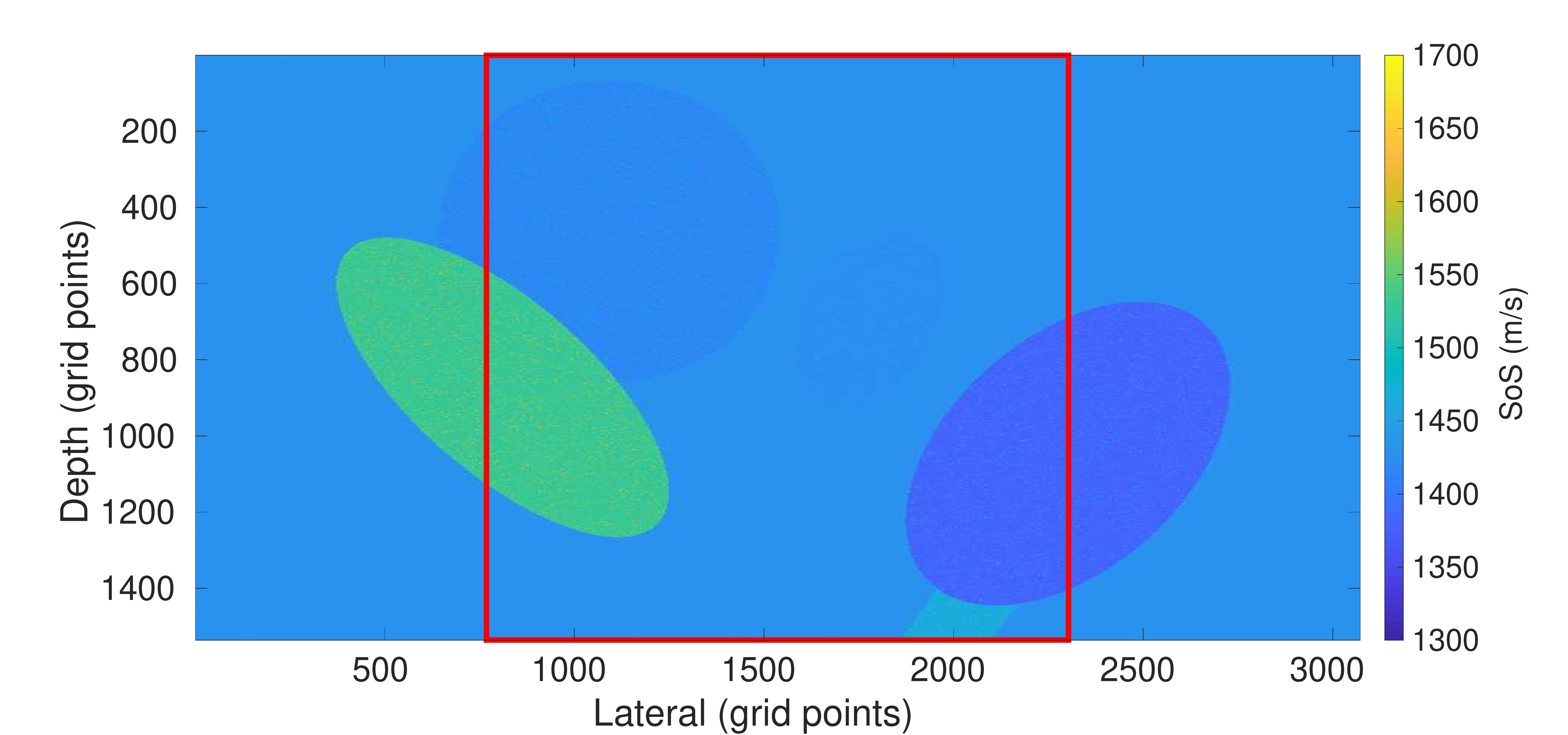}}}
\subfloat[]{\includegraphics[scale=0.3]{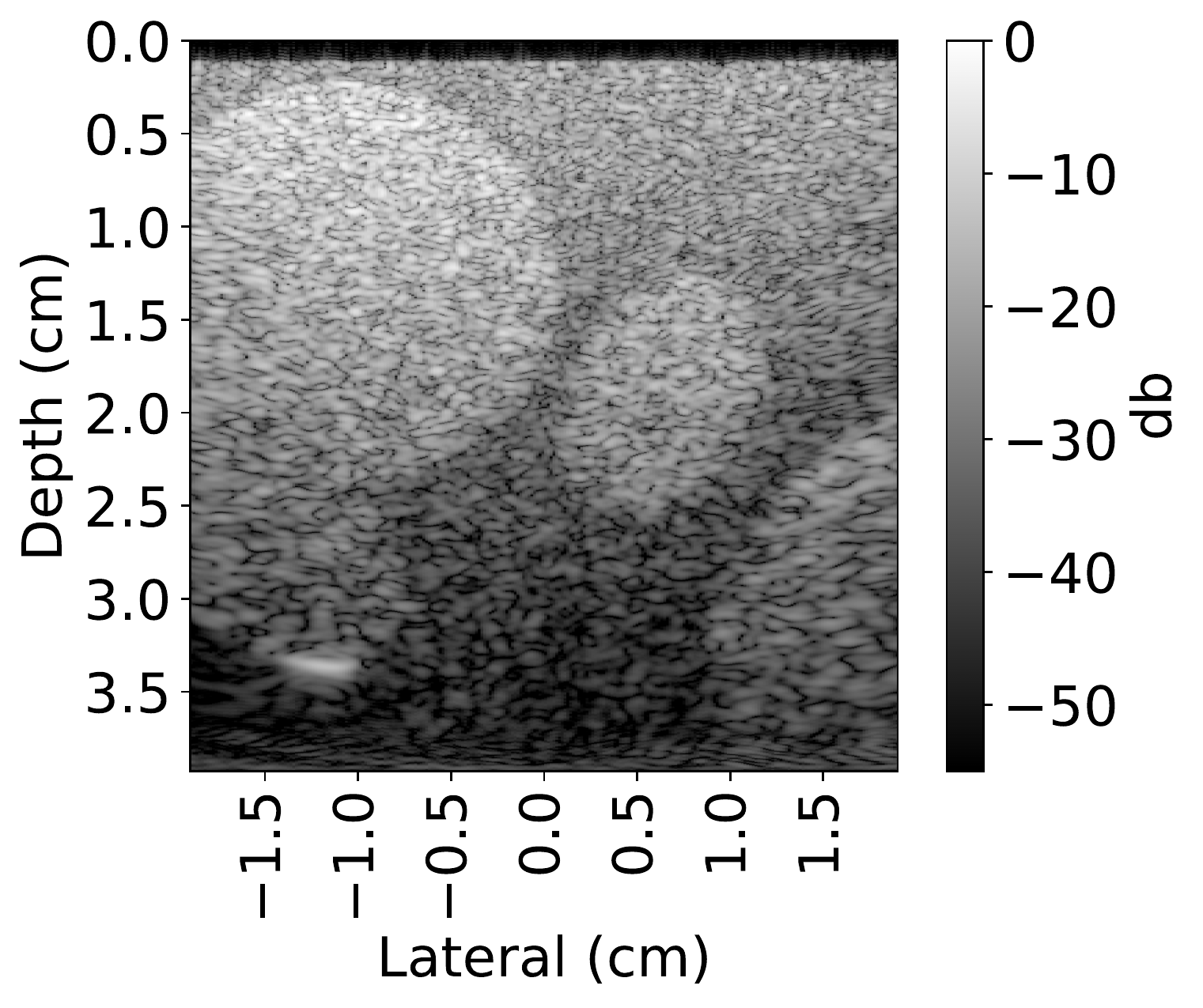}}
\caption{(a): SoS domain of a sample Ellipsoids setup with size \(1536 \times 3072\) grid points, a homogeneous background, and five inclusions with varying SoS values randomly placed inside the medium. The probe head is placed above the central section and the red box shows the corresponding recovered region, (b): reconstructed b-mode image with size \(3.8 \times 3.8\) cm using delay-and-sum beamforming from simulated RF data based on the demonstrated setup in (a).} 
\label{fig:medium_ell}
\end{figure}

The first setup is a setup based on \cite{feigin2019deep} adapted to our transducer properties with 192 active channels and single PW acquisition.
The setup is a simplified model for organs and lesions.
In this setup, the medium consists of a homogeneous background with elliptical inclusions randomly located in the medium where the SoS values are assigned randomly in the range $[1300-1700]$~$m/s$.
Although the expected SoS in the breast tissue is in a smaller range \([1433 - 1565]\)~\(m/s\) \citep{li2009vivo}, the higher range (i.e., \([1300-1700]\)~\(m/s\)) is chosen for better generalization \citep{feigin2019deep}. 
Details of the simulation parameters for this setup are added  in Appendix~A.3.
This setup is considered as a baseline setup for comparison purposes. Henceforth, we refer to this setup as Ellipsoids setup, an example of this setup is shown in Figure~\ref{fig:medium_ell}. 

\subsubsection{Proposed Setup: Tomosynthesis to Ultrasound (T2US)}

Here we introduce a new simulation setup that includes more complicated structures in the mediums. 
The intensity of CT images that are measured in the Hounsfield units (HU) is related to tissue density. SoS and attenuation of the tissues are directly related to the density as well. Therefore, the idea of simulating ultrasound images from CT images was investigated in prior studies \citep{dillenseger2009fast,shams2008real,cong2013fast,kutter2009visualization, sak2017using} but in the context of developing fast ultrasound simulations due to the high computational demands of standard acoustic simulation setups. 
However, since the k-wave toolbox supports GPU and CPU acceleration, it has a decent run time on modern Nvidia GPUs. 
Plus, we use a single PW and 2D grid. 
Therefore, we propose a different approach to create ultrasound images from CT images. 
The proposed method extracts the underlying tissue-like structures from Tomosynthesis images and creates heterogeneous SoS mediums for the standard k-wave simulation. 

14 Tomosynthesis image stacks are used. 
Each image stack contains between 29 to 74 slices.
These images are acquired from patients diagnosed with benign or malignant breast lesions. 

\begin{figure}[!t]
\centering
\begin{tabular}{c @{\hspace{0.8mm}}c @{\hspace{0.8mm}}c }
\multicolumn{2}{c}{\includegraphics[scale=0.55]{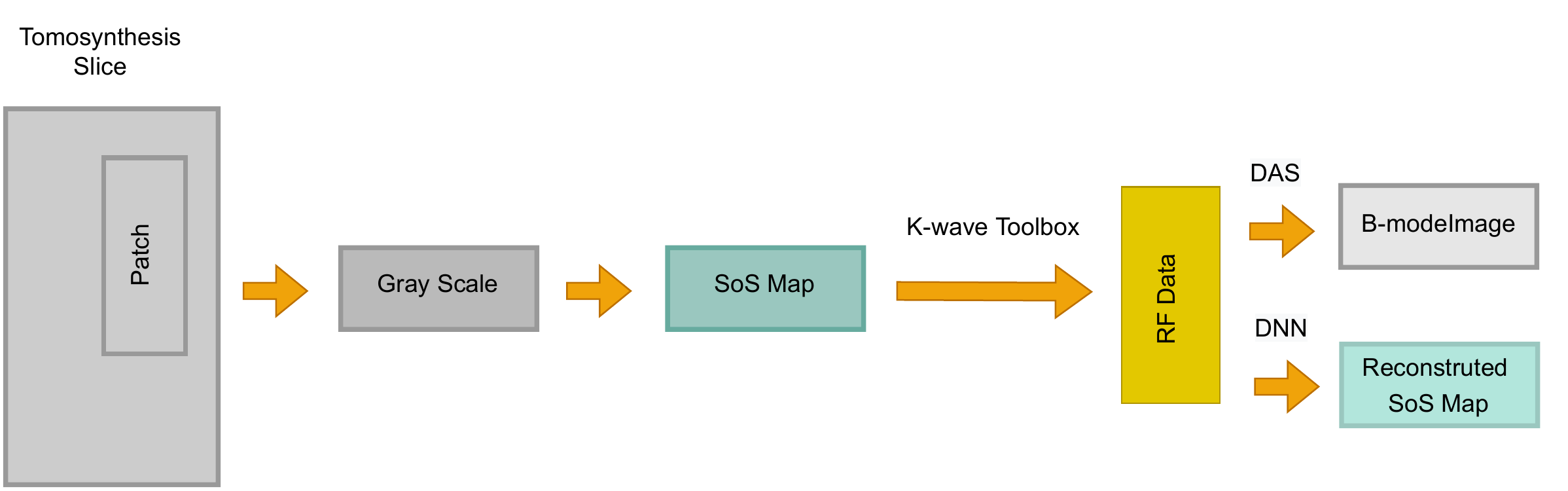}}\\ 
\multicolumn{2}{c}{(a)}\\ 
{\makecell{\includegraphics[scale=0.142]{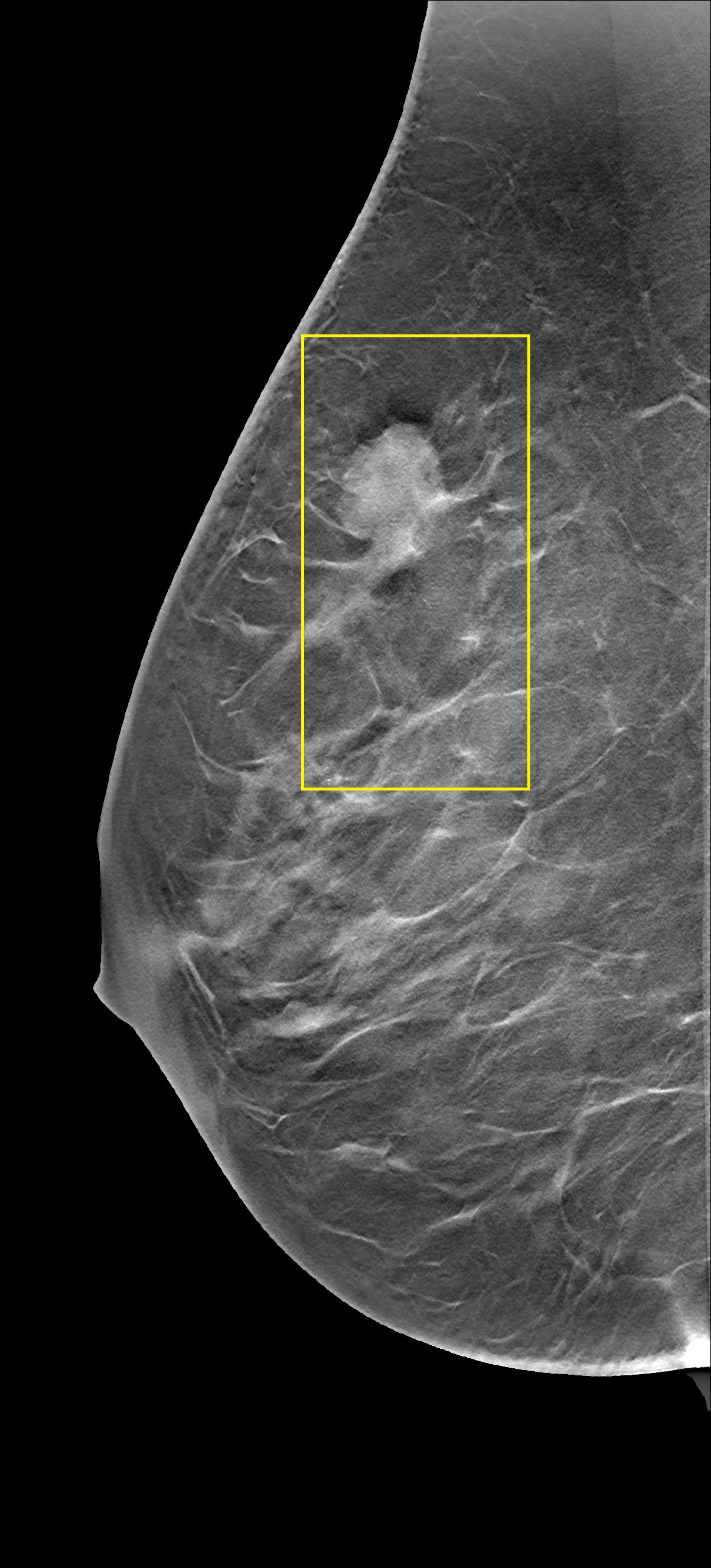} \\ (b)}}&  {\makecell{
\includegraphics[width=12cm]{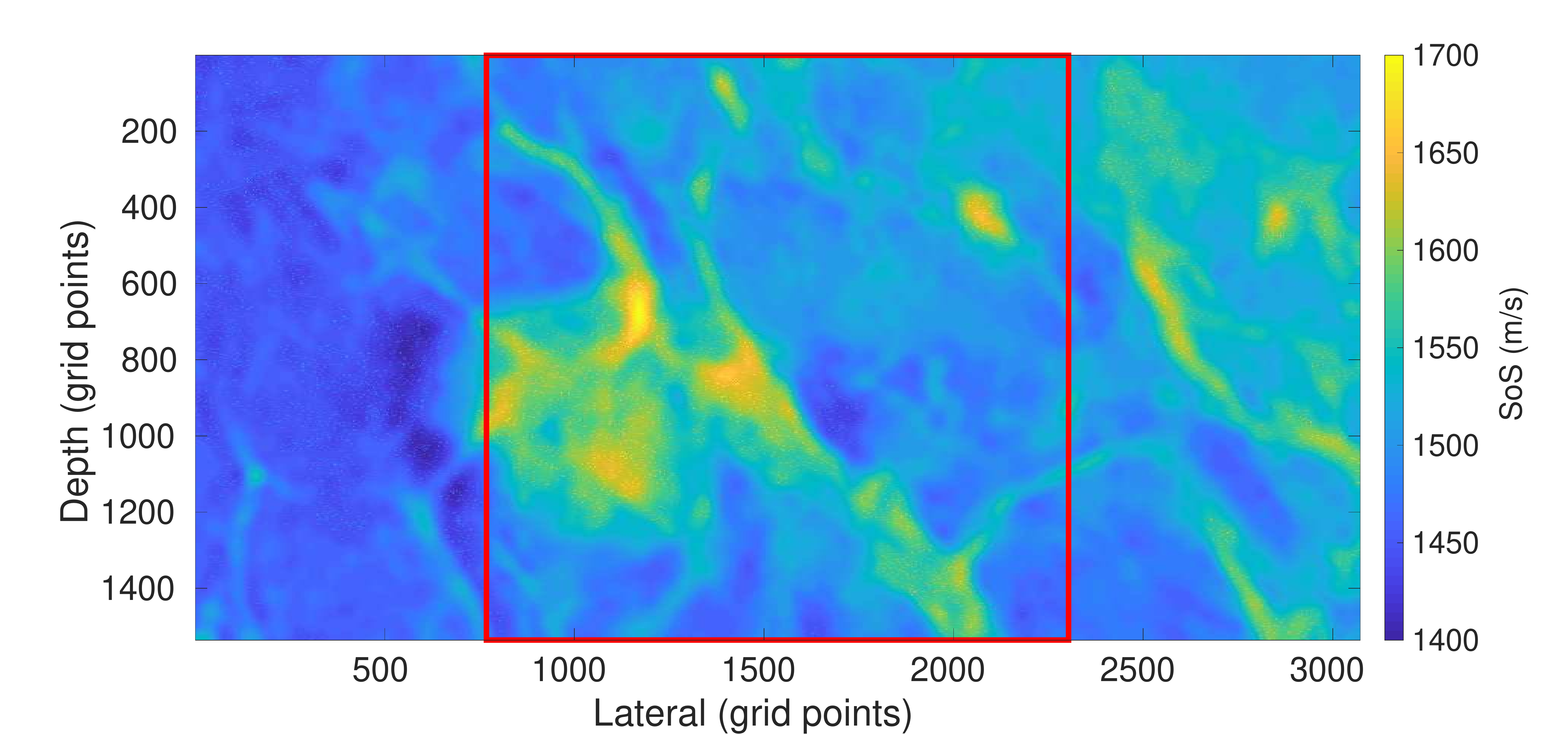} \\ {(c)} \\  \includegraphics[scale=0.35]{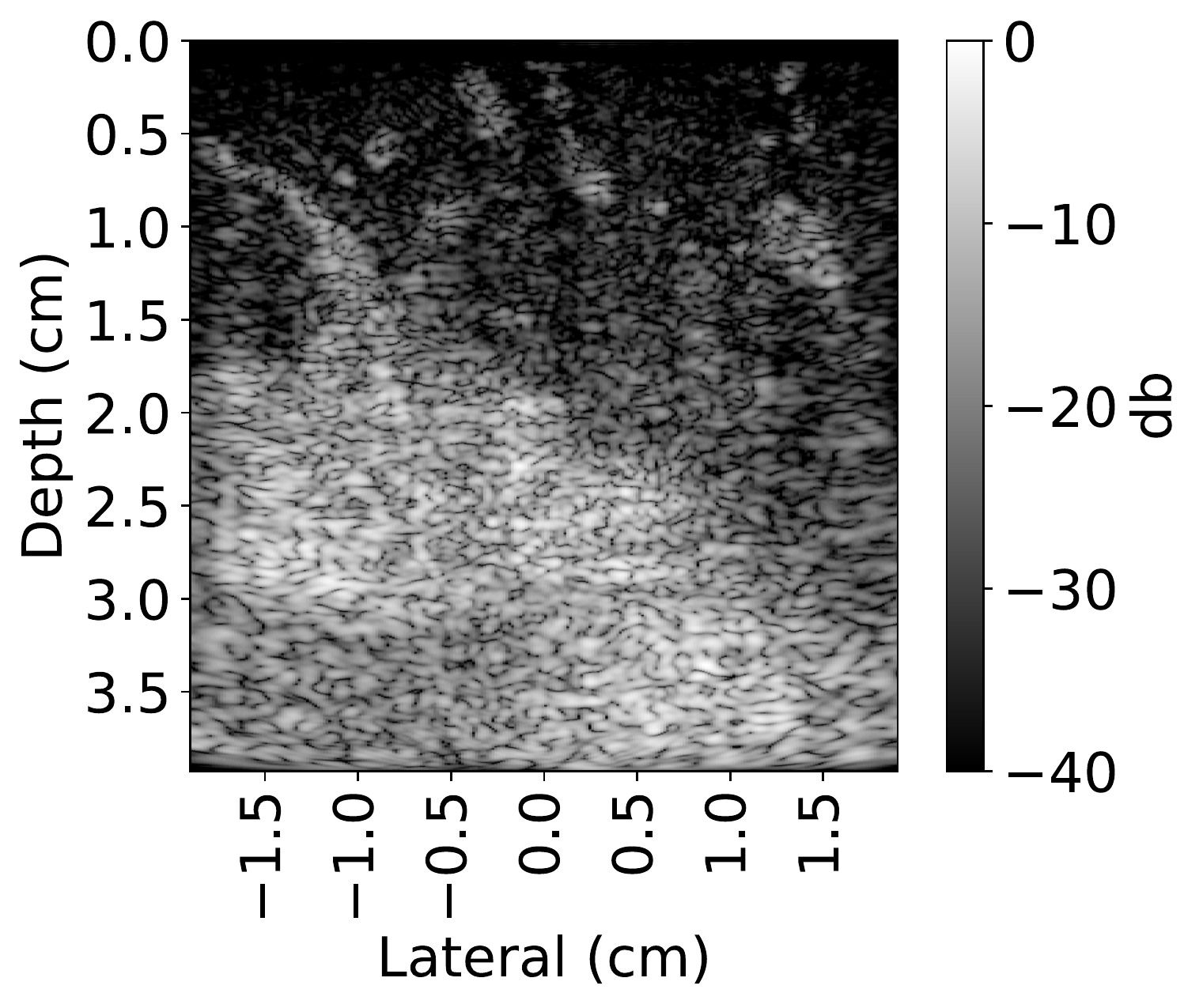} \\ {(d)} }}  

\end{tabular}

\caption{T2US setup: (a): Overview: from Tomosynthesis image to generating the RF data for training which can be used to reconstruct the b-mode image using delay-and-sum (DAS) algorithm or to reconstruct SoS map using a deep neural network (DNN), (b) Random patch: an example random patch (yellow box) from the Tomosynthesis slice, (c): SoS map: the example patch is first converted to grayscale and smoothed, and then the grayscale values are mapped to SoS values in the range \([1300-1700]\)~\(m/s\), the probe is placed above the central section of the medium and the field of view directly below probe head is recovered (red box), (d) B-mode image: the reconstructed b-mode image from simulated RF data using delay-and-sum beamforming.} 

\label{fig:medium_t}
\end{figure}

\paragraph{SoS Map} From each slice, 10 patches with size \(384\times768\) pixels are randomly extracted.
Patches are chosen in a way that the black regions outside the breast area present in the Tomosynthesis slices are excluded.
Overlapping regions in the patches from the same slide are not excluded. In order to reduce the range, HU values are converted to grayscale values. Because of the presence of modality noise in Tomosynthesis images, the patches are smoothed with a 2D Gaussian smoothing kernel with a standard deviation of \(4\).
The patches are resized to fit the medium size \(1536\times3072\) using nearest-neighbor interpolation.
        
The grayscale values are then mapped to SoS in the range \([1300-1700]\)~\(m/s\) (rescale function of MATLAB from Data Import and Analysis, Preprocessing Data toolbox is used). 
Currently, in order to define malignancy, solid breast lesions are examined based on the qualitative analysis of tumor margins and geometry (e.g., speculation, boundary uniformity, ovality). Studies suggest that the echogenicity of the lesion on its own does not provide enough contrast for differentiation \citep{blickstein1995echogenicity,ruby2019breast}. Nevertheless, there is an echogenicity contrast present between lesions and fatty tissues \citep{gokhale2009ultrasound}.
Thus, we included random hyperechogenic regions. 
Based on the mapped SoS, echogenicity of \(10\%\) of the pixels with assigned values higher than \(1550\)~\(m/s\), or lower than \(1450\)~\(m/s\) varies by \([4.4-5.5]\%\) for half of the cases. This intends to create varying echogenicity independent of the presence of a lesion. These patches are given to the simulation as the SoS domain.

It is noteworthy that the assigned SoS values are not intended to reflect real corresponding SoS values to a particular tissue type present in the Tomosynthesis image, but the aim is to include more detailed structures rather than geometric shapes used in the Ellipsoids setup. An overview is shown in Figure~\ref{fig:medium_t}.

\subsection{Deep Neural Network}

\begin{figure}[!t]
\centering
\subfloat[]{\raisebox{+0.03\totalheight}{\includegraphics[height = 5.5cm]{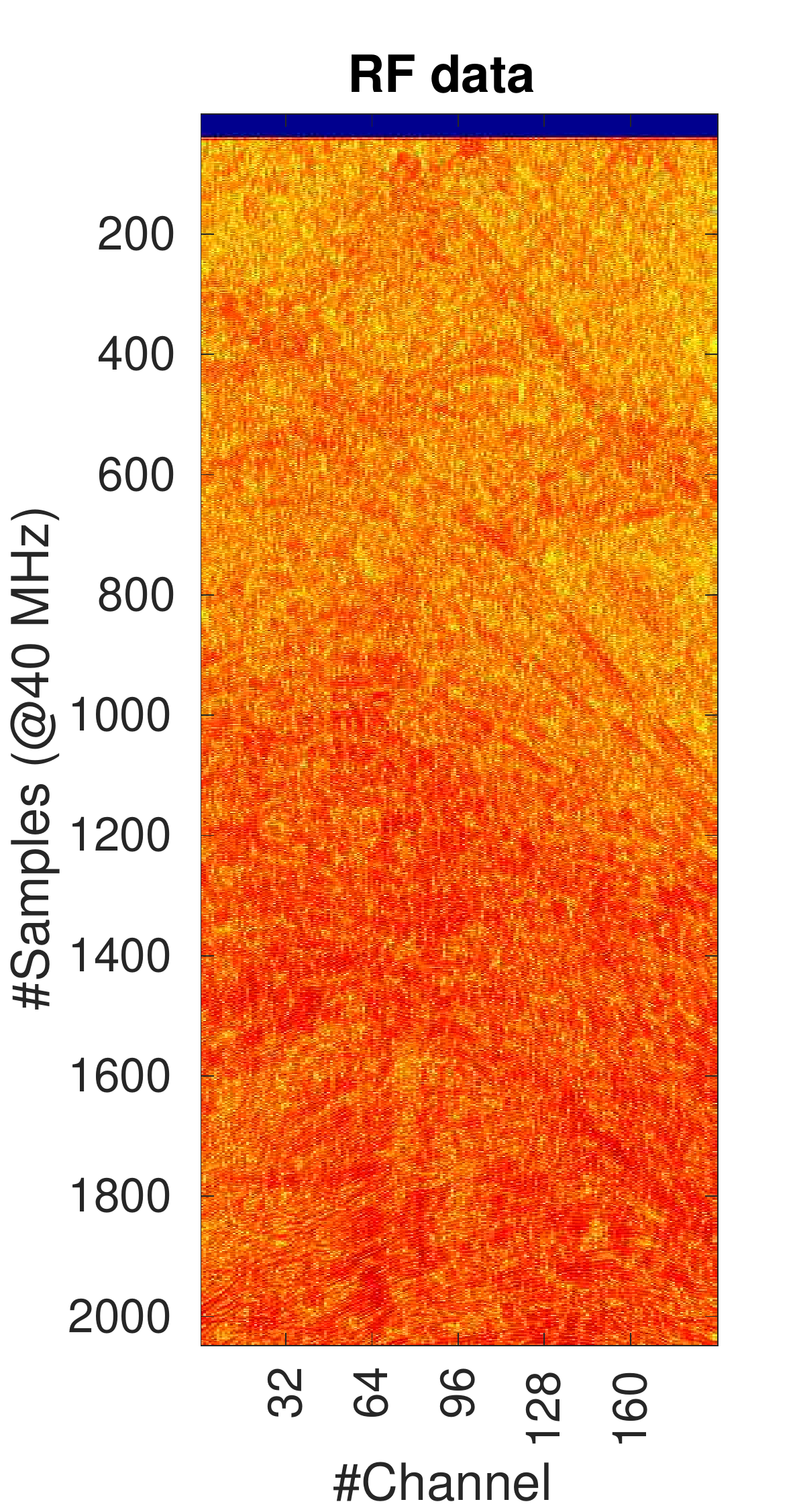}}}
\subfloat[]{\includegraphics[scale=0.5]{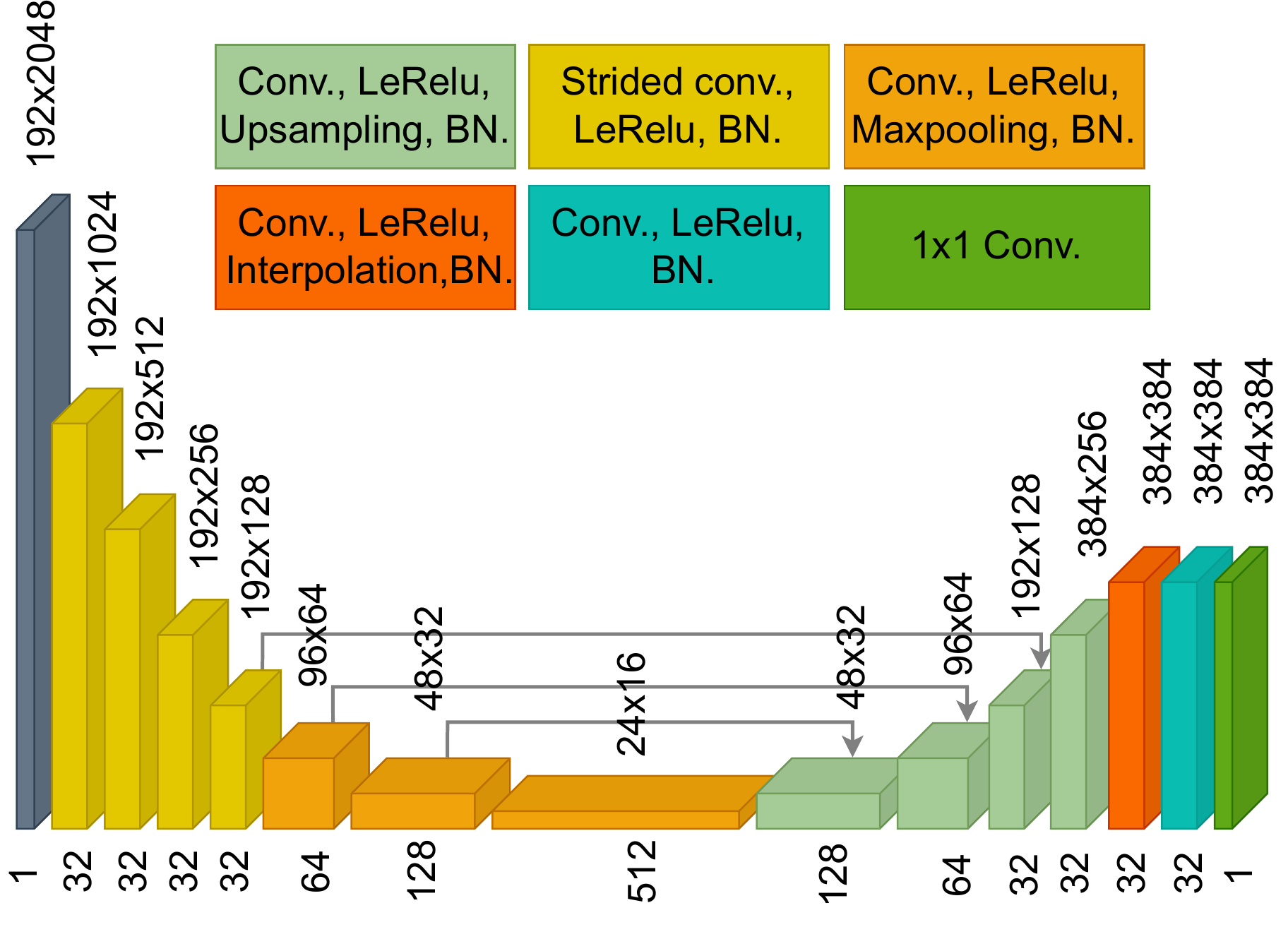}}
\subfloat[]{\raisebox{0.15\totalheight}{\includegraphics[trim={11cm 0.1cm 10cm 0.1cm},clip,width = 3.2cm]{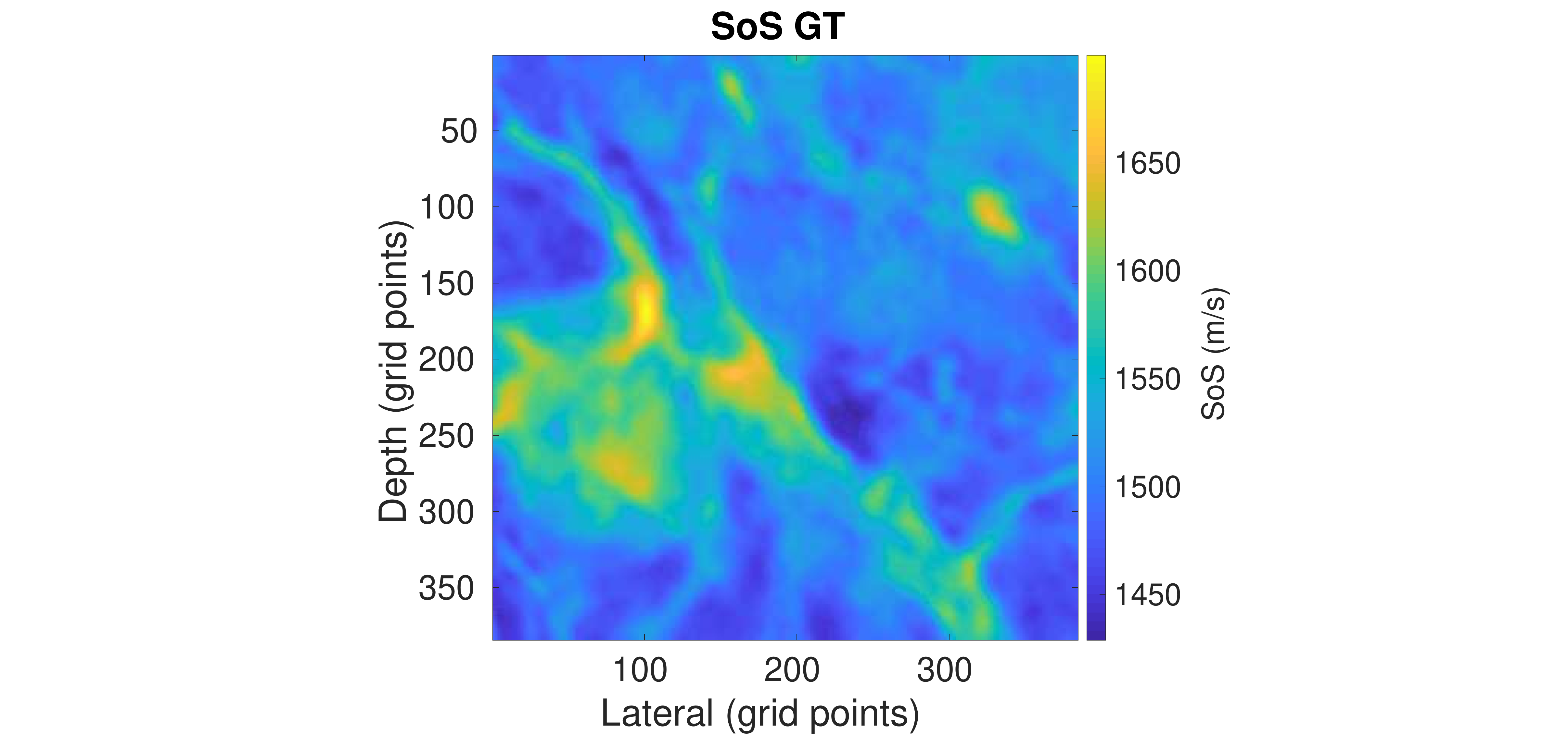}}}
\caption{(a) Input: RF data with 192 channels sampled at $40$~$MHz$ (a matrix of size $192\times2048$) from a single zero-degree PW is the input of the encoder-decoder network, (b) Network architecture: consists of encoding and decoding paths and skip connections. Each layer is color-coded and the operations are shown in the boxes with corresponding colors. During the training, the network takes RF data as input. The input is mapped to the SoS GT created with simulation, (c) Output: a sample SoS GT matrix of size $384\times384$ pixels.} 
\label{fig:skip}
\end{figure}

The network architecture is based on \cite{feigin2019deep,feigin2020detecting} adopted to a transducer geometry with 192 active channels presented in \cite{jush2020dnn}. 
The network is a fully convolutional neural network (FCNN), encoder-decoder network. 
The input of the network is RF data from a single PW acquisition, sampled at \(40~MHz\) (in form of a matrix of \(192\times2048\), \(64\) bit floating point). 
The output of the network is the predicted SoS map presented as a \(384\times384\) matrix with \(0.1~mm\) resolution.
The network architecture is shown in Figure~\ref{fig:skip}. 
The contracting and extracting paths each contain seven layers.
The details of the operations in each layer are color-coded in Figure~\ref{fig:skip} (b).



\subsection{Experimental Setup}

The network is trained only on simulated data and tested both on simulated and measured data acquired from a commercial breast phantom, CIRS multi-modality breast biopsy, and sonographic trainer (Model 073). 
This phantom mimics breast tissue properties with heterogeneous SoS, density, and attenuation. 
The data is acquired with the LightABVS prototype system presented in \cite{hager2019lightabvs,jush2020dnn}.

\section{Results and Discussion}

In this section, first of all, we will show the results of the networks trained with simulated datasets and compare the performance of the encoder-decoder network on different training data pools. 
Afterward, we investigate the stability of the trained networks in the presence of out-of-domain simulated data and noise to pinpoint the sensitivity of the network to different acoustic parameters. 
Moreover, experimental results from the CIRS phantom are included and the reproducibility and stability of the networks are discussed. 

\subsection{Simulated Data}

\subsubsection{Training Data}\label{sec: training}

\begin{figure}[!t]
\centering
\subfloat[]{\includegraphics[scale=0.5]{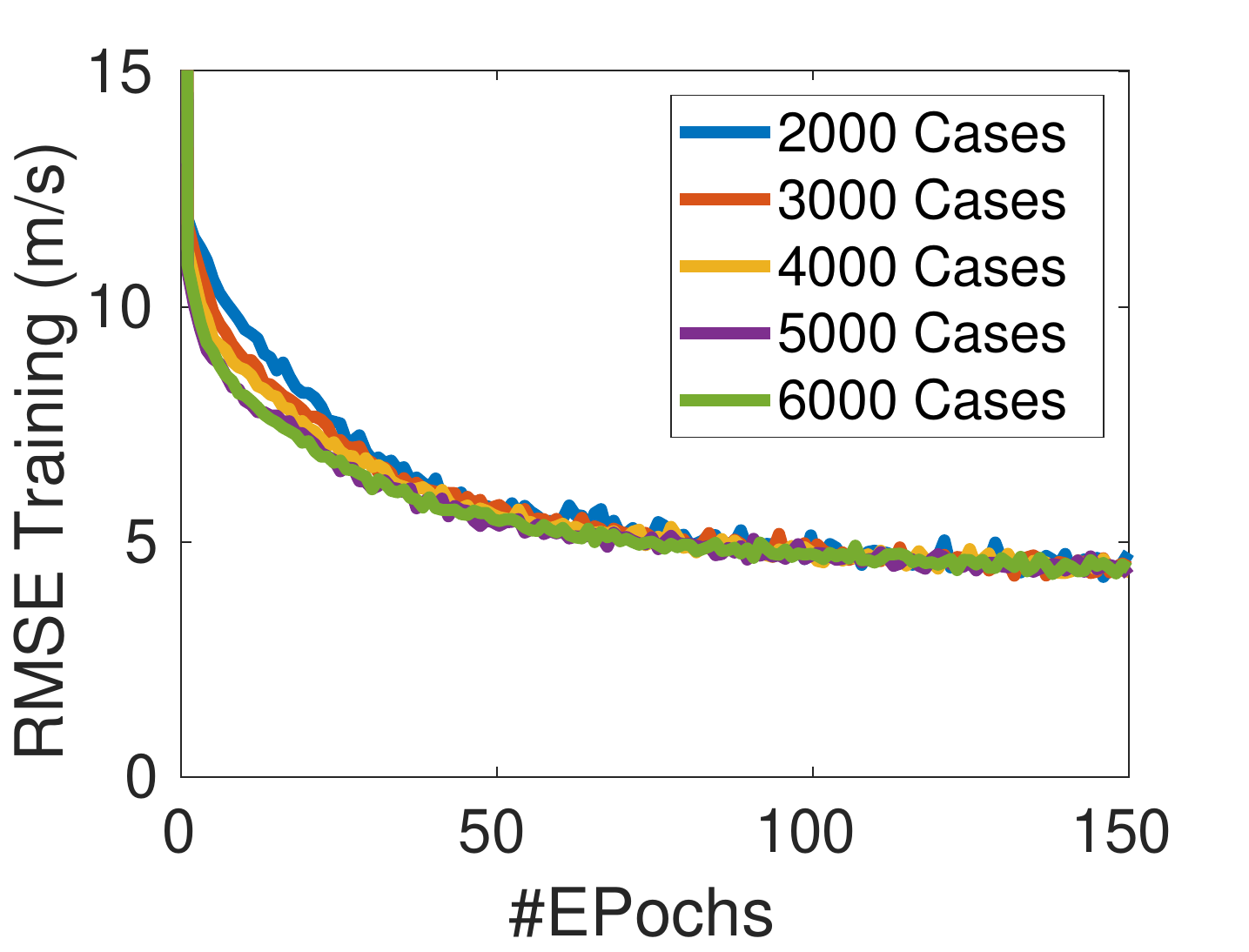}}\hspace{0.5mm}
\subfloat[]{\includegraphics[scale=0.5]{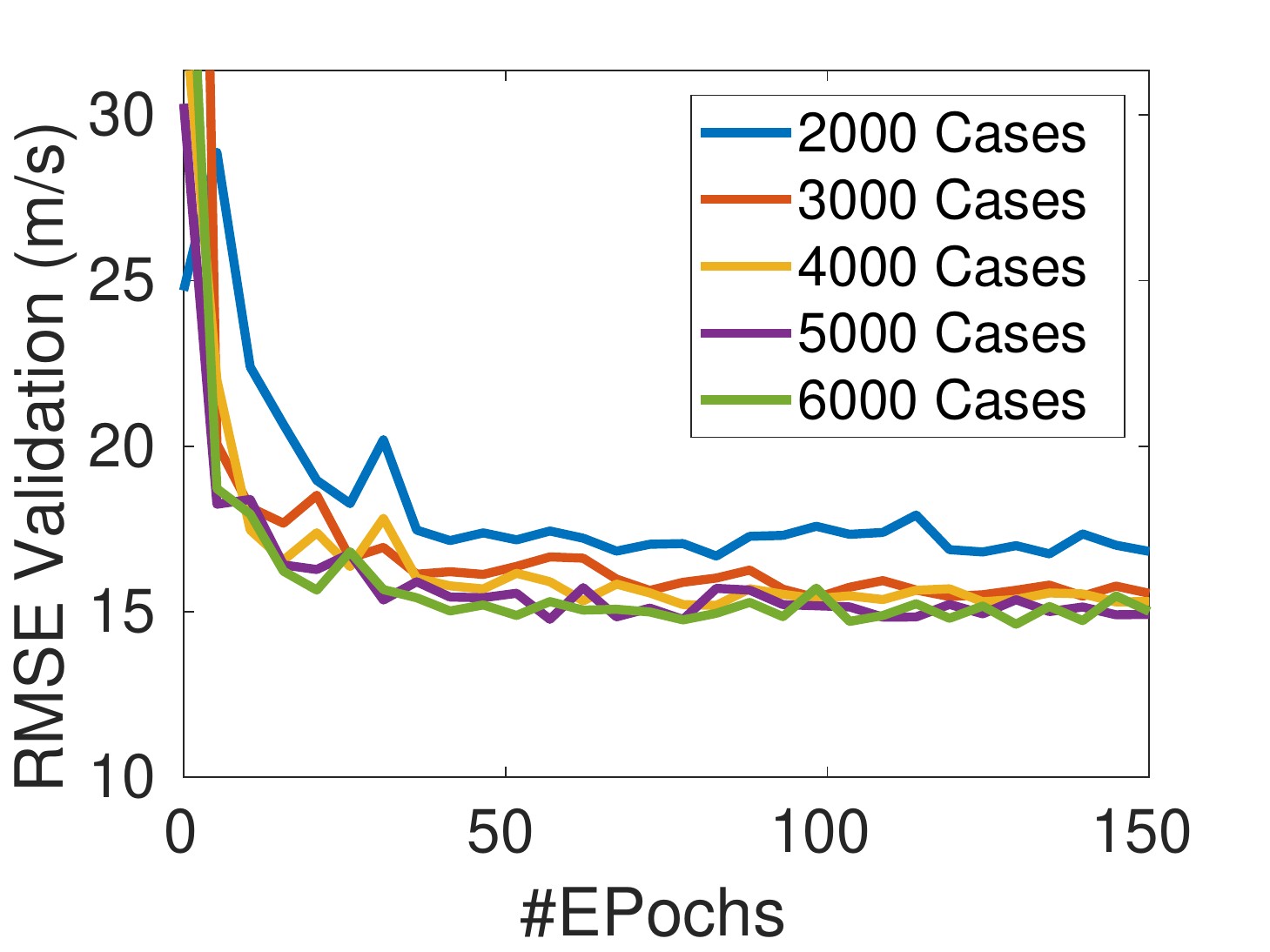}}\hspace{0.5mm}\\
\caption{ (a): Learning curves for training dataset with varying sizes from 2000-6000 with 1000 steps, (b): Learning curve for the validation set for the corresponding training set in (a). The learning curves on the validation set for 5000 and 6000 cases are similar showing a saturation and indicating that increasing the number of training cases of the same kind does not necessarily improve the performance of the network on the unseen data.}
\label{fig: howmany}
\end{figure}

Selecting the size of the training dataset for a network is a non-trivial task. In order to select the required size, we have trained the network on training sets of 2000, 3000, 4000, 5000, and 6000 cases of the Ellipsoids dataset and observed that there is a saturation of the performance and increasing the number of training cases does not improve the performance even on the simulated dataset. Figure \ref{fig: howmany} demonstrates this fact by comparing the learning curves for training and validation sets of varying sizes. 
This implies that increasing the number of training samples of the same kind of simulated data is not the optimal solution to improve the performance of the network. There are two options to improve the performance of the network, one is to modify the network architecture and the other is to improve the quality and diversity of the training set. Here, we focus on the second approach. 

The network is trained with three sets of input datasets: Ellipsoids setup and T2US setup to show that the network is capable of reconstructing each dataset, and combined sets of two aforementioned datasets to show that the network can learn two mappings jointly. The stability study focuses on the combined dataset compared to the baseline simplified model.

For each simulation setup, \(6150\) samples are simulated.  The dataset is divided to \(6000\) train (\(90\%\) for training and \(10\%\) for validation) and \(150\) test cases. For more information regarding training and data processing see Appendix~B.  The network is trained for 150 epochs for Ellipsoids and T2US setups and 200 epochs for the Combined setup using mini-batch stochastic gradient descent and MSE loss function.
Table~\ref{table: errors} shows the corresponding error rates on the test datasets after convergence. 

Sample cases of qualitative results for Ellipsoids, T2US, and Combined setups are shown in Figure~\ref{fig: Sos Prediction Combined}. In order to compare qualitative cases, their corresponding RMSE, MAE, and structural similarity index (a.k.a. SSIM \citep{wang2004image}) values are added below each case in Figure~\ref{fig: Sos Prediction Combined}.

\begin{table}[tbp]
\renewcommand{\arraystretch}{1.1}
\caption{Root Mean Squared Error (RMSE), Mean Absolute Error (MAE), and Mean Absolute Percentage Error (MAPE) for predicted SoS maps on each test dataset trained on the corresponding training set, SD refers to the standard deviation for each reported value over 10 runs over all test cases.}
\label{table: errors}
\centering
\resizebox{\columnwidth}{!}{\begin{tabular}{l|c|c|c } 
 \textbf{Dataset} & \textbf{RMSE \(\pm\) SD (m/s)} & \textbf{MAE  \(\pm\) SD (m/s)} & \textbf{MAPE \(\pm\) SD (\%)} \\
 \hline
 \textbf{Ellipsoids} & \(14.95\pm0.47\) &  \(5.94\pm0.58\)  & \(0.41\pm0.04\) \\ 
 \textbf{T2US} & \(22.74\pm0.90\)  & \(16.80\pm0.60\) &   \(1.09\pm0.03\) \\ 
 \textbf{Combined}  & \(20.96\pm0.80\)  & \(14.03\pm1.33\) &   \(0.81\pm0.11\) \\ 
\end{tabular}}
\end{table}

\begin{figure*}[!t]
\renewcommand{\arraystretch}{0.05}
\centering
\begin{tabular}{c @{\hspace{0.5mm}} c @{\hspace{0.5mm}}c @{\hspace{0.5mm}}c  @ {\hspace{0.5mm}}c @{\hspace{0.5mm}}c @{\hspace{0.5mm}}c @{\hspace{0.5mm}}c @{\hspace{0.5mm}}c @{\hspace{0.5mm}}c @{\hspace{0.5mm}}c @{\hspace{0.5mm}}l}

&\scriptsize{Case~1}& \scriptsize{Case~2}& \scriptsize{Case~3}& \scriptsize{Case~4}& \scriptsize{Case~5}& \scriptsize{Case~6}& \scriptsize{Case~7}& \scriptsize{Case~8}& \scriptsize{Case~9}& \scriptsize{Case~10}& \scriptsize[m/s] \\

\raisebox{-.5\totalheight}{\scriptsize \rot{GT}}& 
\raisebox{-.5\totalheight}{\includegraphics[width = 1.3cm]{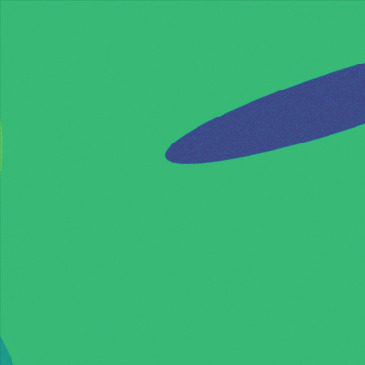}}&
\raisebox{-.5\totalheight}{\includegraphics[width = 1.3cm]{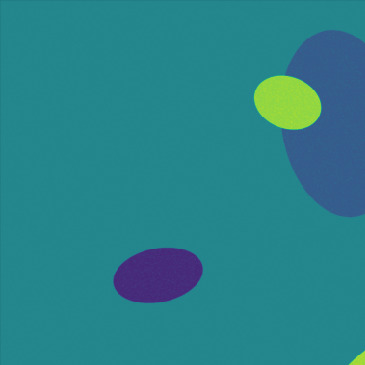}}&
\raisebox{-.5\totalheight}{\includegraphics[width = 1.3cm]{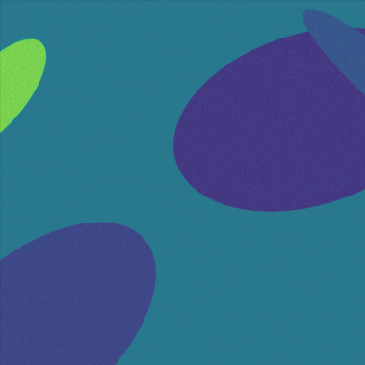}}&
\raisebox{-.5\totalheight}{\includegraphics[width = 1.3cm]{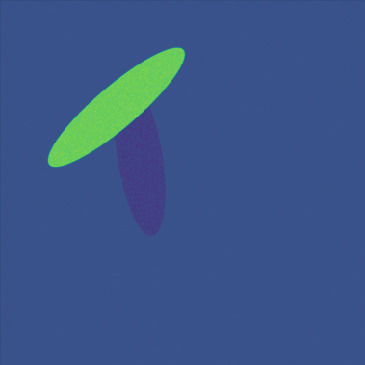}}&
\raisebox{-.5\totalheight}{\includegraphics[width = 1.3cm]{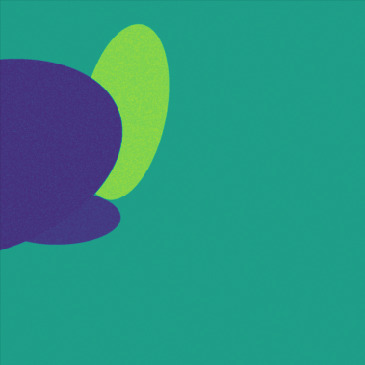}}&
\raisebox{-.5\totalheight}{\includegraphics[width = 1.3cm]{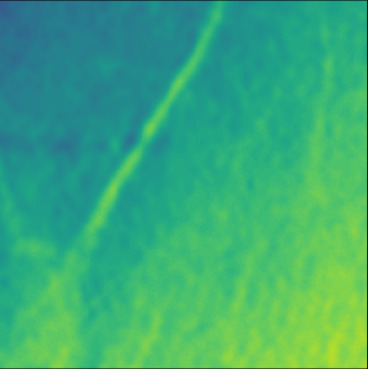}}  &
\raisebox{-.5\totalheight}{\includegraphics[width = 1.3cm]{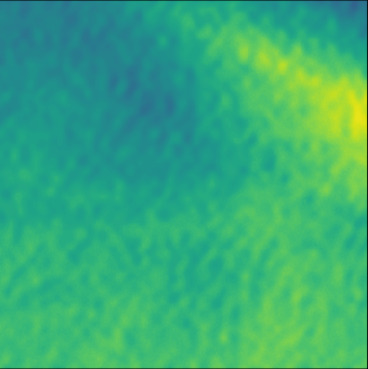}}&
\raisebox{-.5\totalheight}{\includegraphics[width = 1.3cm]{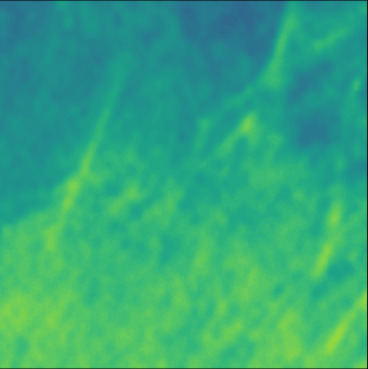}}&
\raisebox{-.5\totalheight}{\includegraphics[width = 1.3cm]{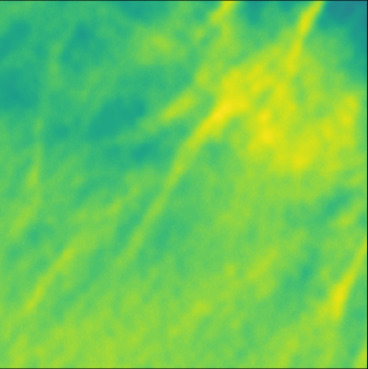}}&
\raisebox{-.5\totalheight}{\includegraphics[width = 1.3cm]{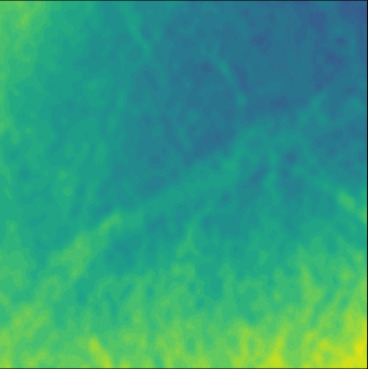}}&
\raisebox{-.5\totalheight}{\includegraphics[height =1.3cm]{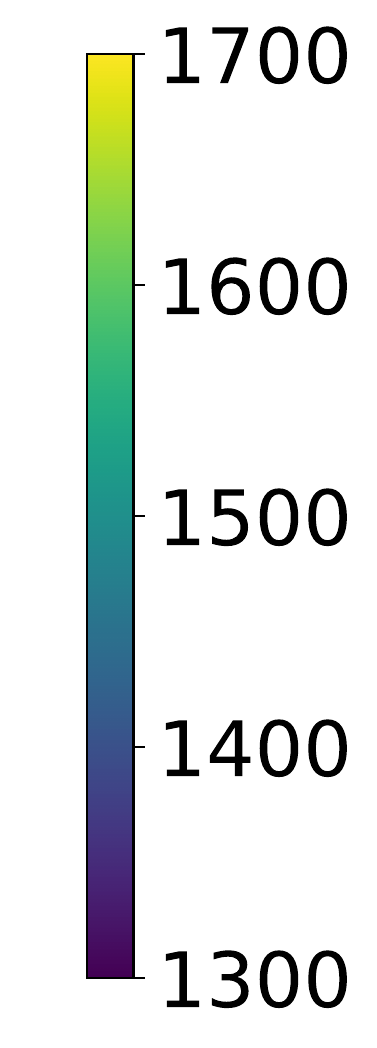}}\\

\raisebox{-.5\totalheight}{\scriptsize \rot {\makecell{{Predicted} \\ {{SoS}}  }}} & 
\tikzmark{top left 1}\raisebox{-.5\totalheight}{\includegraphics[width = 1.3cm]{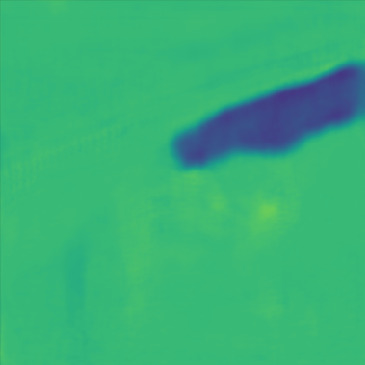}}&
\raisebox{-.5\totalheight}{\includegraphics[width = 1.3cm]{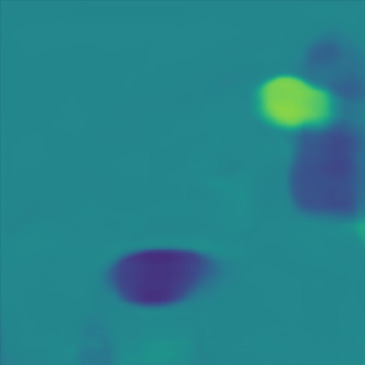}}&
\raisebox{-.5\totalheight}{\includegraphics[width = 1.3cm]{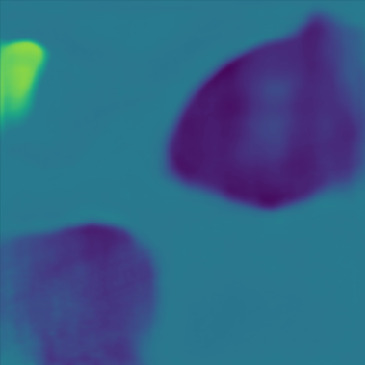}}&
\raisebox{-.5\totalheight}{\includegraphics[width = 1.3cm]{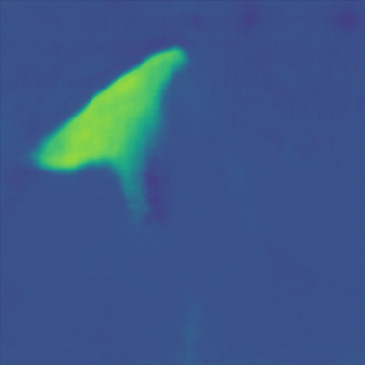}}&
\raisebox{-.5\totalheight}{\includegraphics[width = 1.3cm]{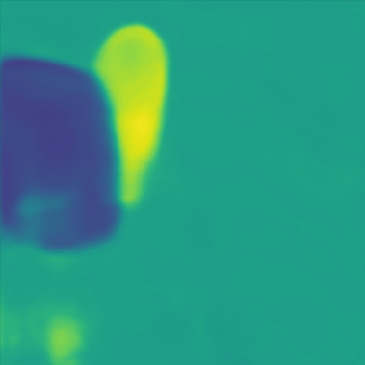}}& 
\tikzmark{top left 2}\raisebox{-.5\totalheight}{\includegraphics[width = 1.3cm]{img/Tomos/111_GT.jpg}}&
\raisebox{-.5\totalheight}{\includegraphics[width = 1.3cm]{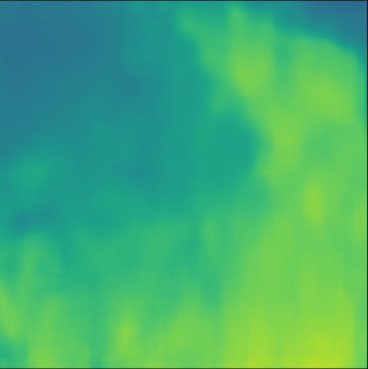}}&
\raisebox{-.5\totalheight}{\includegraphics[width = 1.3cm]{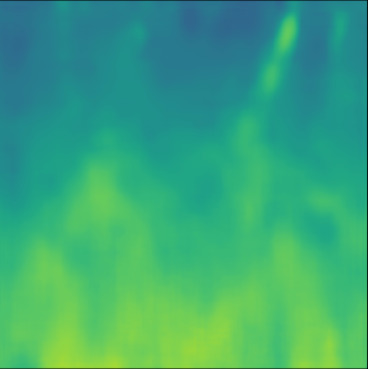}}&
\raisebox{-.5\totalheight}{\includegraphics[width = 1.3cm]{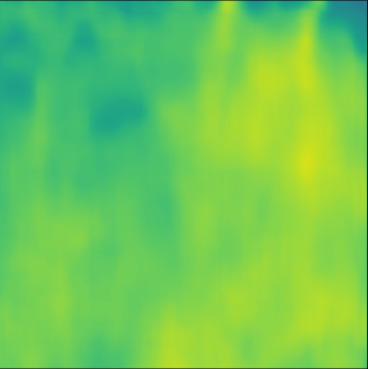}}&
\raisebox{-.5\totalheight}{\includegraphics[width = 1.3cm]{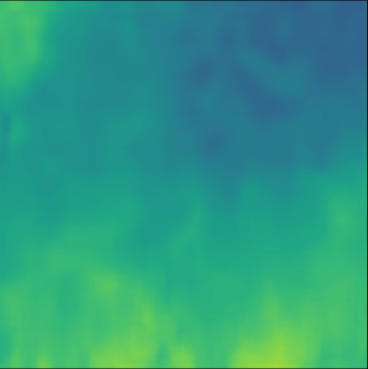}}&
\raisebox{-.5\totalheight}{\includegraphics[height =1.3cm]{img/colormaps/colorbar_simulated_nolable.pdf}}\\

\raisebox{-.5\totalheight}{\scriptsize \rot {\makecell{{Absolute} \\ {Difference} }}} & 
\raisebox{-.5\totalheight}{\includegraphics[width = 1.3cm]{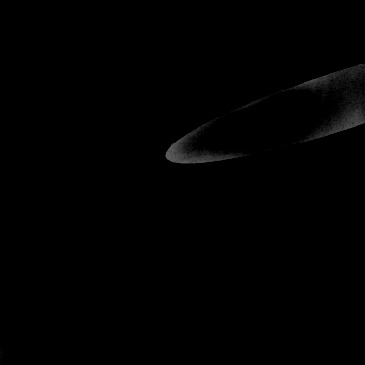}}&
\raisebox{-.5\totalheight}{\includegraphics[width = 1.3cm]{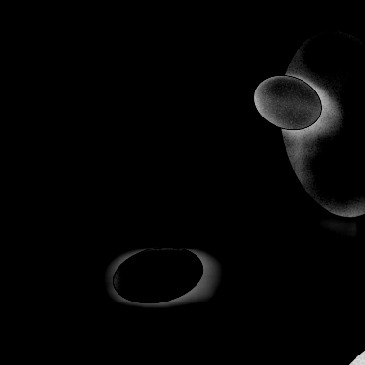}}&
\raisebox{-.5\totalheight}{\includegraphics[width = 1.3cm]{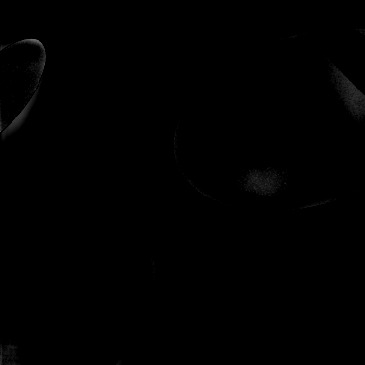}}&
\raisebox{-.5\totalheight}{\includegraphics[width = 1.3cm]{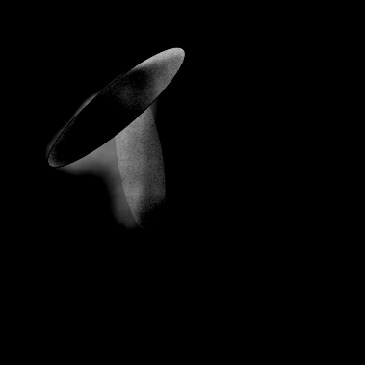}}&
\raisebox{-.5\totalheight}{\includegraphics[width = 1.3cm]{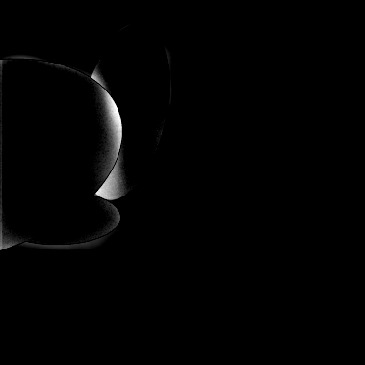}}&
\raisebox{-.5\totalheight}{\includegraphics[width = 1.3cm]{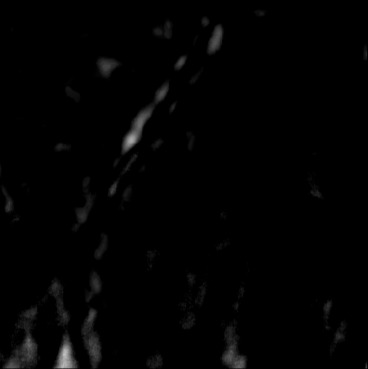}}&
\raisebox{-.5\totalheight}{\includegraphics[width = 1.3cm]{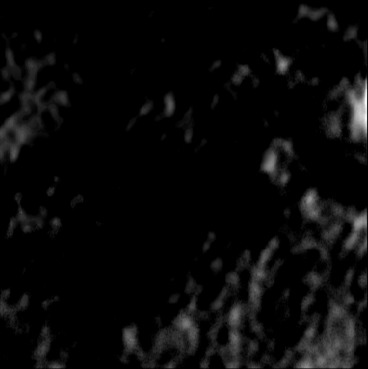}}&
\raisebox{-.5\totalheight}{\includegraphics[width = 1.3cm]{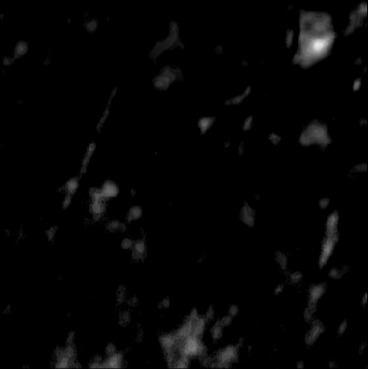}}&
\raisebox{-.5\totalheight}{\includegraphics[width = 1.3cm]{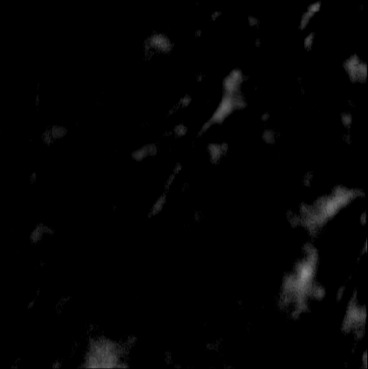}}&
\raisebox{-.5\totalheight}{\includegraphics[width = 1.3cm]{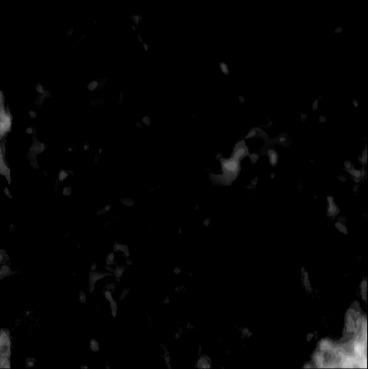}}&
\raisebox{-.5\totalheight}{\includegraphics[height =1.3cm]{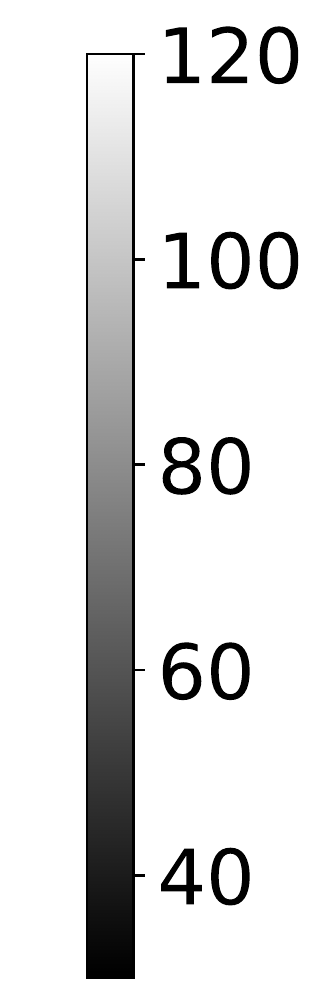}}\\

\addlinespace[1ex]
{\scriptsize RMSE} & 
{\scriptsize 5.50} &  {\scriptsize 13.45} &  {\scriptsize 11.26} & {\scriptsize 9.83} &  {\scriptsize 15.48} & \scriptsize {21.46} & \scriptsize {29.29} & \scriptsize {23.93} & \scriptsize {31.73} & \scriptsize {21.36} & \scriptsize[m/s]  \\ 
{\scriptsize MAE}  & 
\scriptsize 2.17 & \scriptsize 5.43 & \scriptsize 6.44 & \scriptsize 4.00 & \scriptsize 6.89 & \scriptsize 16.12 & \scriptsize 22.20 & \scriptsize 18.83 & \scriptsize 25.33 & \scriptsize 16.68 & \scriptsize[m/s] \\ 
\scriptsize SSIM & 
\scriptsize 0.89 & \scriptsize 0.82 & \scriptsize 0.72 & \scriptsize 0.85 & \scriptsize 0.80 \tikzmark{bottom right 1} &\scriptsize 0.73 & \scriptsize 0.71 & \scriptsize 0.64 & \scriptsize 0.65 &  \scriptsize 0.72 \tikzmark{bottom right 2}  \\  


\addlinespace[2ex]
\raisebox{-.5\totalheight}{\scriptsize \rot {\makecell{{Predicted} \\ {{SoS}}  }}} & 
\tikzmark{top left 3}\raisebox{-.5\totalheight}{\includegraphics[width = 1.3cm]{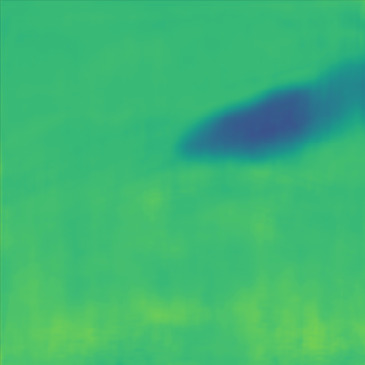}}&
 \raisebox{-.5\totalheight}{\includegraphics[width = 1.3cm]{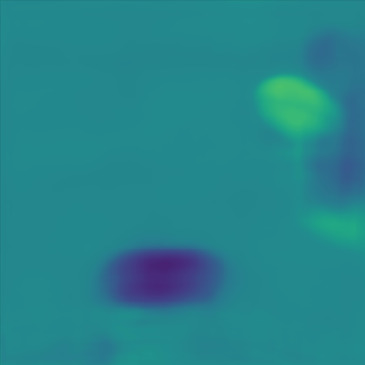}}&
 \raisebox{-.5\totalheight}{\includegraphics[width = 1.3cm]{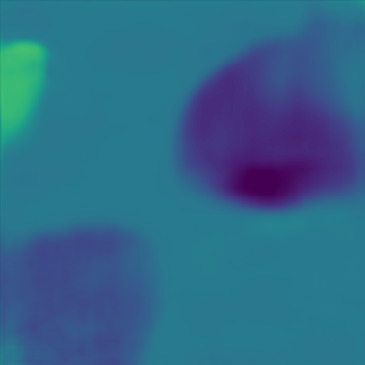}}&
 \raisebox{-.5\totalheight}{\includegraphics[width = 1.3cm]{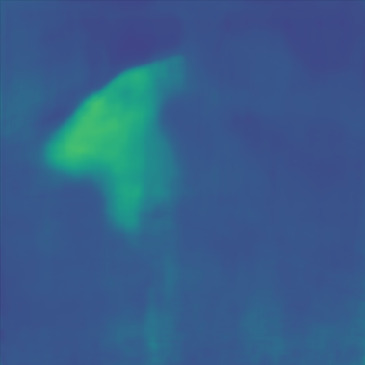}}&
 \raisebox{-.5\totalheight}{\includegraphics[width = 1.3cm]{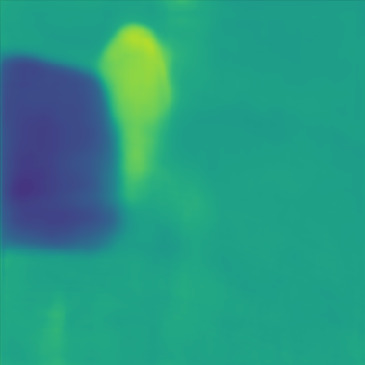}}&
  \raisebox{-.5\totalheight}{\includegraphics[width = 1.3cm]{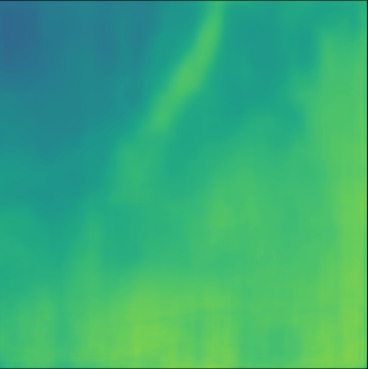}}&
 \raisebox{-.5\totalheight}{\includegraphics[width = 1.3cm]{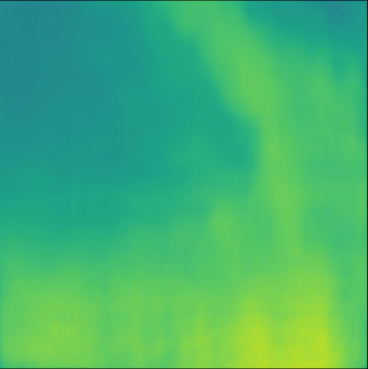}}&
 \raisebox{-.5\totalheight}{\includegraphics[width = 1.3cm]{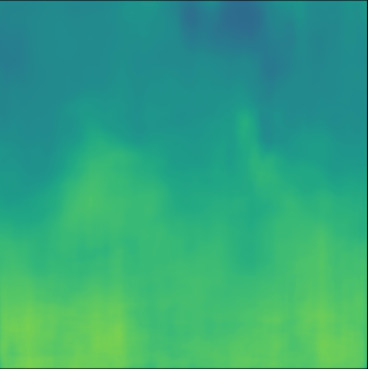}}&
 \raisebox{-.5\totalheight}{\includegraphics[width = 1.3cm]{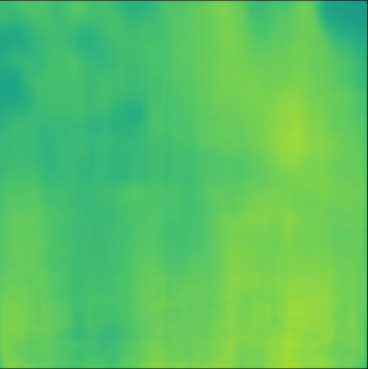}}&
 \raisebox{-.5\totalheight}{\includegraphics[width =1.3cm]{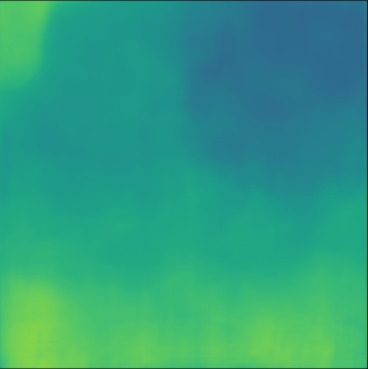}}&
\raisebox{-.5\totalheight}{\includegraphics[height =1.3cm]{img/colormaps/colorbar_simulated_nolable.pdf}}\\

\raisebox{-.5\totalheight}{\scriptsize \rot {\makecell{{Absolute} \\ {Difference} }}} & 
 \raisebox{-.5\totalheight}{\includegraphics[width = 1.3cm]{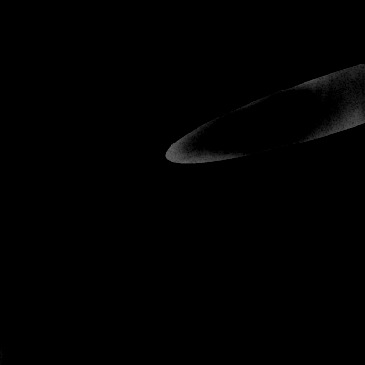}}&
\raisebox{-.5\totalheight}{\includegraphics[width = 1.3cm]{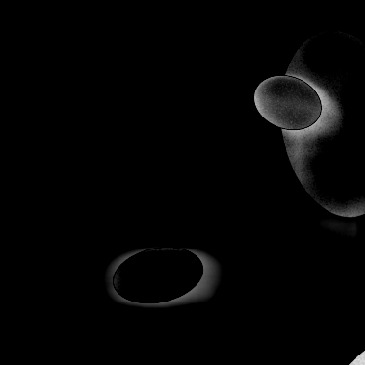}}&
 \raisebox{-.5\totalheight}{\includegraphics[width = 1.3cm]{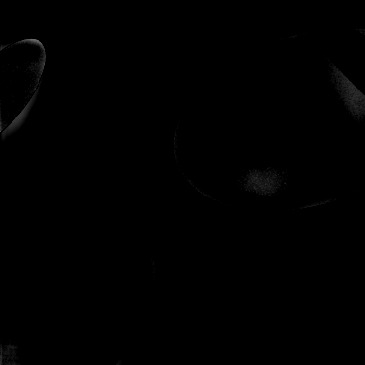}}&
 \raisebox{-.5\totalheight}{\includegraphics[width = 1.3cm]{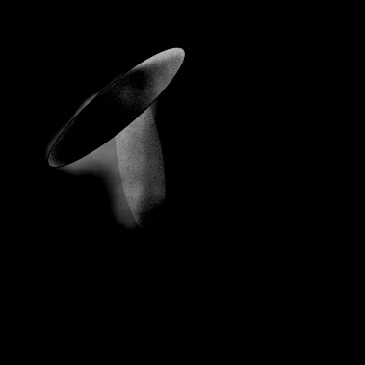}}&
\raisebox{-.5\totalheight}{\includegraphics[width = 1.3cm]{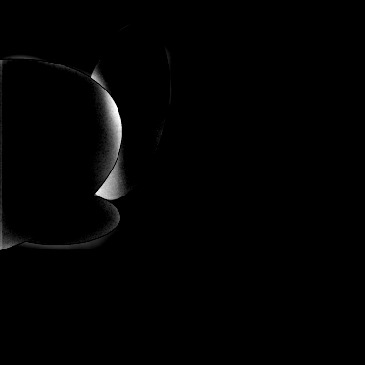}}&
\raisebox{-.5\totalheight}{\includegraphics[width = 1.3cm]{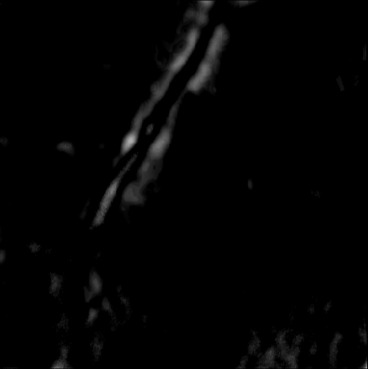}}&
 \raisebox{-.5\totalheight}{\includegraphics[width = 1.3cm]{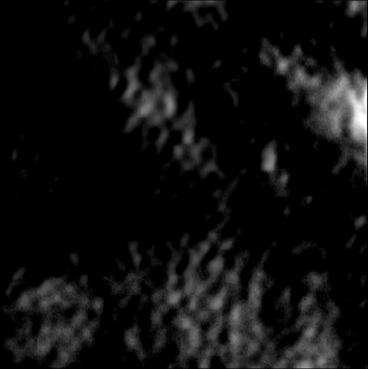}}&
 \raisebox{-.5\totalheight}{\includegraphics[width = 1.3cm]{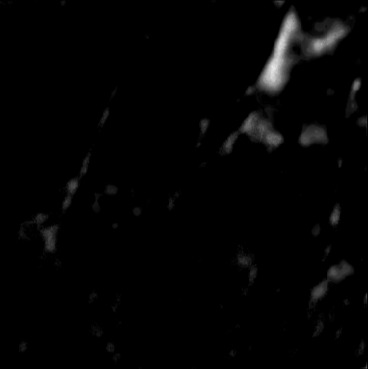}}&
 \raisebox{-.5\totalheight}{\includegraphics[width = 1.3cm]{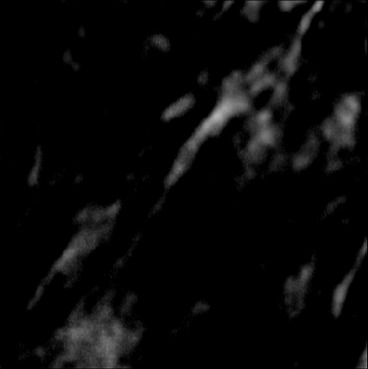}}&
\raisebox{-.5\totalheight}{\includegraphics[width = 1.3cm]{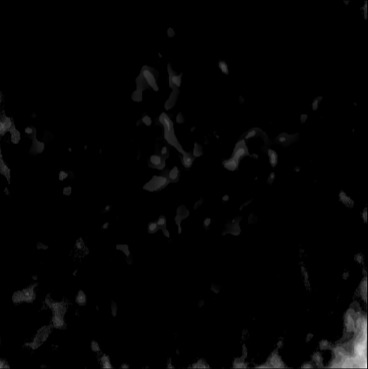}}&
\raisebox{-.5\totalheight}{\includegraphics[height =1.3cm]{img/colormaps/colorbar_diff_nolable.pdf}}\\
\addlinespace[1ex]
{\scriptsize RMSE} & 
\scriptsize 8.69 & \scriptsize 16.16 & \scriptsize 10.87 & \scriptsize 15.78 & \scriptsize 13.57 & \scriptsize 17.49 & \scriptsize 34.35 & \scriptsize 21.25 & \scriptsize 28.63 & \scriptsize 20.45 & \scriptsize[m/s]  \\ 
{\scriptsize MAE}  & 
\scriptsize4.79 & \scriptsize7.48 & \scriptsize6.62 & \scriptsize8.08 & \scriptsize7.12 & \scriptsize13.58 & \scriptsize26.08 & \scriptsize15.02 & \scriptsize22.64 & \scriptsize15.84 & \scriptsize[m/s] \\ 
\scriptsize SSIM & 
\scriptsize0.88 & \scriptsize0.91 & \scriptsize0.84 & \scriptsize0.88 & \scriptsize0.89 & \scriptsize0.76 & \scriptsize0.73 & \scriptsize0.66 & \scriptsize0.70 & \scriptsize0.74 \tikzmark{bottom right 3}  & \\  

\end{tabular}

\DrawBox[ultra thick, draw=green, dotted]{top left 1}{bottom right 1}
\DrawBox[ultra thick, draw=blue, dotted]{top left 2}{bottom right 2}
\DrawBox[ultra thick, draw=red, dotted]{top left 3}{bottom right 3}

\caption{Predicted SoS maps of ten test cases where the network is trained with Ellipsoids (green box), T2US (blue box) or Combined training data (red box); 1st row: GT, 2nd and 7th row: Predicted SoS maps, 3rd and 8th row: Absolute error between GT and predicted SoS, row 4-6 and 9-11: corresponding RMSE, MAE and SSIM values for each case.}
\label{fig: Sos Prediction Combined}
\end{figure*}

\paragraph{Ellipsoids Setup}

In the quantitative evaluation, the network converges to average RMSE, MAE, and MAPE of $14.95\pm 0.47$~$m/s$ (mean~$\pm$~SD), $5.94\pm 0.5$~$m/s$ and $0.41\pm 0.04$\%, respectively (over 10 runs). SD is the standard deviation value.  
The typical RMSE for these kinds of simulated datasets tested on encoder-decoder networks are reported in the range $[18.9-22.4]$~$m/s$ \citep{feigin2019deep} and $[28.7-48]$~$m/s$ \citep{oh2021neural}. 

Figure~\ref{fig: Sos Prediction Combined}, green box, shows that the network can handle cases with single and separated multiple lesions (Cases  1,  2, and  3), overlapping lesions (Cases 4 and 5), and lesions placed partially inside another lesion (Case  2).
As the number of inclusions and the complexity of the medium increases the corresponding error rates increase. A deviation from the GT is present in all cases at the boundary of the lesion and background medium which can be seen as line distortions in the boundaries (Cases 1, 2, and 3) and brighter regions in absolute difference images (Cases 1, 2, 4 and 5). Note that for easier visual interpretation of SoS difference at the edges, only in absolute difference figures, errors lower than \(30\)~\(m/s\) are excluded \citep{feigin2019deep}. The network is more likely to over/underestimate values when the lesion is placed in the shadow of another lesion or in the case of overlapping lesions (Cases 4, 2, and 5).

\paragraph{T2US Setup}

The network trained with T2US dataset converges to average RMSE, MAE and MAPE of \(22.74 \pm 0.9\)~\(m/s\), \(16.80 \pm 0.6\)~\(m/s\) and \(1.09 \pm 0.03\)\%, respectively, over 10 runs (Table~\ref{table: errors}). 
Although these error rates are higher compared to the aforementioned Ellipsoids setup, they are still relevant, since they are in the range of the RMSE values reported by other groups that used simulated data for training (RMSE of $[18.9-22.4]$~$m/s$ \citep{feigin2019deep} and  $[28.7-48]$~$m/s$ \citep{oh2021neural}). 

Figure~\ref{fig: Sos Prediction Combined}, blue box, demonstrates qualitative results. 
The network can handle the gradual variation in the medium even when the boundaries are not clear (Cases  7 and 10).
Additionally, the network can reconstruct some of the fine structures (Cases  6 and  8). 
However, in some cases, when the variation between the SoS values for the fine structures and the background is low some of the fine structures are missed by the network (Cases 7 and 9).

\paragraph{Combined Setup}

First of all, we tested the trained networks with their dissimilar test set, meaning, the network trained on the Ellipsoids setup is tested on the T2US test set, and the network trained on the T2US setup is tested on the Ellipsoids test set. 
Although networks trained with each set of simulated data were capable of performing well on their corresponding test dataset, testing them using the dissimilar test set resulted in failure and high RMSE and MAE values shown in Table~\ref{table: RMSE on cross simulation}. 
Therefore, as expected the network only generalizes on the same distribution on which it is trained.

\begin{table}[tbp]
\renewcommand{\arraystretch}{1.3}
\caption{RMSE and MAE for predicted SoS maps on each test dataset trained on the dissimilar training set, SD refers to the standard deviation for each reported value for multiple runs.}
\label{table: RMSE on cross simulation}
\centering
\begin{tabular}{l|c|c } 
 \hline
 \rowcolor{lightgray!20}
 \textbf{RMSE (\(m/s\)) } & \textbf{Ellipsoids (train)} & \textbf{T2US (train)} \\ %
  \hline
 \textbf{Ellipsoids (test) } & \(14.95 \pm 0.47\) & \(59.61 \pm 3.15 \)  \\
 \textbf{T2US (test)} & \(140.38 \pm 6.80\) & \(22.74 \pm 0.90\)  \\
 \hline
 \rowcolor{lightgray!20}
 \textbf{MAE (\(m/s\)) } & \textbf{Ellipsoids (train)} & \textbf{T2US (train)} \\ %
  \hline
 \textbf{Ellipsoids (test) } &  \(5.94\pm 0.58\)  & \(45.14 \pm 2.78 \) \\
 \textbf{T2US (test)} & \(111.45 \pm 4.6\) & \(16.80\pm0.60\) \\

\end{tabular}
\end{table}


The aim of the study is to improve the performance and stability of the existing approach by including more complicated, tissue-like structures. 
Therefore, both sets are needed to be presented to the network during the training phase. 
We have chosen a multi-task learning approach to train the network on both datasets jointly. Multi-task learning is an approach where the data from task A and task B are interleaved so the weights can be jointly optimized \citep{kirkpatrick2017overcoming}.



The test dataset as well as the training dataset is a half-half combination of Ellipsoids and T2US setups.
Using the Combined dataset as shown in Table~\ref{table: errors}, the network converges to mean RMSE of \(20.96 \pm 0.80\)~\(m/s\), MAE of \(14.03 \pm 1.33\)~\(m/s\) and MAPE of \(0.81 \pm 0.11\)\% (over 10 runs). 
The error rates infer that the network is jointly optimized on both datasets and is capable of performing on both data representations. 

Figure~\ref{fig: Sos Prediction Combined}, red box, shows 10 predictions from the network trained with the Combined setup. 
For the shown cases in Figure~\ref{fig: Sos Prediction Combined}, some of the cases show improvements (Cases 3, 5, 6, and 8) but on average, in comparison to singles sets, the errors are slightly higher when trained on the Combined setup. 
Nevertheless, the network can still handle multiple inclusions (Cases 2, 3, and 5), overlapping inclusions (Cases 2, 4, and 5), gradual variations (Cases  7 and  10) and  
even reconstruct fine structures (Case 6). 

\paragraph{Clinical Relevance}\label{clinical} Based on the study by \cite{li2009vivo}, fatty tissue and breast parenchyma have mean SoS values of $1422\pm9$~$m/s$ and $1487\pm21$~$m/s$, respectively. 
Whereas the mean SoS for lesions is higher, e.g., for malignant breast lesions the mean value is $1548\pm17$~$m/s$, and for benign lesions is $1513\pm27$~$m/s$.
\cite{ruby2019breast} demonstrated that breast lesions create SoS contrast ($\Delta$SoS) compared to the background tissue. They reported $\Delta$SoS in range $[14-118]$~$m/s$ and $[7-41]$~m/s for malignant and benign lesions, respectively. Malignant lesions show significantly higher $\Delta$SoS compared to benign lesions $\Delta SoS > 41.64 $~$m/s$. Therefore, theoretically, the achieved error measured shown in Table~\ref{table: errors} are clinically relevant.

\subsubsection{Out-of-Domain Data}

\begin{figure}[!t]
\centering
\includegraphics[trim={0cm 14cm 5cm 0cm},clip,scale=0.35]{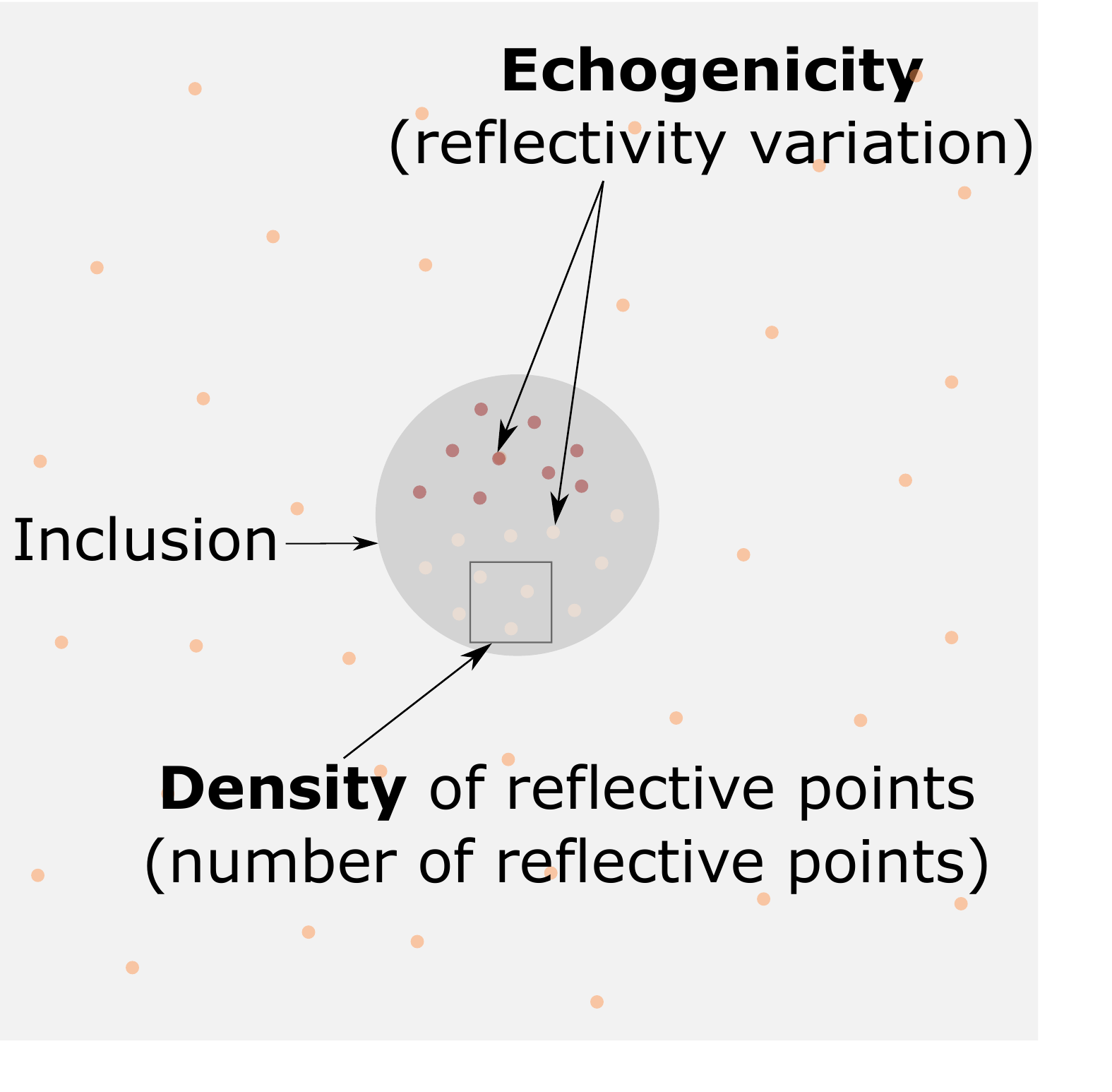}
\caption{SoS domain of a sample medium setup with an embedded inclusion (shown as a circle with the darker gray color in the center of the medium). SoS contrast is shown by different gray shades. The dots inside the inclusion and in the background represent the scatterers. The scatterers can have higher (red), equal (orange), or lower (pink) echogenicity compared to the background. The number of scatterers is called the density of the reflective points. The inclusion can have echogenicity contrast and/or SoS contrast and/or density of scatterers contrast compared to the background.} 
\label{fig: SoS_domain, inclusion}
\end{figure} 

In section~\ref{sec: training}, we showed that the network trained with the Combined dataset can learn both representations. Nevertheless, the network can be sensitive to the variation of acoustic parameters, e.g., echogenicity, the density of the scatterers, noise, and geometry. A comprehensive analysis of the sensitivity of the network to these parameters is currently missing from the literature that can provide insights to improve the quality of the simulated training data for future studies. Additionally, we argue that the diverse training set, i.e., combining the baseline dataset (Ellipsoids setup) with our proposed setup (T2US) improves the stability of the network. Thus, in this section, we test the networks trained with the Ellipsoids dataset and the Combined dataset on simulated test sets that have different properties compared to the training dataset, e.g., echogenicity, the density of reflective scatterers, noise characteristics, and geometry.

An example of a simulation medium is shown in Figure \ref{fig: SoS_domain, inclusion}. Consider a medium with a homogeneous background with scatterers randomly distributed inside the medium. 
Different echogenicities can be created by increasing or decreasing the impedance contrast of the small scatterers or speckles included in the heterogeneity (pink and red dots inside the inclusion in Figure~\ref{fig: SoS_domain, inclusion}).
The anechoic region can be modeled by removing the speckles inside the inclusion. 
The following properties are kept constant through the simulation of the training sets:
\begin{enumerate*}[label=(\roman*)]
\item Echogenicity: only hyperechogenic regions are modeled (by using speckles with a higher standard deviation (i.e., more echoic) than the background for \(10\)\% of the scatterers). 
\item Density of scatterers: the number of scatterers shown as 'Density of reflective points' in Figure~\ref{fig: SoS_domain, inclusion} is kept constant and equal to \(10\)\% of the grid points for all training sets. 
\end{enumerate*}
In the following section, these parameters are altered to investigate the stability of the trained networks.

\begin{figure}[!t]
\centerline{
\subfloat{\includegraphics[scale=0.3]{ 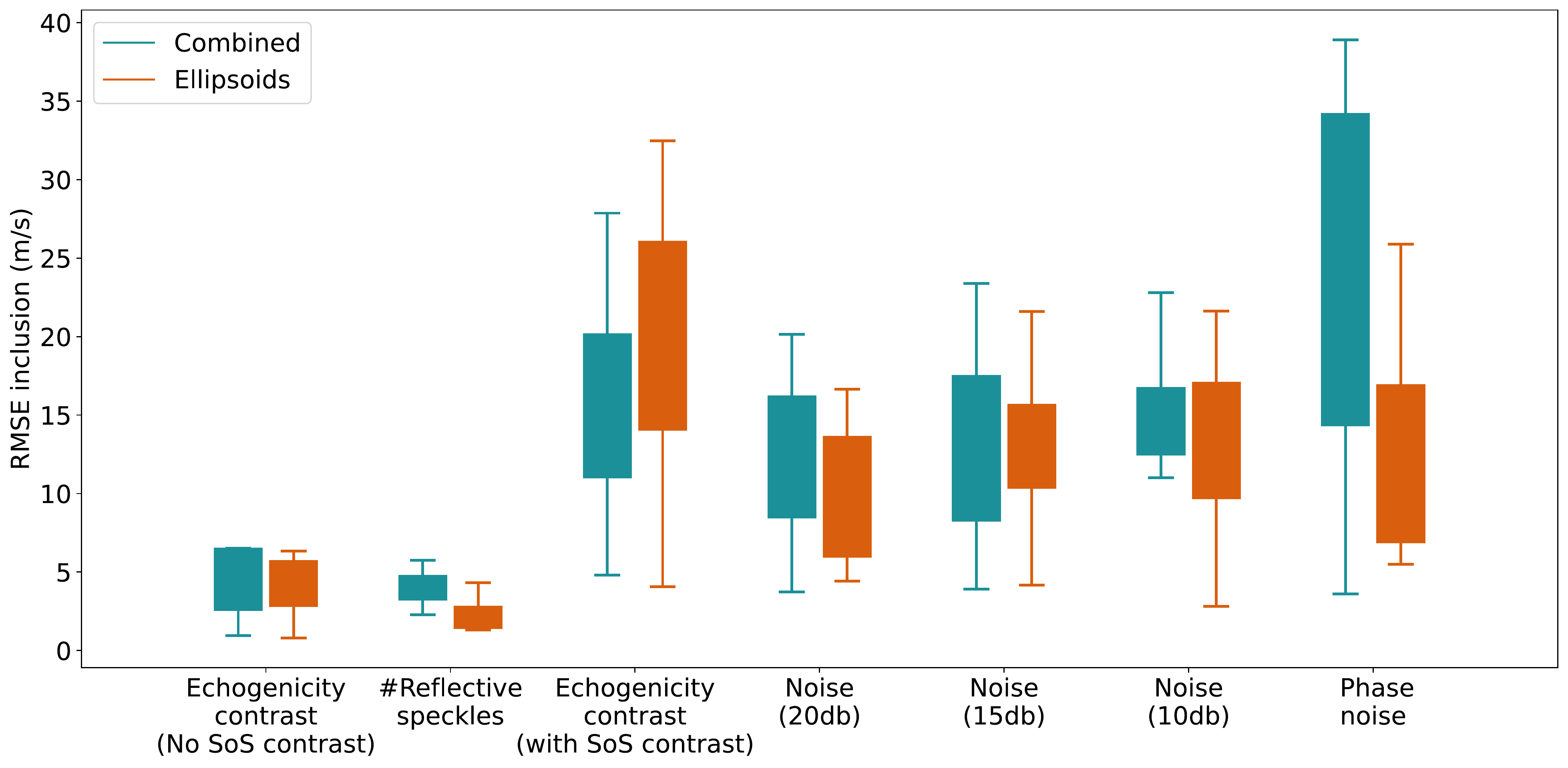}}} 
\centerline{
\subfloat{\includegraphics[scale=0.3]{ 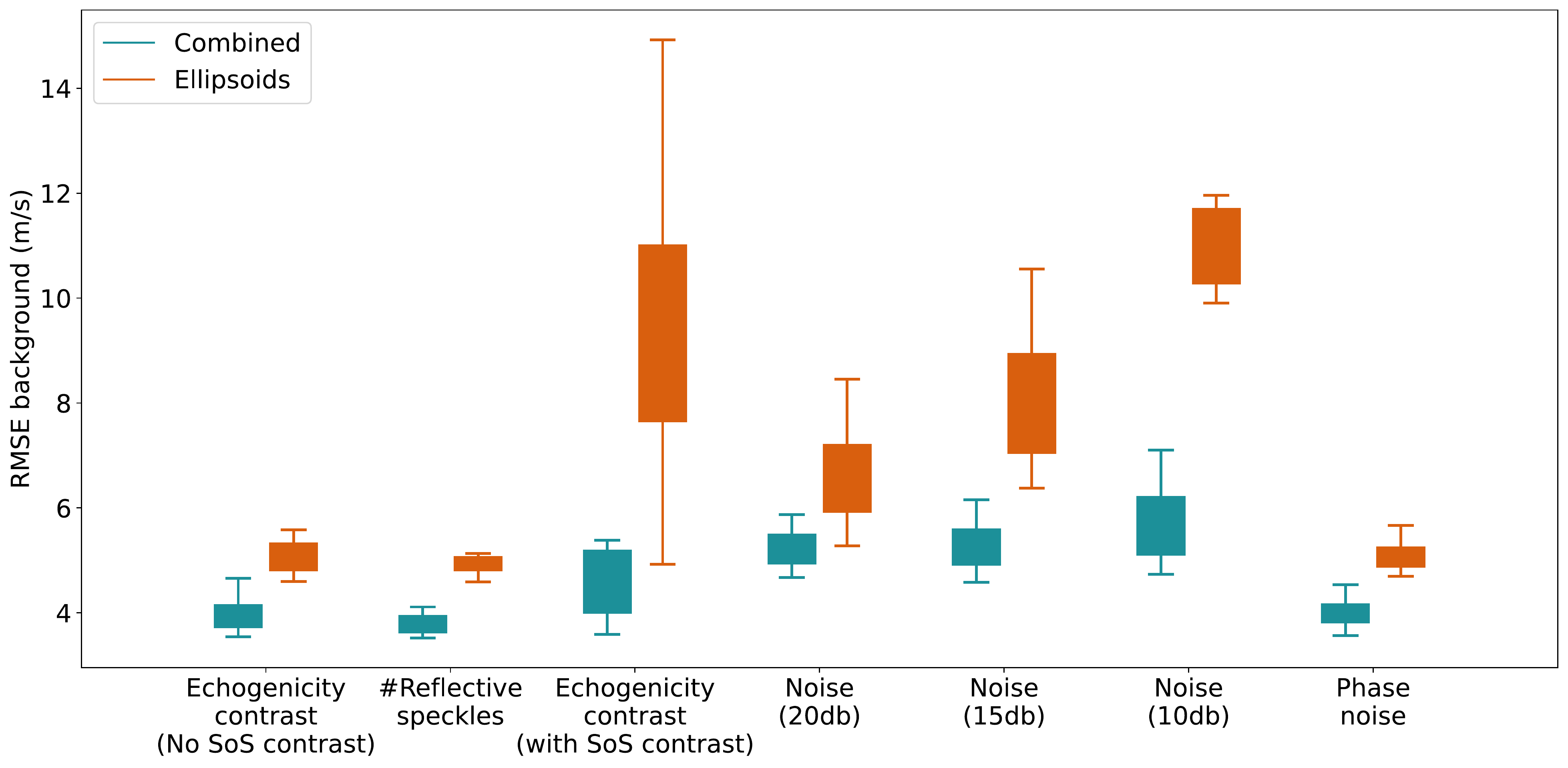}}}
\caption{(above): Comparison between RMSE values computed \underline{inside the inclusion} on out-of-domain simulated data for the network trained with the Combined dataset and the Ellipsoids dataset only, (below): Comparison between RMSE values computed \underline{outside the inclusion} on out-of-domain simulated data for the network trained with the Combined dataset and the Ellipsoids dataset only; indicating that both networks perform mostly similarly inside the inclusion area but the network trained with the Combined dataset performs better outside the inclusion area especially in the presence of noise, meaning, there are fewer over/underestimations outside the inclusion area for the network trained on the Combined dataset.}
\label{fig: rmse_in}
\end{figure}

For out-of-domain test sets we simulated a simple homogeneous medium with an inclusion placed in the central section of the medium but introduced variations in  echogenicity, the number of reflective scatters, noise compared to the training set, and computed RMSE of two networks inside the inclusion area and in the background. 
The results are presented in Figure~\ref{fig: rmse_in}.
Example cases of each variation scenario are shown in Figure~\ref{fig: digital phantom, ECHO NO sOs}, Figure~\ref{fig: digital phantom, SoS contrast} and Figure~\ref{fig: digital phantom, noise} for quantitative evaluations.

\paragraph{Echogenicity and Speckles}

\begin{figure*}[!t]
\centering 

\renewcommand{\arraystretch}{0.05}
\begin{tabular}{@{\hspace{0.5mm}} c @{\hspace{0.5mm}}c @{\hspace{0.5mm}}c @{\hspace{0.5mm}}c @ {\hspace{0.5mm}}c @{\hspace{0.5mm}}c  @{\hspace{0.5mm}}c  @{\hspace{0.5mm}}c  @{\hspace{0.5mm}}c @{\hspace{0.5mm}}l}
& \scriptsize {Case 1} & \scriptsize{Case 2} & \scriptsize{Case 3} & \scriptsize{Case 4} & \scriptsize{Case 5}  & \scriptsize{ Case  6} & \scriptsize{Case 7} & \scriptsize{Case 8}  & \\

\raisebox{-.5\totalheight}{\scriptsize \rot {B-mode}} & 
\raisebox{-.5\totalheight}{\includegraphics[trim={3cm 1.5cm 4cm 1cm},clip, width = 1.7cm]{ 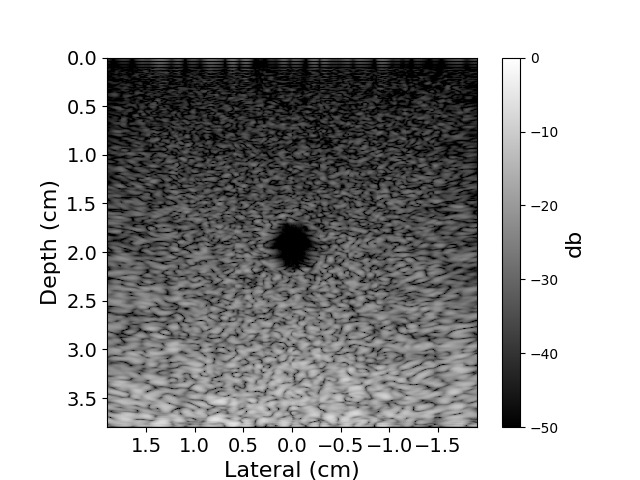}}&
\raisebox{-.5\totalheight}{\includegraphics[trim={3cm 1.5cm 4cm 1cm},clip, width = 1.7cm]{ 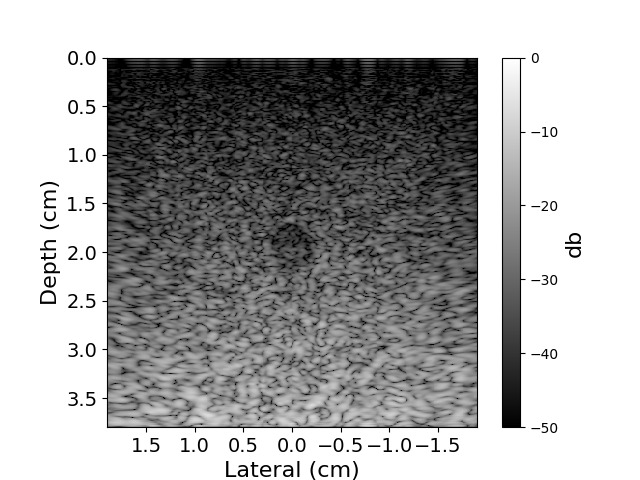}}&
\raisebox{-.5\totalheight}{\includegraphics[trim={3cm 1.5cm 4cm 1cm},clip, width = 1.7cm]{ 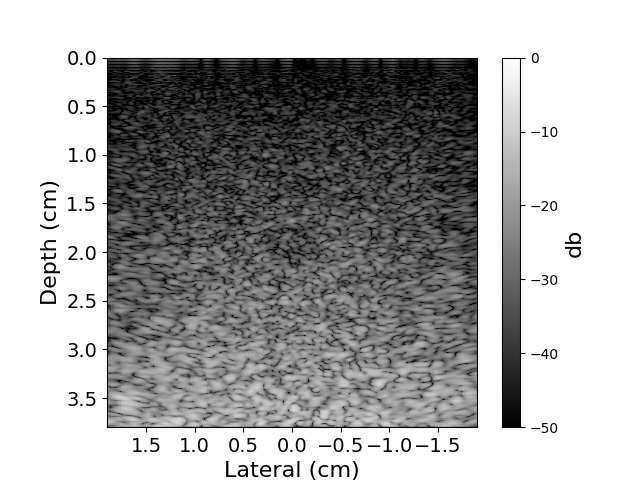}}&
\raisebox{-.5\totalheight}{\includegraphics[trim={3cm 1.5cm 4cm 1cm},clip, width = 1.7cm]{ 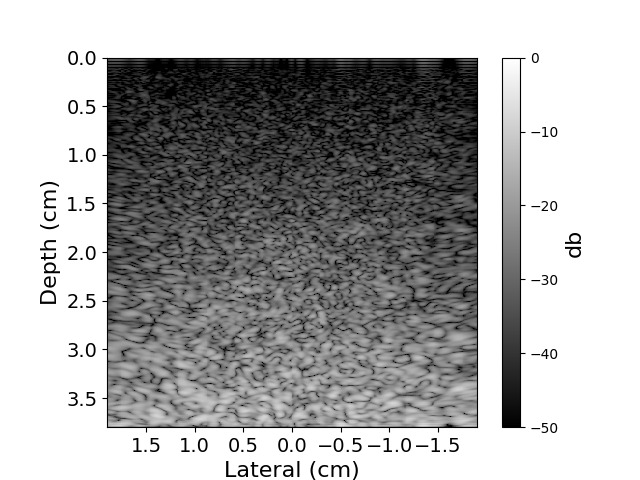}}&
\raisebox{-.5\totalheight}{\includegraphics[trim={3cm 1.5cm 4cm 1cm},clip, width = 1.7cm]{ 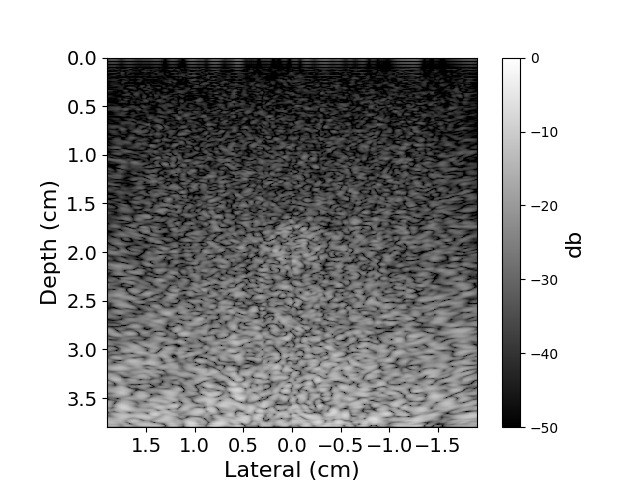}}&
\raisebox{-.5\totalheight}{\includegraphics[trim={3cm 1.5cm 4cm 1cm},clip, width =1.7cm]{ 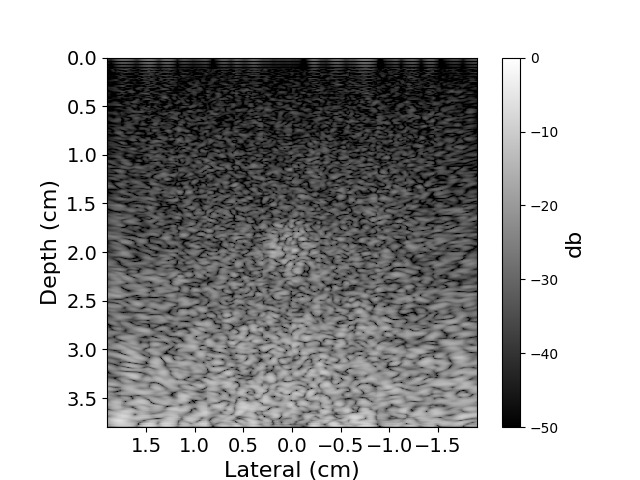}}&
\raisebox{-.5\totalheight}{\includegraphics[trim={3cm 1.5cm 4cm 1cm},clip, width =1.7cm]{ 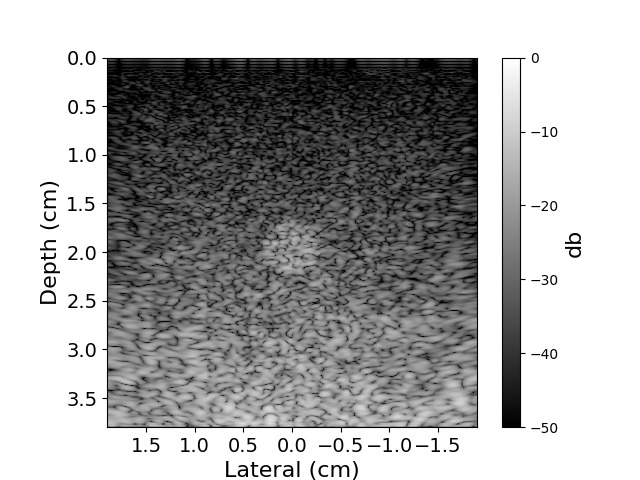}}&
\raisebox{-.5\totalheight}{\includegraphics[trim={3cm 1.5cm 4cm 1cm},clip, width =1.7cm]{ 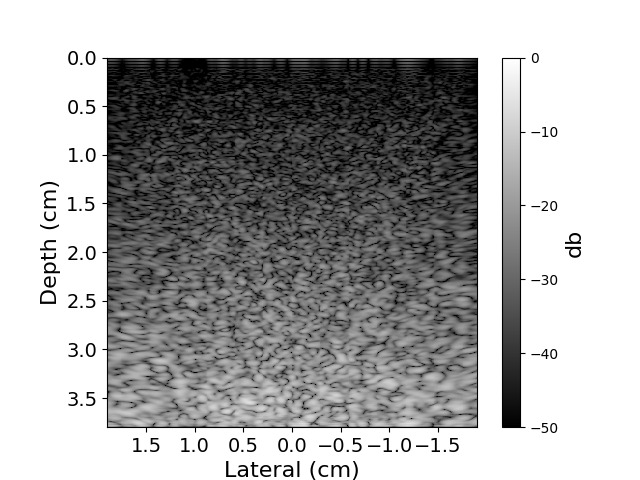}}&
\raisebox{-.5\totalheight}{\includegraphics[height =1.7cm]{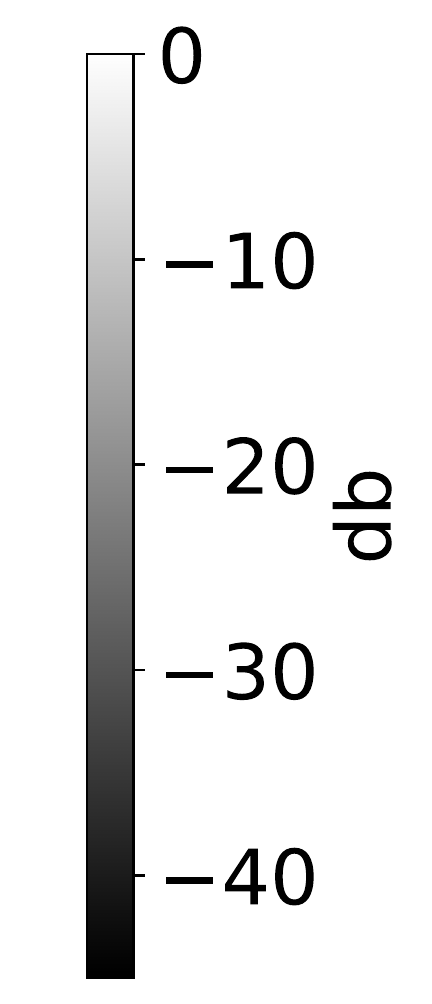}}\\
 
\raisebox{-.5\totalheight}{\scriptsize \rot{GT}} & 
\raisebox{-.5\totalheight}{\includegraphics[trim={3cm 1.5cm 4cm 1.5cm},clip, width = 1.7cm]{ 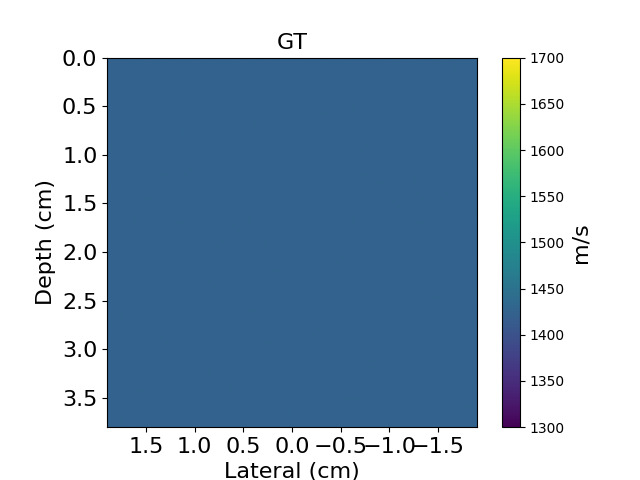}}&
\raisebox{-.5\totalheight}{\includegraphics[trim={3cm 1.5cm 4cm 1.5cm},clip, width = 1.7cm]{ 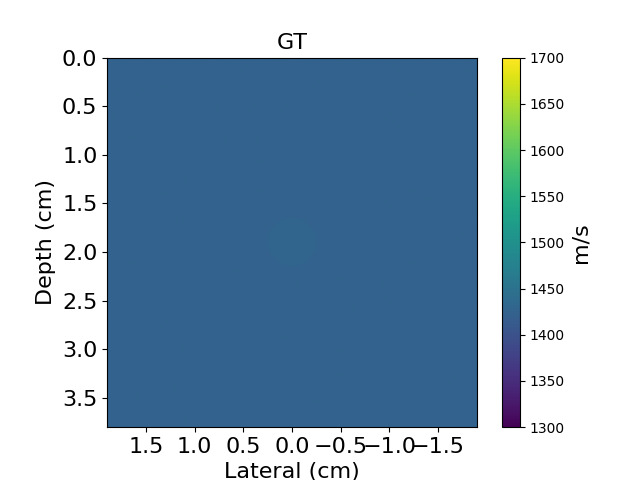}}&
\raisebox{-.5\totalheight}{\includegraphics[trim={3cm 1.5cm 4cm 1.5cm},clip, width = 1.7cm]{ 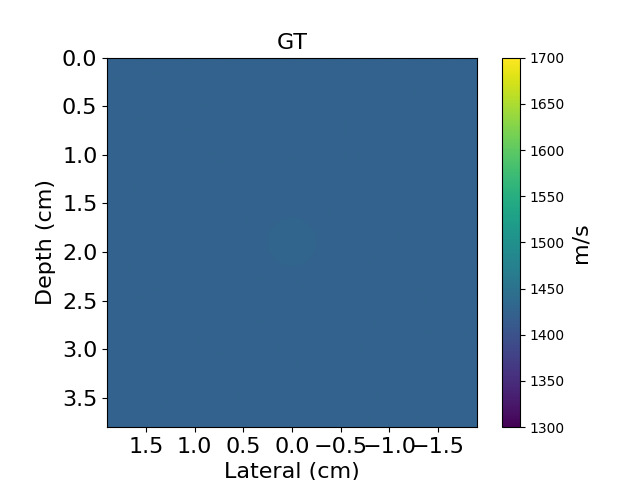}}&
\raisebox{-.5\totalheight}{\includegraphics[trim={3cm 1.5cm 4cm 1.5cm},clip, width = 1.7cm]{ 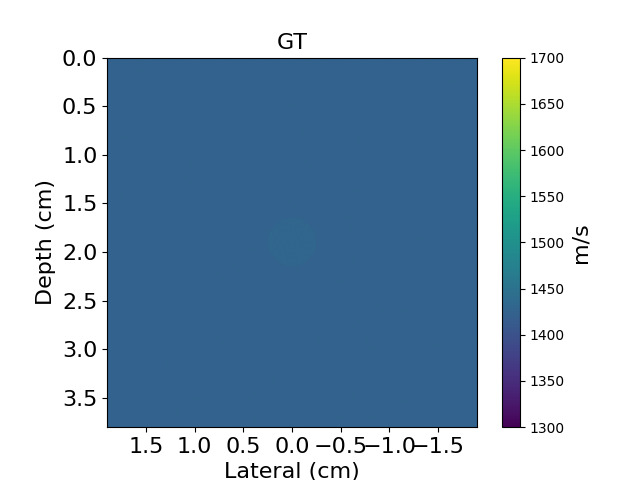}}&
\raisebox{-.5\totalheight}{\includegraphics[trim={3cm 1.5cm 4cm 1.5cm},clip, width = 1.7cm]{ 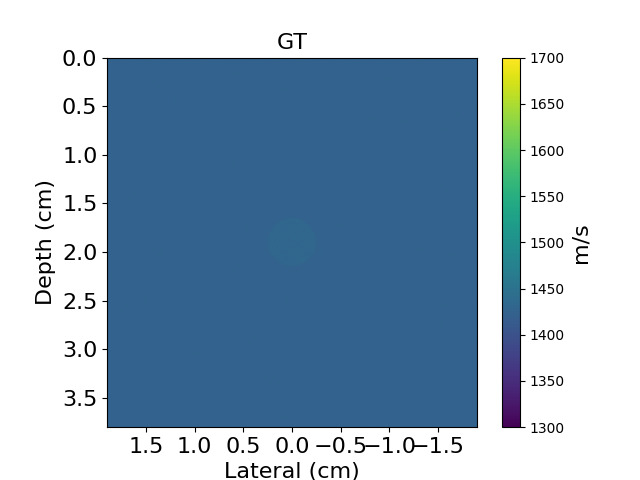}}&
\raisebox{-.5\totalheight}{\includegraphics[trim={3cm 1.5cm 4cm 1.5cm},clip, width =1.7cm]{ 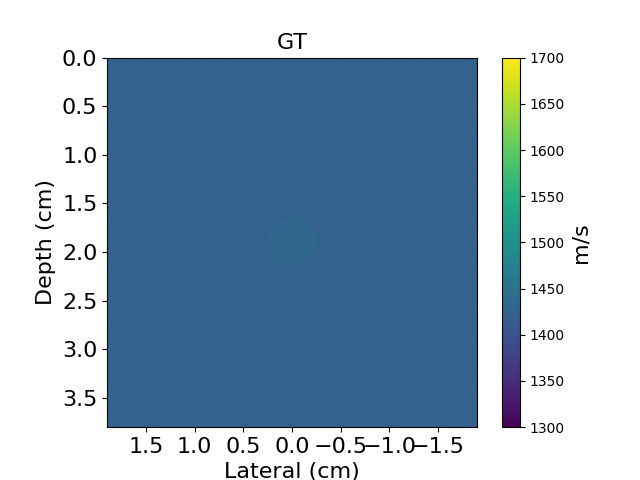}}&
\raisebox{-.5\totalheight}{\includegraphics[trim={3cm 1.5cm 4cm 1.5cm},clip, width =1.7cm]{ 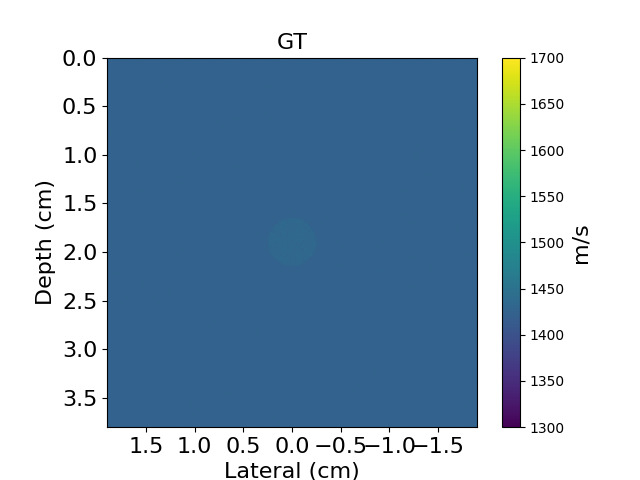}}&
\raisebox{-.5\totalheight}{\includegraphics[trim={3cm 1.5cm 4cm 1.5cm},clip, width =1.7cm]{ 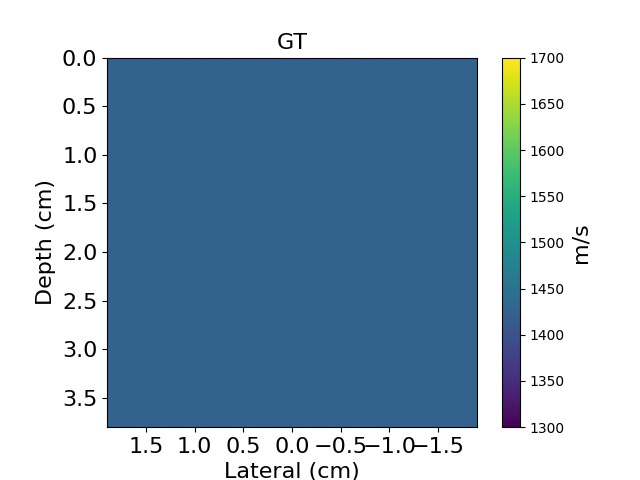}}&
\raisebox{-.5\totalheight}{\includegraphics[height =1.7cm]{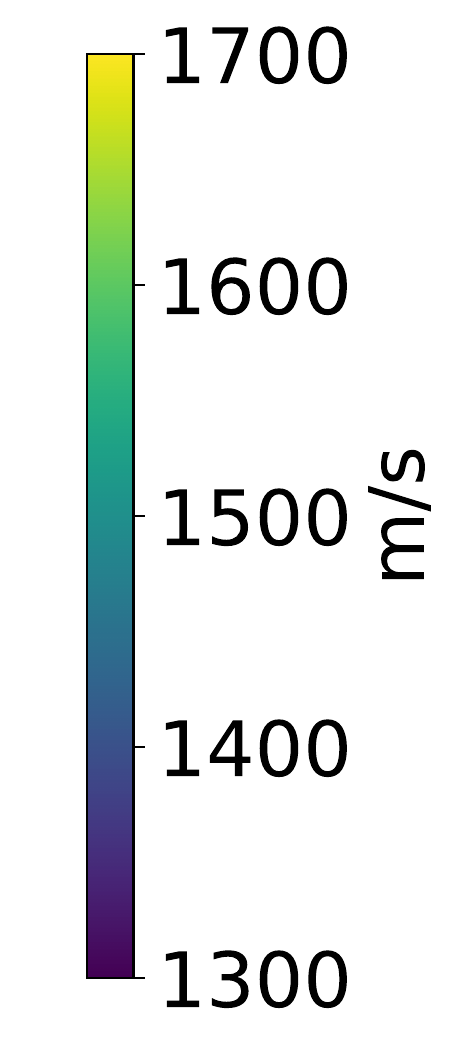}}\\

 \raisebox{-.5\totalheight}{\scriptsize \rot {\makecell{{Predicted} \\ {{SoS}} \\ {Combined} }}} & 
 \raisebox{-.5\totalheight}{\includegraphics[trim={3cm 1.5cm 4cm 1.5cm},clip, width = 1.7cm]{ 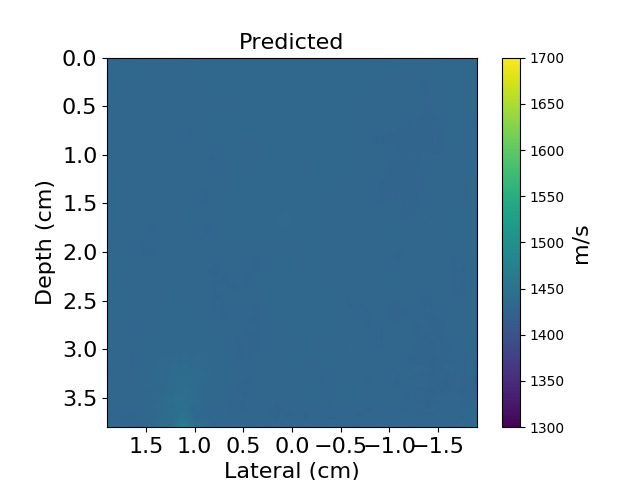}}&
 \raisebox{-.5\totalheight}{\includegraphics[trim={3cm 1.5cm 4cm 1.5cm},clip, width = 1.7cm]{ 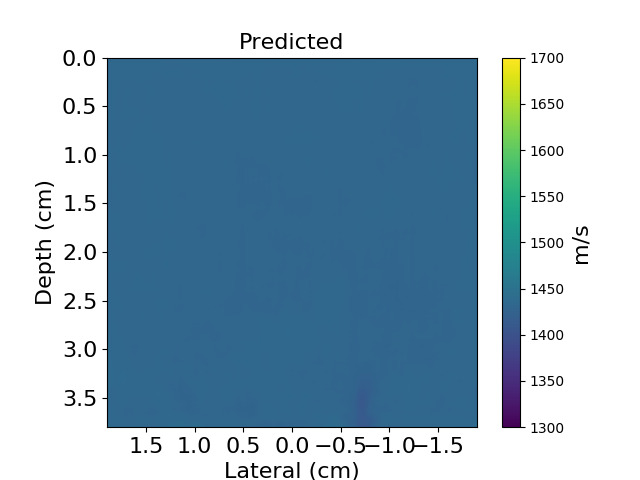}}&
 \raisebox{-.5\totalheight}{\includegraphics[trim={3cm 1.5cm 4cm 1.5cm},clip, width = 1.7cm]{ 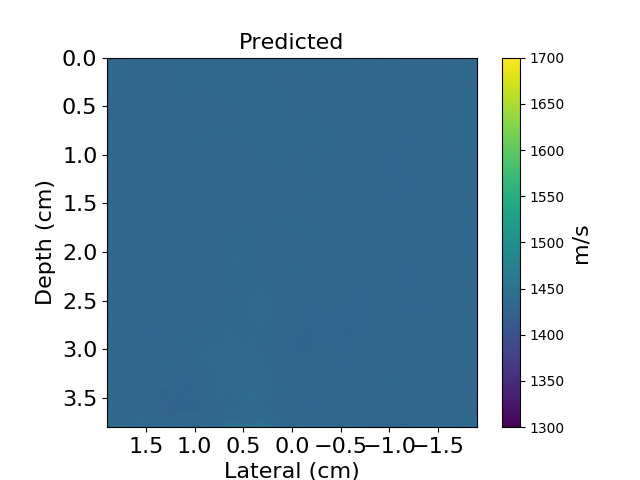}}&
 \raisebox{-.5\totalheight}{\includegraphics[trim={3cm 1.5cm 4cm 1.5cm},clip, width = 1.7cm]{ 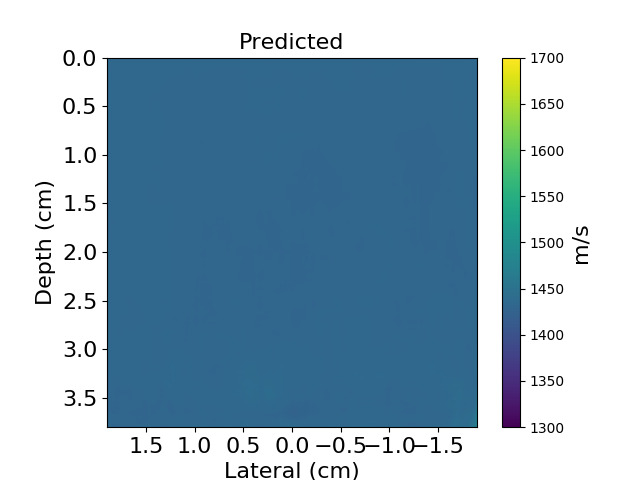}}&
 \raisebox{-.5\totalheight}{\includegraphics[trim={3cm 1.5cm 4cm 1.5cm},clip, width = 1.7cm]{ 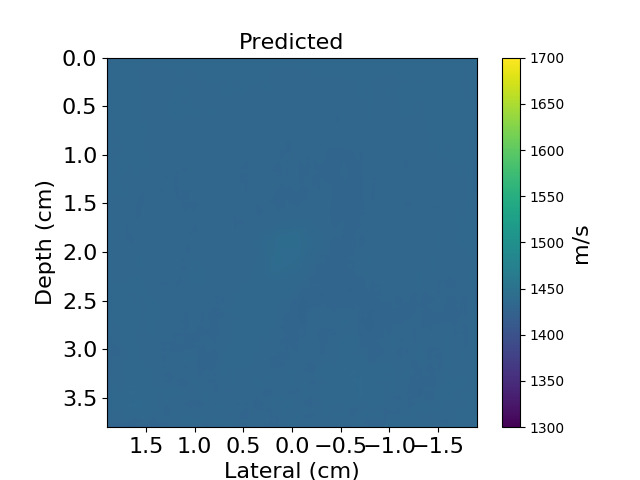}}&
 \raisebox{-.5\totalheight}{\includegraphics[trim={3cm 1.5cm 4cm 1.5cm},clip, width =1.7cm]{ 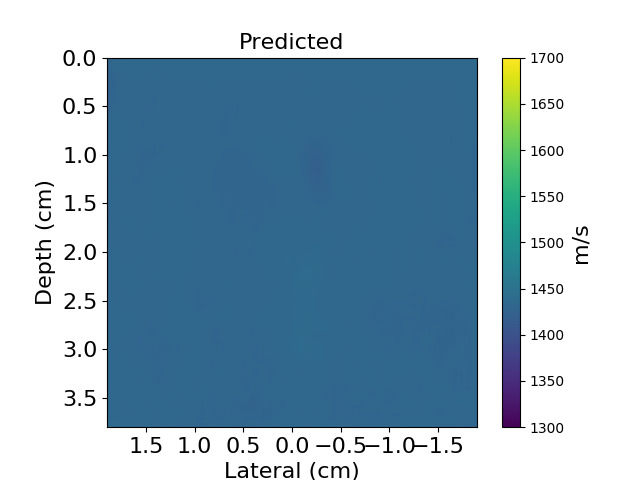}}&
 \raisebox{-.5\totalheight}{\includegraphics[trim={3cm 1.5cm 4cm 1.5cm},clip, width =1.7cm]{ 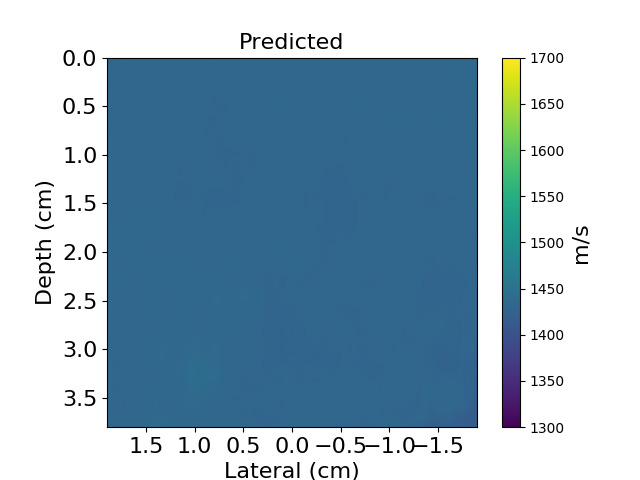}}&
 \raisebox{-.5\totalheight}{\includegraphics[trim={3cm 1.5cm 4cm 1.5cm},clip, width =1.7cm]{ 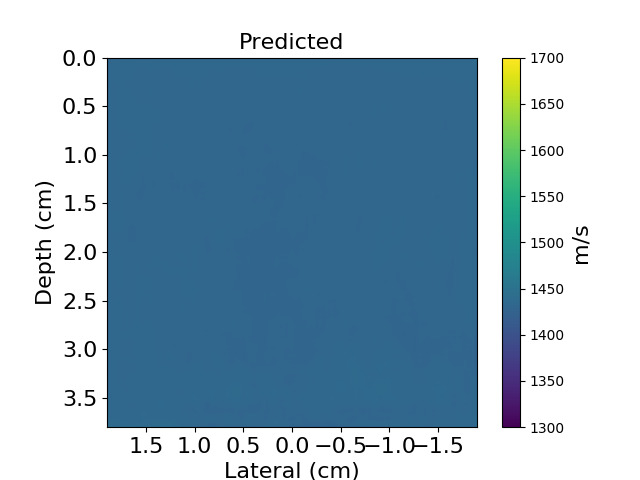}}&
\raisebox{-.5\totalheight}{\includegraphics[height =1.7cm]{img/colormaps/colorbar_simulated.pdf}}\\
 
  \raisebox{-.5\totalheight}{\scriptsize \rot{ \makecell{{Absolute} \\ {{Difference}} }}} & 
 \raisebox{-.5\totalheight}{\includegraphics[trim={3cm 1.5cm 4cm 1.5cm},clip, width = 1.7cm]{ 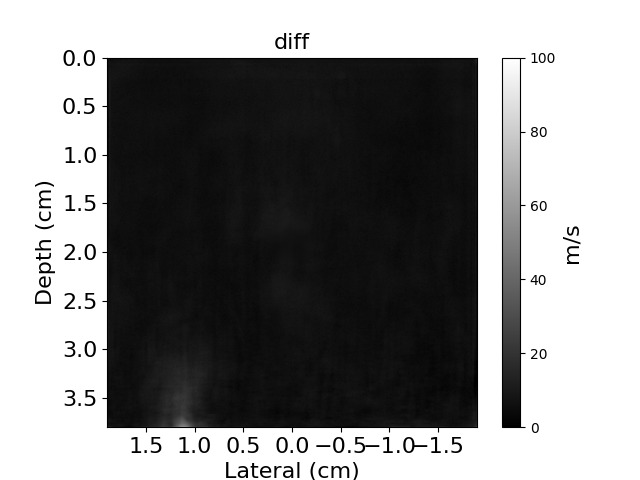}}&
 \raisebox{-.5\totalheight}{\includegraphics[trim={3cm 1.5cm 4cm 1.5cm},clip, width = 1.7cm]{ 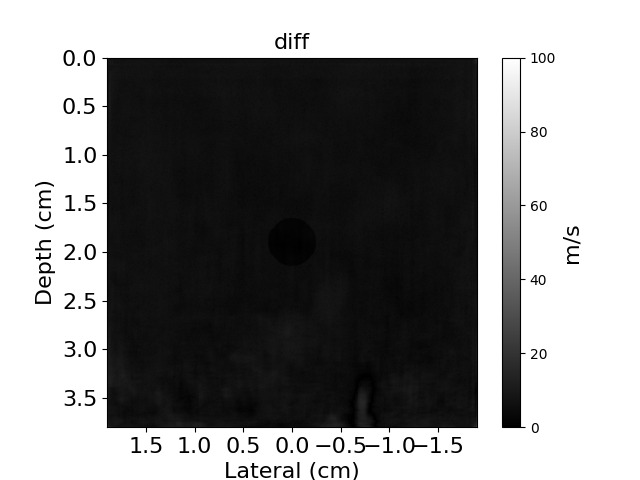}}&
 \raisebox{-.5\totalheight}{\includegraphics[trim={3cm 1.5cm 4cm 1.5cm},clip, width = 1.7cm]{ 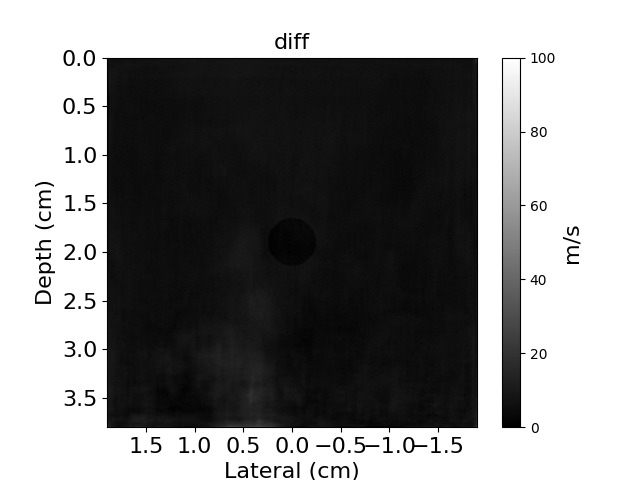}}&
 \raisebox{-.5\totalheight}{\includegraphics[trim={3cm 1.5cm 4cm 1.5cm},clip, width = 1.7cm]{ 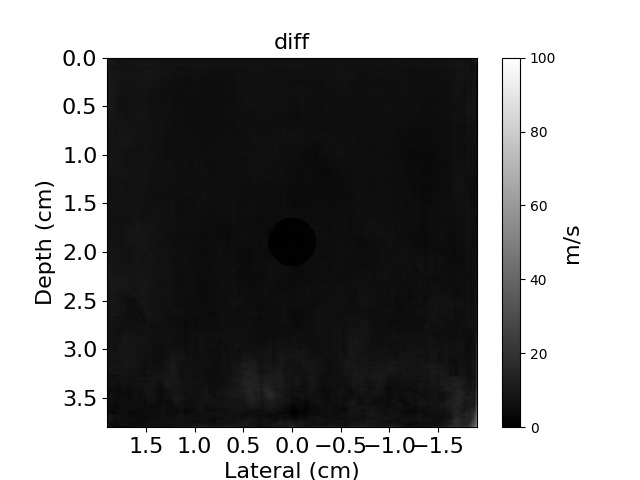}}&
 \raisebox{-.5\totalheight}{\includegraphics[trim={3cm 1.5cm 4cm 1.5cm},clip, width = 1.7cm]{ 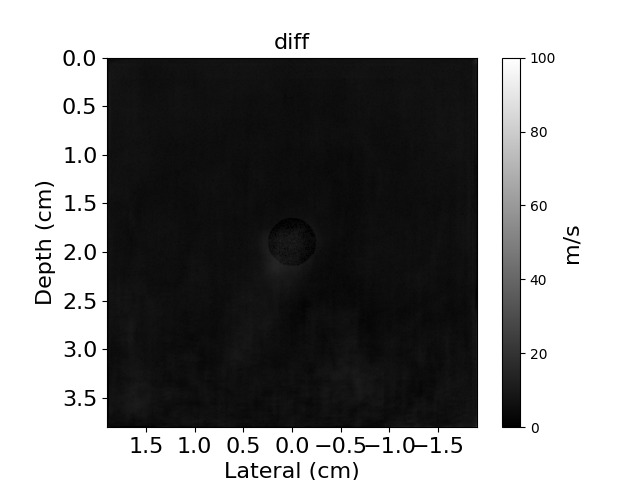}}&
 \raisebox{-.5\totalheight}{\includegraphics[trim={3cm 1.5cm 4cm 1.5cm},clip, width =1.7cm]{ 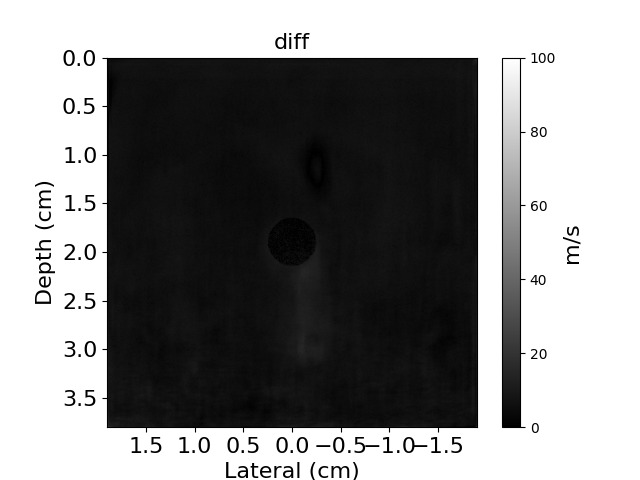}}&
 \raisebox{-.5\totalheight}{\includegraphics[trim={3cm 1.5cm 4cm 1.5cm},clip, width =1.7cm]{ 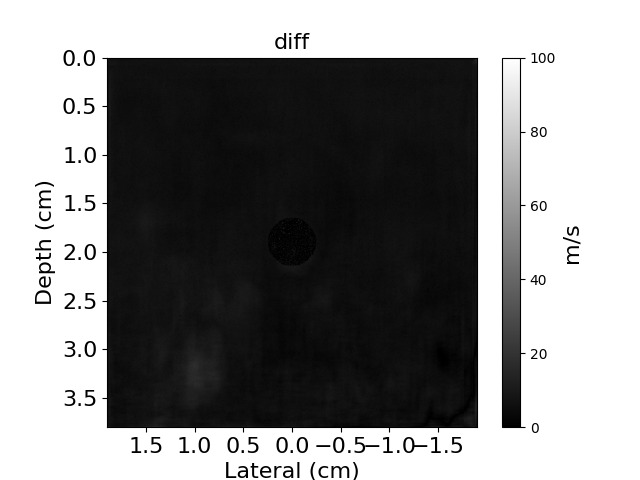}}&
 \raisebox{-.5\totalheight}{\includegraphics[trim={3cm 1.5cm 4cm 1.5cm},clip, width =1.7cm]{ 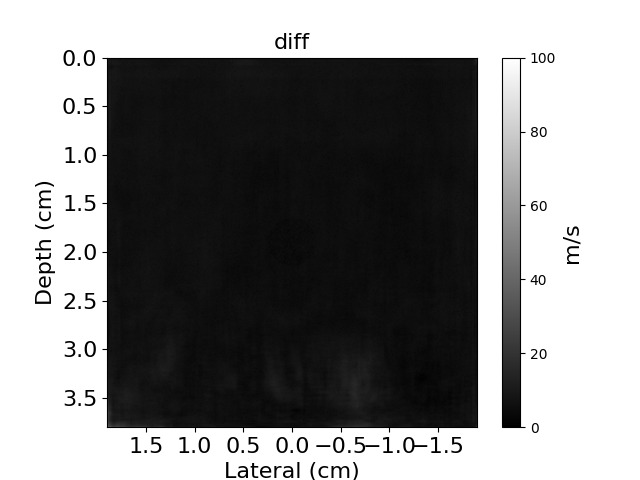}}&
\raisebox{-.5\totalheight}{\includegraphics[height =1.7cm]{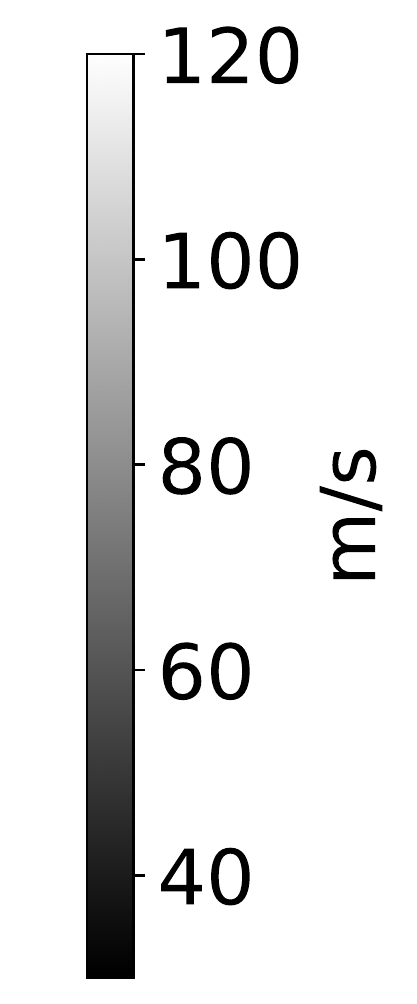}}\\

 \raisebox{-.5\totalheight}{\scriptsize \rot{ \makecell{{Predicted} \\ {{SoS}} \\ {Ellipsoids} }}} & 
 \raisebox{-.5\totalheight}{\includegraphics[trim={3cm 1.5cm 4cm 1.5cm},clip, width = 1.7cm]{ 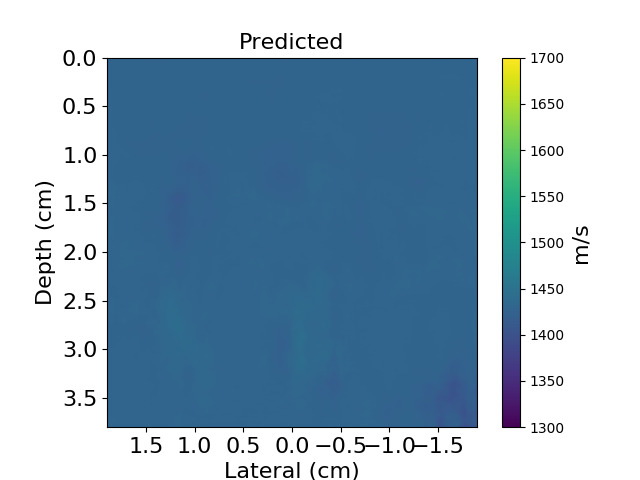}}&
 \raisebox{-.5\totalheight}{\includegraphics[trim={3cm 1.5cm 4cm 1.5cm},clip, width = 1.7cm]{ 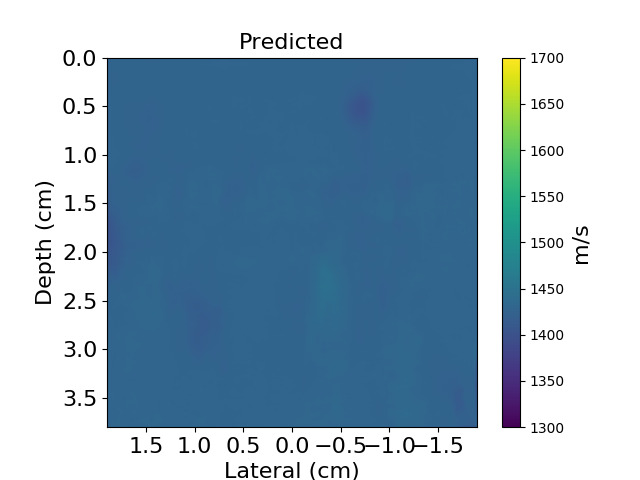}}&
 \raisebox{-.5\totalheight}{\includegraphics[trim={3cm 1.5cm 4cm 1.5cm},clip, width = 1.7cm]{ 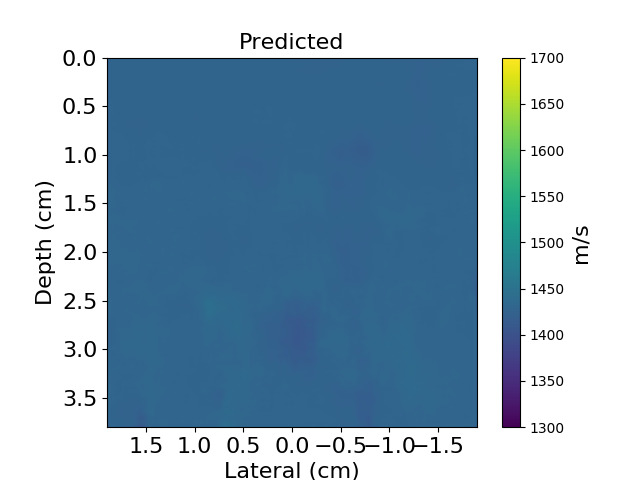}}&
 \raisebox{-.5\totalheight}{\includegraphics[trim={3cm 1.5cm 4cm 1.5cm},clip, width = 1.7cm]{ 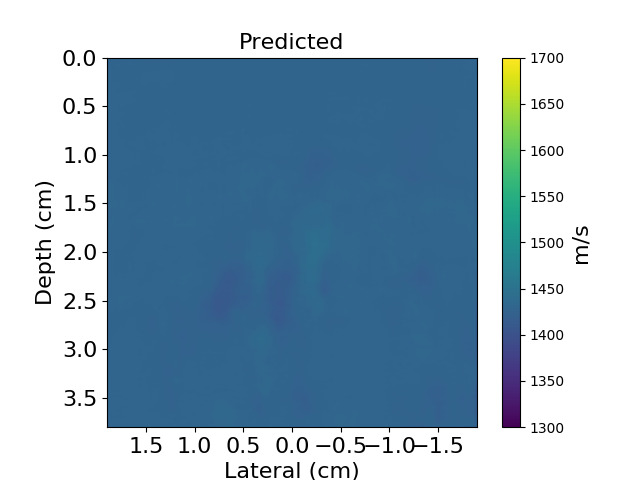}}&
 \raisebox{-.5\totalheight}{\includegraphics[trim={3cm 1.5cm 4cm 1.5cm},clip, width = 1.7cm]{ 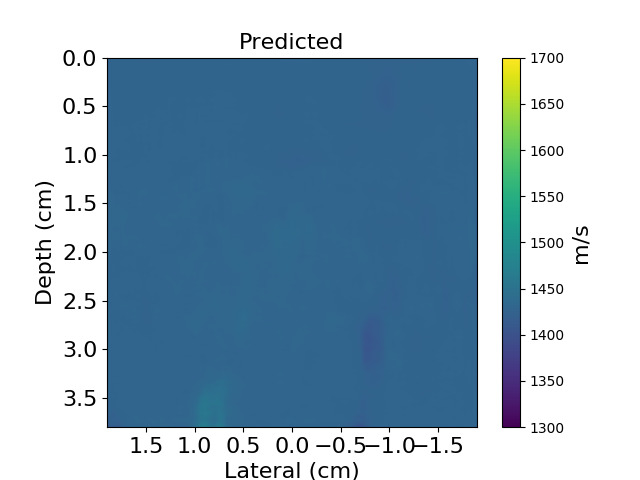}}&
 \raisebox{-.5\totalheight}{\includegraphics[trim={3cm 1.5cm 4cm 1.5cm},clip, width =1.7cm]{ 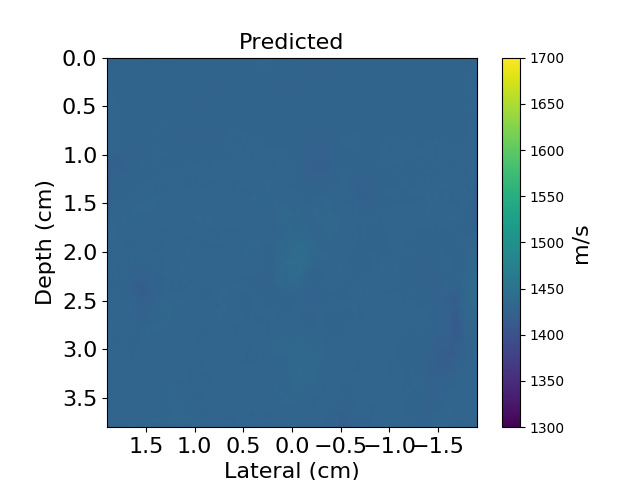}}&
 \raisebox{-.5\totalheight}{\includegraphics[trim={3cm 1.5cm 4cm 1.5cm},clip, width =1.7cm]{ 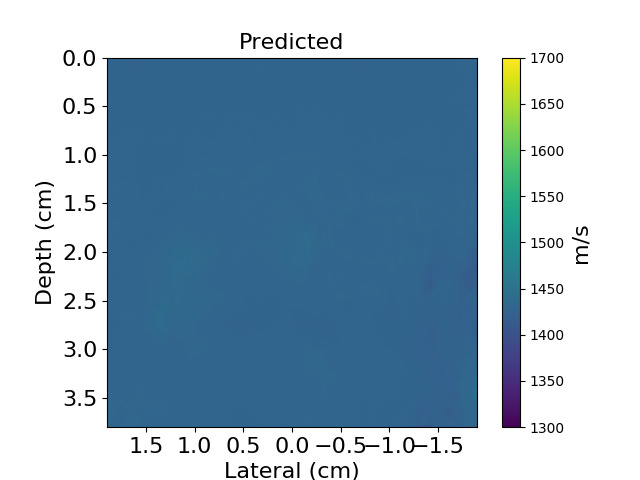}}&
 \raisebox{-.5\totalheight}{\includegraphics[trim={3cm 1.5cm 4cm 1.5cm},clip, width =1.7cm]{ 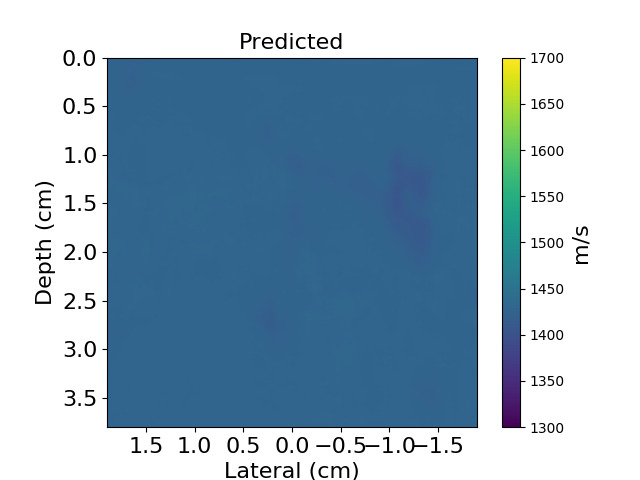}}&
\raisebox{-.5\totalheight}{\includegraphics[height =1.7cm]{img/colormaps/colorbar_simulated.pdf}}\\
 
\raisebox{-.5\totalheight}{\scriptsize \rot{\makecell{{Absolute} \\  {Difference} }}}& 
\raisebox{-.5\totalheight}{\includegraphics[trim={3cm 1.5cm 4cm 1.5cm},clip, width = 1.7cm]{ 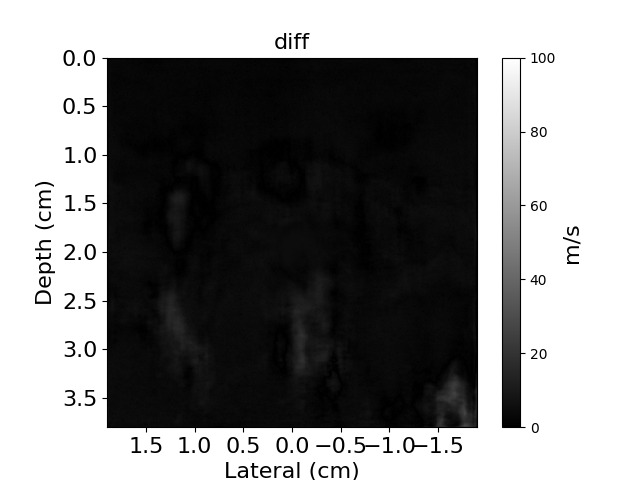}}&
 \raisebox{-.5\totalheight}{\includegraphics[trim={3cm 1.5cm 4cm 1.5cm},clip, width = 1.7cm]{ 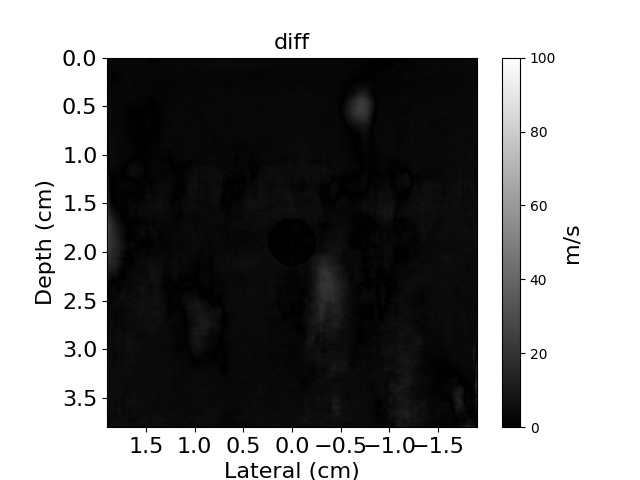}}&
 \raisebox{-.5\totalheight}{\includegraphics[trim={3cm 1.5cm 4cm 1.5cm},clip, width = 1.7cm]{ 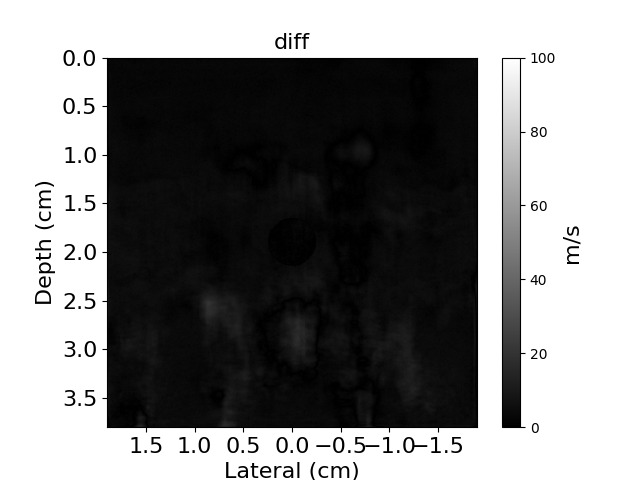}}&
 \raisebox{-.5\totalheight}{\includegraphics[trim={3cm 1.5cm 4cm 1.5cm},clip, width = 1.7cm]{ 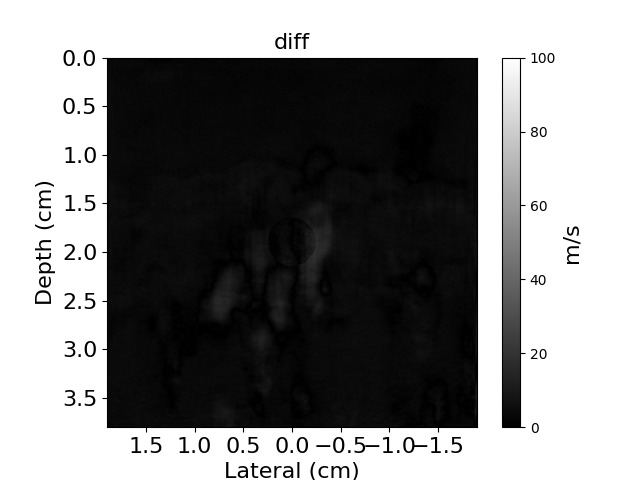}}&
 \raisebox{-.5\totalheight}{\includegraphics[trim={3cm 1.5cm 4cm 1.5cm},clip, width = 1.7cm]{ 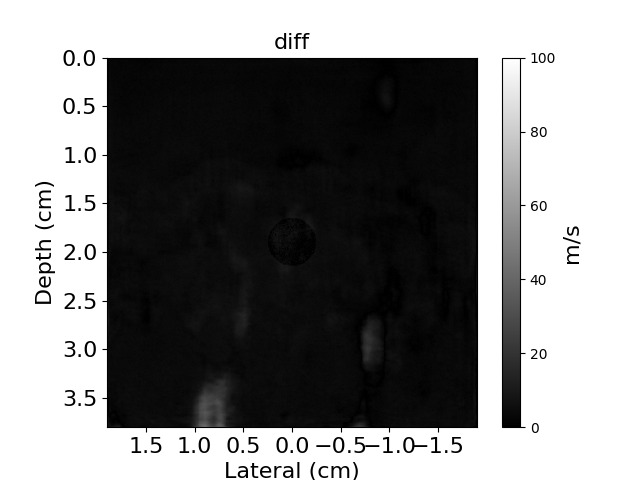}}&
 \raisebox{-.5\totalheight}{\includegraphics[trim={3cm 1.5cm 4cm 1.5cm},clip, width =1.7cm]{ 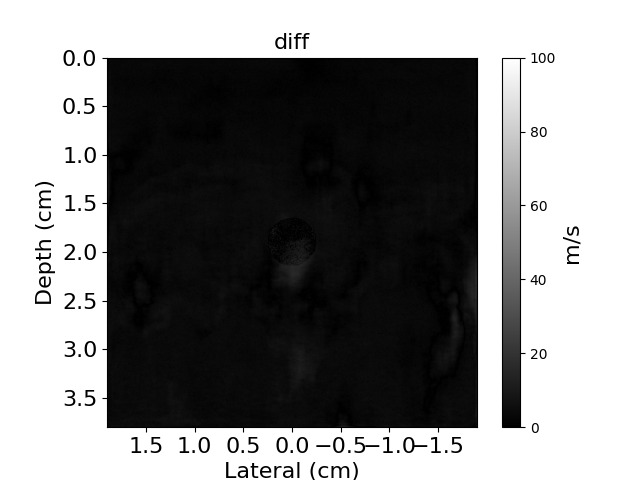}}&
 \raisebox{-.5\totalheight}{\includegraphics[trim={3cm 1.5cm 4cm 1.5cm},clip, width =1.7cm]{ 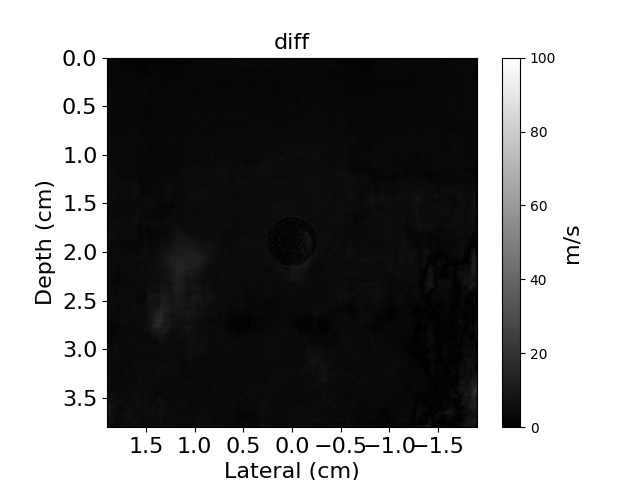}}&
 \raisebox{-.5\totalheight}{\includegraphics[trim={3cm 1.5cm 4cm 1.5cm},clip, width =1.7cm]{ 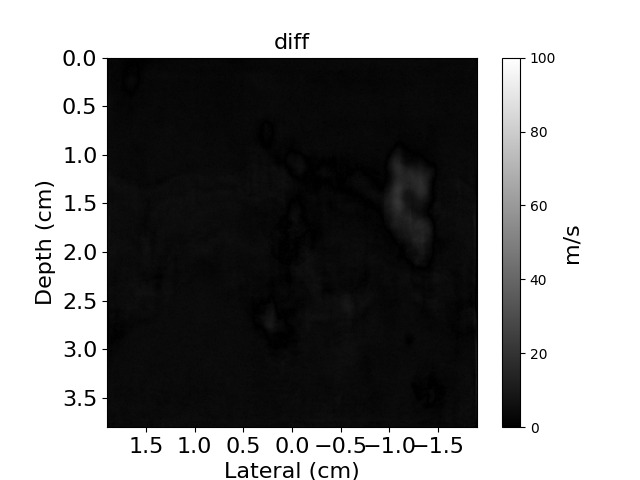}}&
\raisebox{-.5\totalheight}{\includegraphics[height =1.7cm]{img/colormaps/colorbar_diff.pdf}}\\
 
\end{tabular}
\caption{ \textbf{Echogenicity Contrast without SoS Contrast (Case 1-7)}: Comparison between the predicted SoS maps by the networks trained with the Combined and the Ellipsoids datasets for homogeneous mediums with anechoic, hypoechoic, isoechoic, and hyperechoic inclusions with \underline{\textit{no SoS contrast}}; indicating that in the presence of inclusions without SoS contrast and only echogenicity contrast both networks do not indicate any false positive inclusion-shaped areas. However, the network trained with the Ellipsoids setup shows artifacts outside the inclusion area in comparison.\\ 
\textbf{Number of Reflective Speckles (Case 8)}: Case 8 similar to Case 4 shows an isoechoic region but Case 8 has the double number of reflective speckles inside the inclusion area; nevertheless the number of reflective speckles does not affect the indication of inclusion area in the figure. }
\label{fig: digital phantom, ECHO NO sOs}
\end{figure*}

\subparagraph{Echogenicity Contrast without SoS Contrast} The aim of this investigation is to demonstrate that the networks are only sensitive to SoS variations, thus, the presence of inclusions in the b-mode images (created by echogenicity contrast) does not necessarily indicate SoS contrast. As a result, the network is not supposed to show any SoS contrast in the predictions.  

In Figure~\ref{fig: rmse_in}, \say{Echogenicity contrast (No SoS contrast)} boxes show the RMSE values inside the inclusion and in the background for 20 cases where there is an inclusion present in the medium but the inclusion only has an echogenicity contrast and the SoS value inside the inclusion is same as the background.
For instance, Figure \ref{fig: digital phantom, ECHO NO sOs}, Case 1-7 shows examples of such setup, homogeneous mediums with anechoic, hypoechoic, isoechoic, and hyperechoic inclusions with no SoS contrast. 

Based on Figure~\ref{fig: rmse_in}, both networks show comparable RMSE inside the inclusions. 
The RMSE of the background is higher for the Ellipsoids setup compared to the Combined setup. 
Qualitatively, this can be seen as more over/underestimations in the background region in Figure \ref{fig: digital phantom, ECHO NO sOs}, e.g., Cases 2, 3, and 4. 

Nevertheless, both networks despite the presence of inclusion with echogenicity contrast, due to lack of SoS contrast, as expected, do not predict any inclusion-shaped area in the SoS domain. This investigation shows that the network learns the SoS reconstruction. 

\subparagraph{Number of Reflective Speckles} To the best of our knowledge, currently there is no reference method available to determine the optimum numbers of scatterers for training data generation. Thus, the aim of this investigation is to evaluate the sensitivity of the network to the number of scatterers modeled in the heterogeneity. 

Figure~\ref{fig: rmse_in} \say{\#Reflective speckle} boxes demonstrate the RMSE values for 20 cases where the number of scatterers varies from \(5\)\% to \(55\)\% of grid points with the step of \(2.5\)\%. 
Figure~\ref{fig: digital phantom, ECHO NO sOs}, Case 8 shows an example where the number of scatterers increased from \(10\)\% to \(20\)\%. 

Based on Figure~\ref{fig: rmse_in}, for 20 cases, the RMSE of the Combined network inside the inclusion has a similar distribution but is slightly higher than the RMSE of the Ellipsoids setup. 
On the other hand, the RMSE of the Combined network for the background region is lower than the Ellipsoids network.
For none of the evaluated cases (similar to Figure~\ref{fig: digital phantom, ECHO NO sOs}, Case 8) modifying the number of scatterers did not affect the network predictions regarding the presence of an inclusion. Consequently, varied numbers of scatterers can be modeled for the simulation of the training data. 

\begin{figure*}[!t]
\centering 
\renewcommand{\arraystretch}{0.05}
\begin{tabular}{@{\hspace{0.5mm}} c @{\hspace{0.5mm}}c @{\hspace{0.5mm}}c @{\hspace{0.5mm}}c @ {\hspace{0.5mm}}c @{\hspace{0.5mm}}c  @{\hspace{0.5mm}}c  @{\hspace{0.5mm}}c  @{\hspace{0.5mm}}c @{\hspace{0.5mm}}l }
& \scriptsize{ Case  1} & \scriptsize{ Case  2} & \scriptsize{Case  3} & \scriptsize{Case  4} & \scriptsize{Case  5}  & \scriptsize{ Case  6} & \scriptsize{Case  7} & \scriptsize{Case 8} &  \\

\raisebox{-.5\totalheight}{\scriptsize \rot {B-mode}} & 
\raisebox{-.5\totalheight}{\includegraphics[trim={3cm 1.5cm 4cm 1cm},clip, width = 1.7cm]{ 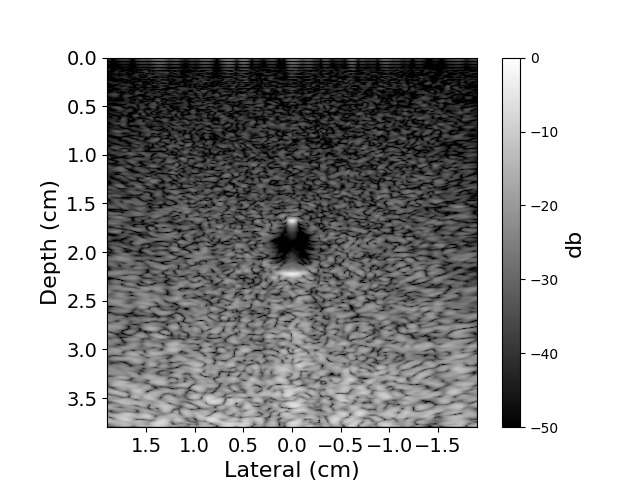}}&
\raisebox{-.5\totalheight}{\includegraphics[trim={3cm 1.5cm 4cm 1cm},clip, width = 1.7cm]{ 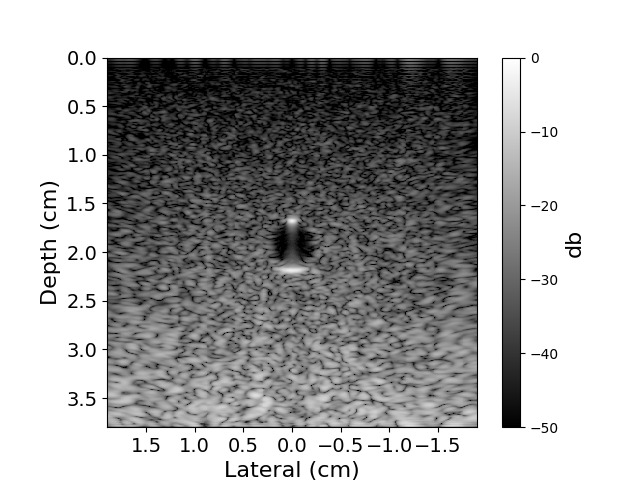}}&
\raisebox{-.5\totalheight}{\includegraphics[trim={3cm 1.5cm 4cm 1cm},clip, width = 1.7cm]{ 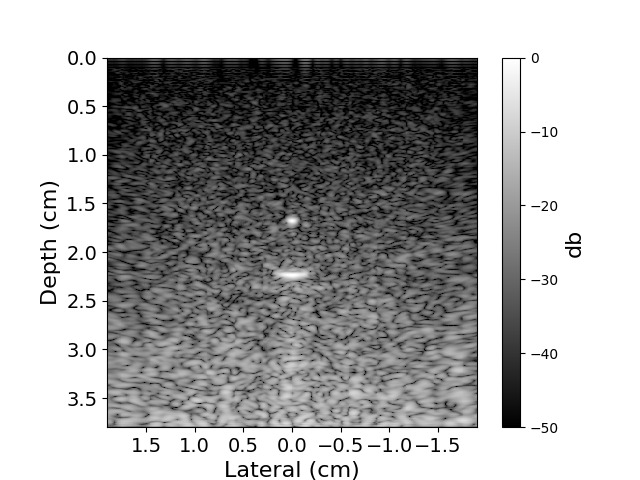}}&
\raisebox{-.5\totalheight}{\includegraphics[trim={3cm 1.5cm 4cm 1cm},clip, width = 1.7cm]{ 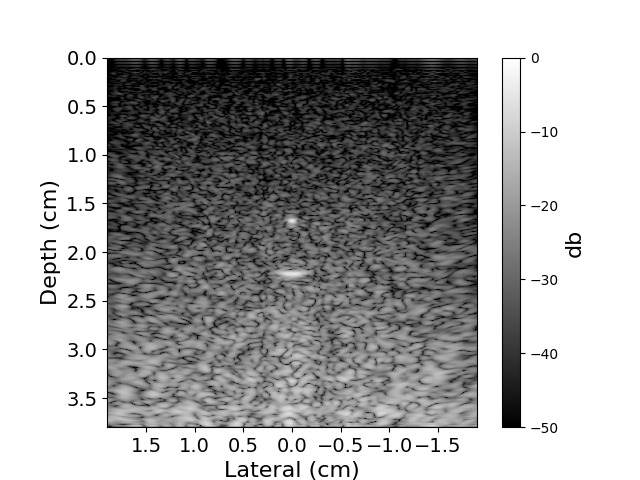}}&
\raisebox{-.5\totalheight}{\includegraphics[trim={3cm 1.5cm 4cm 1cm},clip, width = 1.7cm]{ 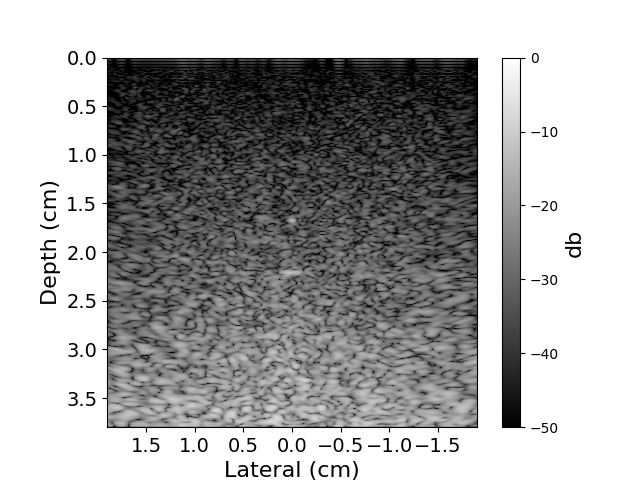}}&
\raisebox{-.5\totalheight}{\includegraphics[trim={3cm 1.5cm 4cm 1cm},clip, width = 1.7cm]{ 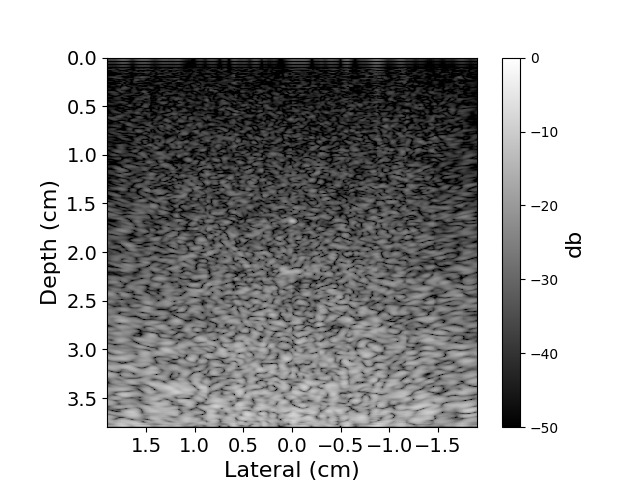}}&
\raisebox{-.5\totalheight}{\includegraphics[trim={3cm 1.5cm 4cm 1cm},clip, width =1.7cm]{ 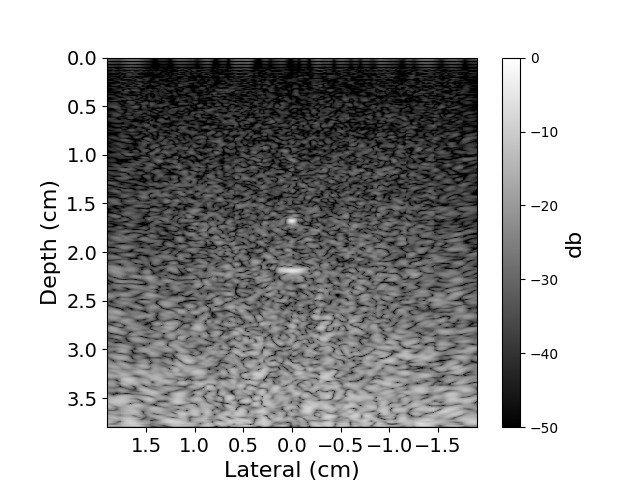}}&
\raisebox{-.5\totalheight}{\includegraphics[trim={3cm 1.5cm 4cm 1cm},clip, width =1.7cm]{ 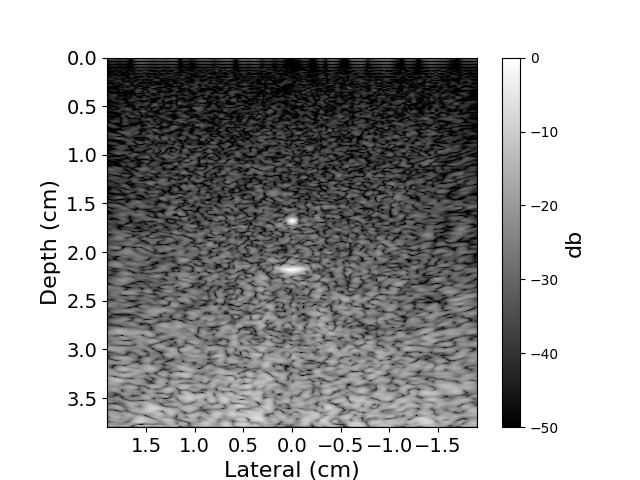}}&
\raisebox{-.5\totalheight}{\includegraphics[height =1.7cm]{img/colormaps/colorbar_bmode.pdf}}\\

\raisebox{-.5\totalheight}{\scriptsize \rot{GT}} & 
\raisebox{-.5\totalheight}{\includegraphics[trim={3cm 1.5cm 4cm 1.5cm},clip, width = 1.7cm]{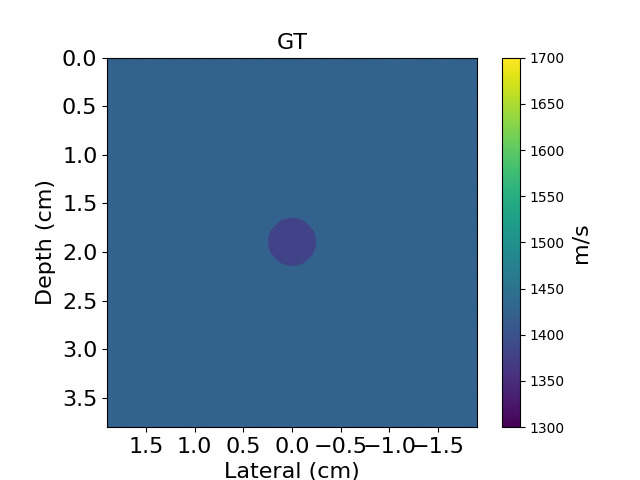}}&
\raisebox{-.5\totalheight}{\includegraphics[trim={3cm 1.5cm 4cm 1.5cm},clip, width = 1.7cm]{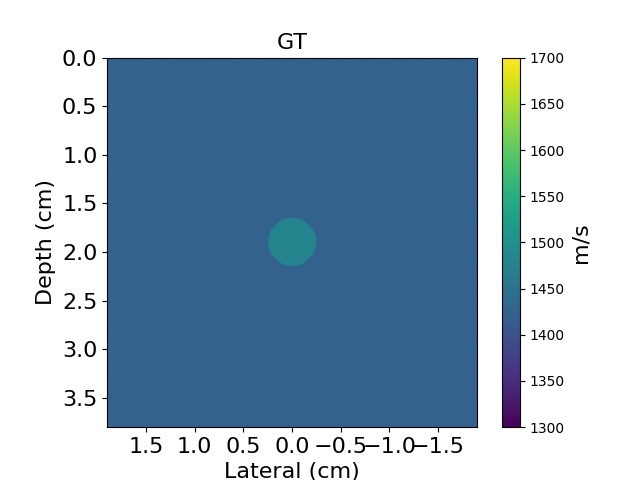}}&
\raisebox{-.5\totalheight}{\includegraphics[trim={3cm 1.5cm 4cm 1.5cm},clip, width = 1.7cm]{ 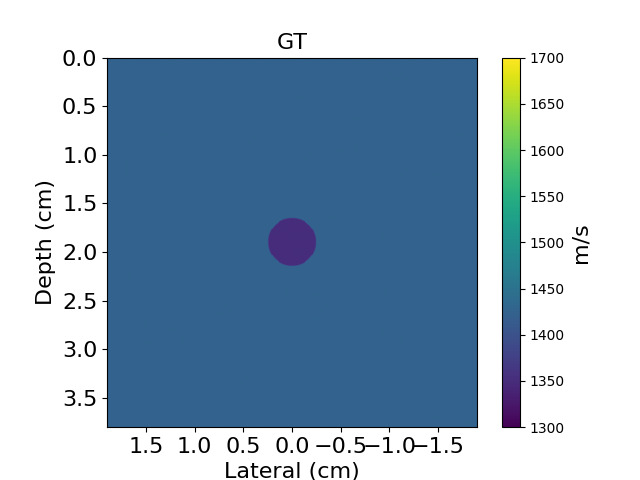}}&
\raisebox{-.5\totalheight}{\includegraphics[trim={3cm 1.5cm 4cm 1.5cm},clip, width = 1.7cm]{ 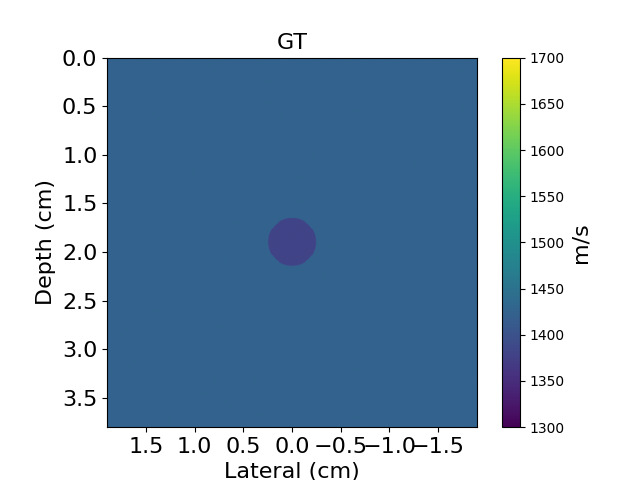}}&
\raisebox{-.5\totalheight}{\includegraphics[trim={3cm 1.5cm 4cm 1.5cm},clip, width = 1.7cm]{ 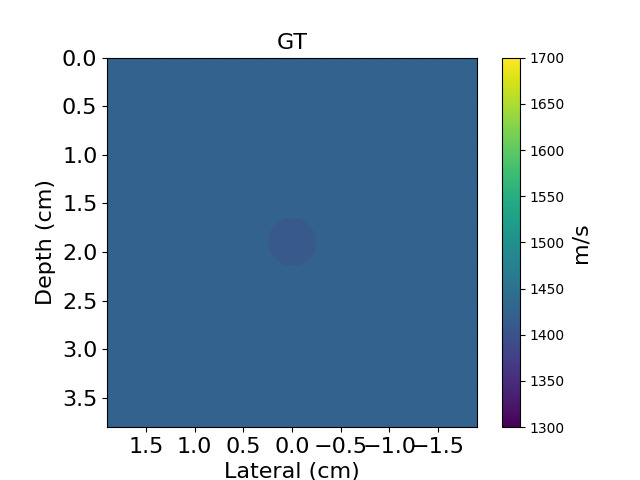}}&
\raisebox{-.5\totalheight}{\includegraphics[trim={3cm 1.5cm 4cm 1.5cm},clip, width = 1.7cm]{ 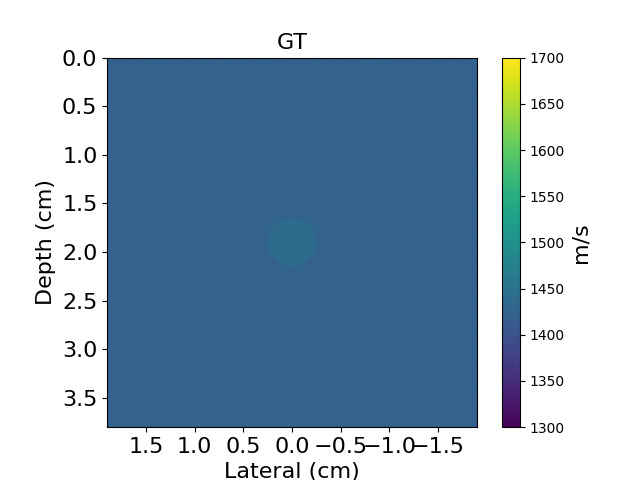}}&
\raisebox{-.5\totalheight}{\includegraphics[trim={3cm 1.5cm 4cm 1.5cm},clip, width =1.7cm]{ 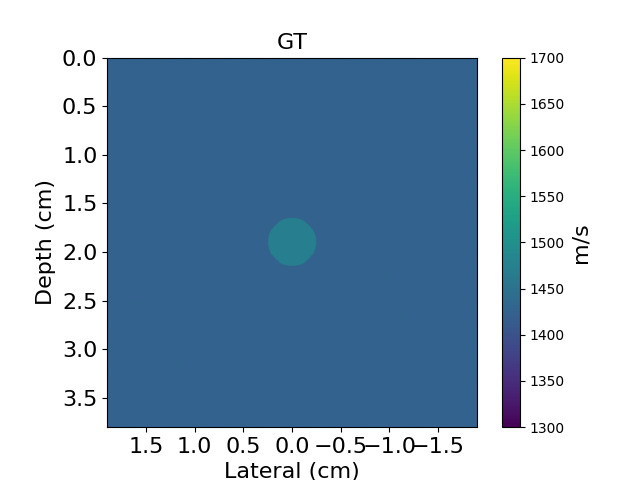}}&
\raisebox{-.5\totalheight}{\includegraphics[trim={3cm 1.5cm 4cm 1.5cm},clip, width =1.7cm]{ 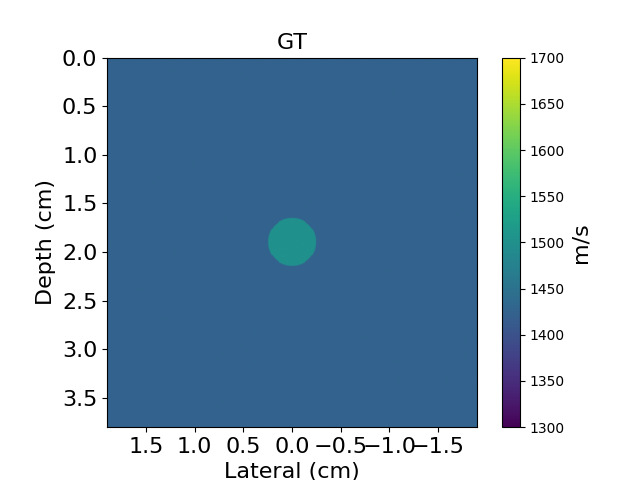}}&
\raisebox{-.5\totalheight}{\includegraphics[height =1.7cm]{img/colormaps/colorbar_simulated.pdf}}\\
 
\raisebox{-.5\totalheight}{\scriptsize \rot {\makecell{{Predicted} \\ {{SoS}} \\ {Combined} }}} & 
\raisebox{-.5\totalheight}{\includegraphics[trim={3cm 1.5cm 4cm 1.5cm},clip, width = 1.7cm]{ 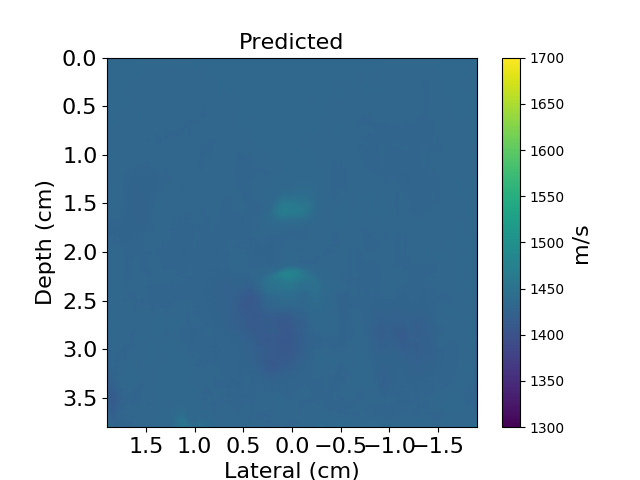}}&
\raisebox{-.5\totalheight}{\includegraphics[trim={3cm 1.5cm 4cm 1.5cm},clip, width = 1.7cm]{ 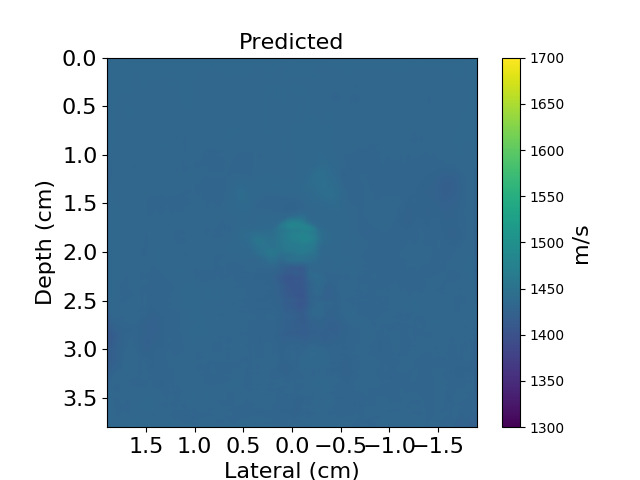}}&
\raisebox{-.5\totalheight}{\includegraphics[trim={3cm 1.5cm 4cm 1.5cm},clip, width = 1.7cm]{ 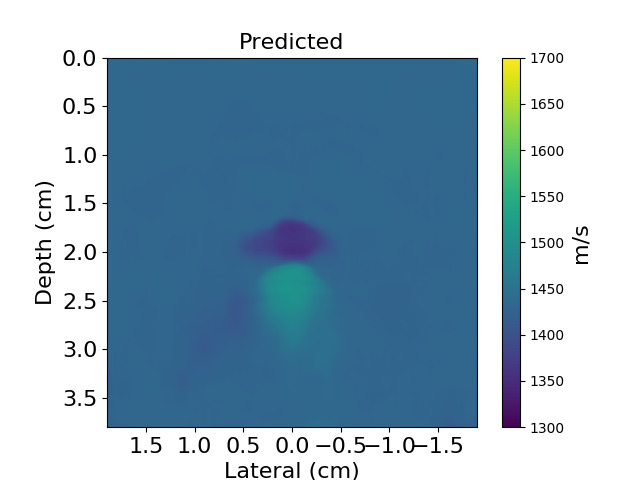}}&
\raisebox{-.5\totalheight}{\includegraphics[trim={3cm 1.5cm 4cm 1.5cm},clip, width = 1.7cm]{ 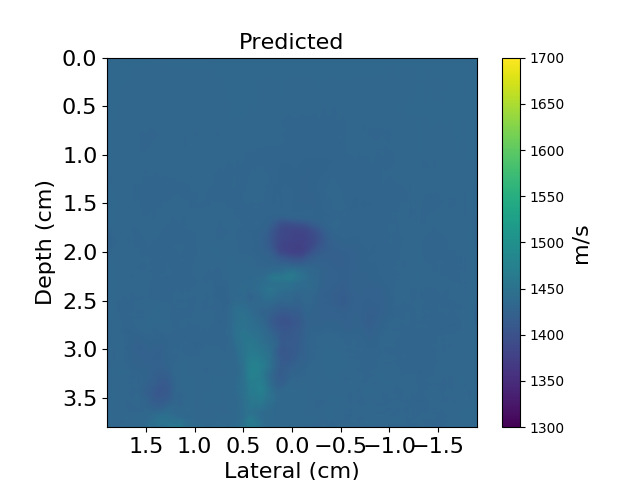}}&
\raisebox{-.5\totalheight}{\includegraphics[trim={3cm 1.5cm 4cm 1.5cm},clip, width = 1.7cm]{ 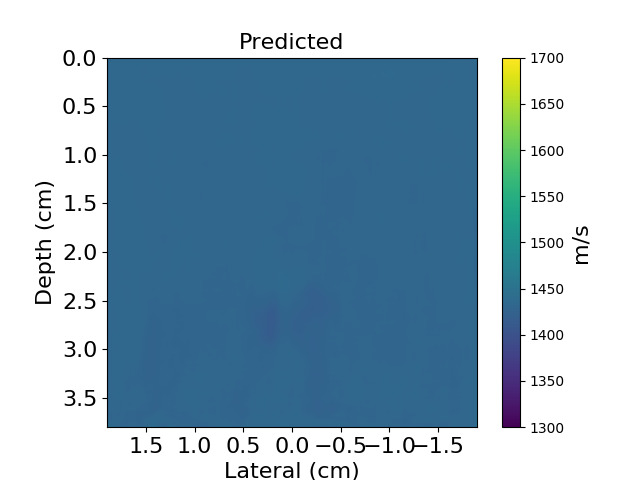}}&
\raisebox{-.5\totalheight}{\includegraphics[trim={3cm 1.5cm 4cm 1.5cm},clip, width = 1.7cm]{ 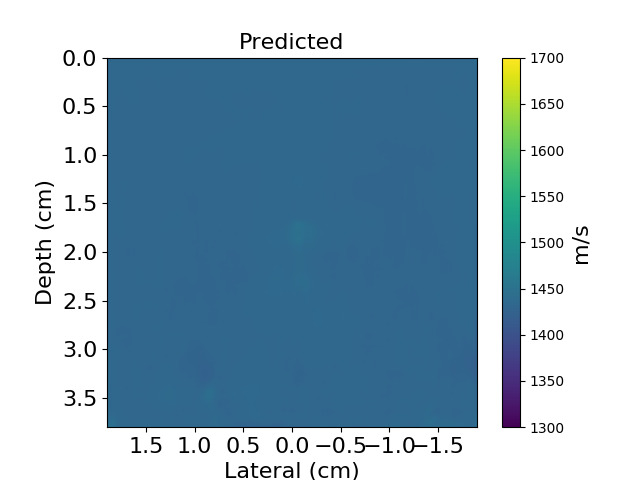}}&
\raisebox{-.5\totalheight}{\includegraphics[trim={3cm 1.5cm 4cm 1.5cm},clip, width =1.7cm]{ 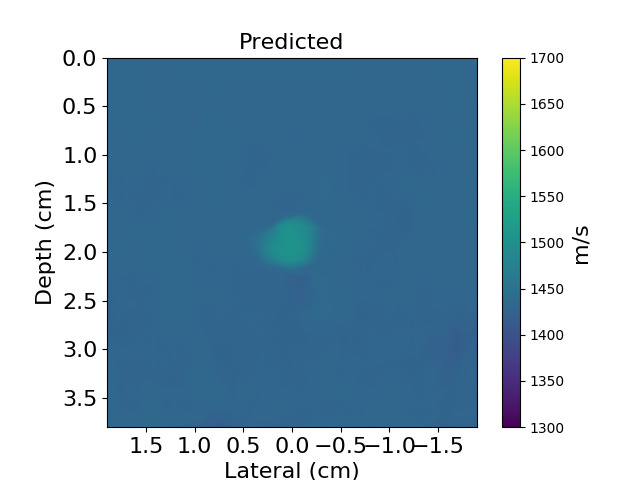}}&
\raisebox{-.5\totalheight}{\includegraphics[trim={3cm 1.5cm 4cm 1.5cm},clip, width =1.7cm]{ 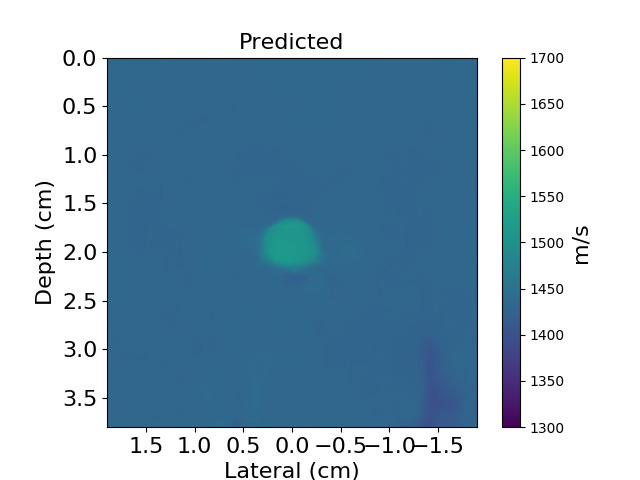}}&
\raisebox{-.5\totalheight}{\includegraphics[height =1.7cm]{img/colormaps/colorbar_simulated.pdf}}\\

\raisebox{-.5\totalheight}{\scriptsize \rot{ \makecell{{Absolute} \\ {{Difference}} }}} & 
\raisebox{-.5\totalheight}{\includegraphics[trim={3cm 1.5cm 4cm 1.5cm},clip, width = 1.7cm]{ 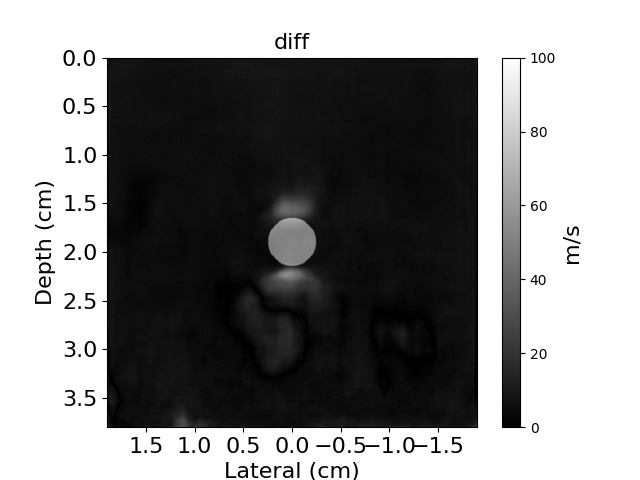}}&
\raisebox{-.5\totalheight}{\includegraphics[trim={3cm 1.5cm 4cm 1.5cm},clip, width = 1.7cm]{ 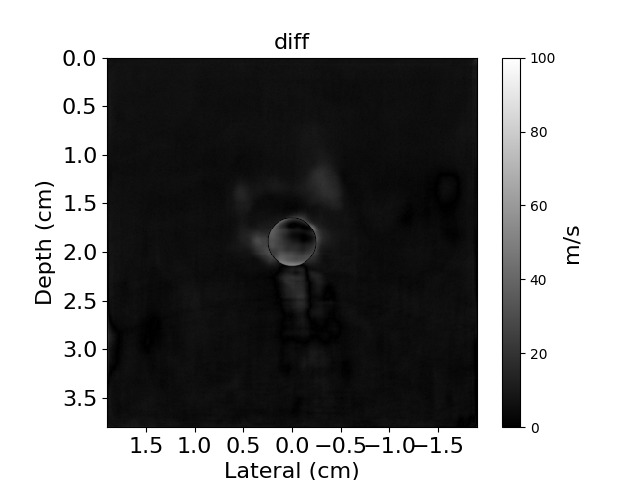}}&
\raisebox{-.5\totalheight}{\includegraphics[trim={3cm 1.5cm 4cm 1.5cm},clip, width = 1.7cm]{ 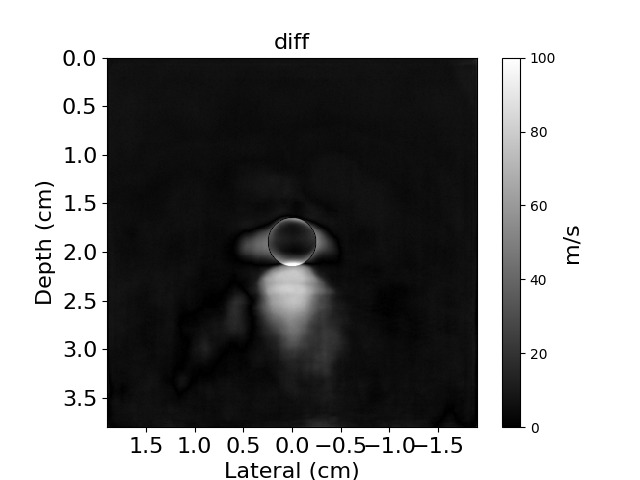}}&
\raisebox{-.5\totalheight}{\includegraphics[trim={3cm 1.5cm 4cm 1.5cm},clip, width = 1.7cm]{ 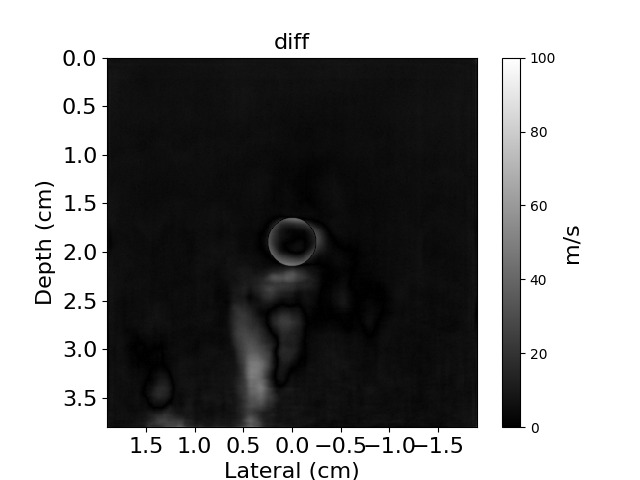}}&
\raisebox{-.5\totalheight}{\includegraphics[trim={3cm 1.5cm 4cm 1.5cm},clip, width = 1.7cm]{ 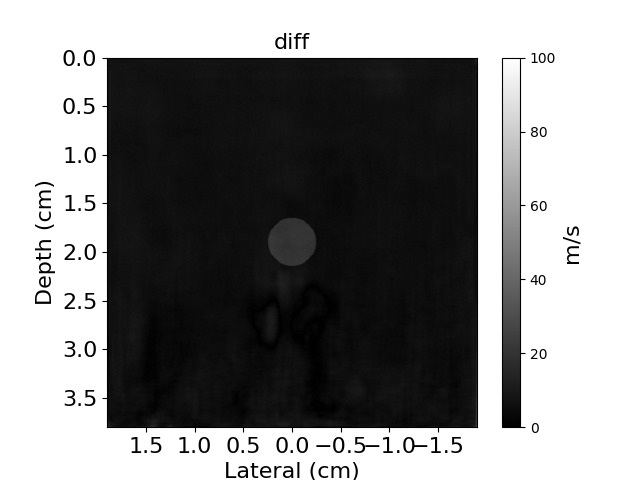}}&
\raisebox{-.5\totalheight}{\includegraphics[trim={3cm 1.5cm 4cm 1.5cm},clip, width = 1.7cm]{ 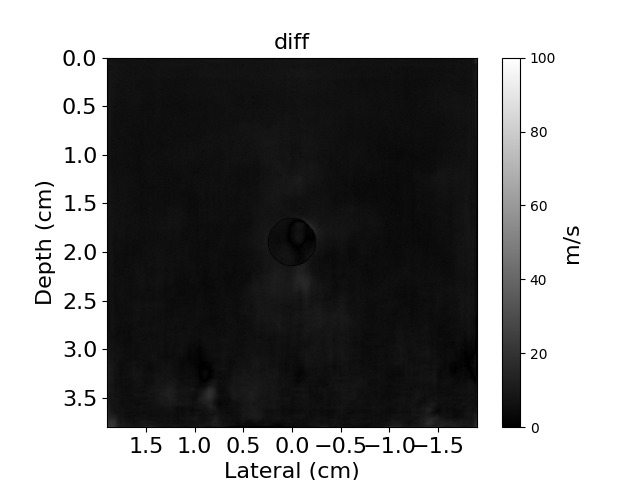}}&
\raisebox{-.5\totalheight}{\includegraphics[trim={3cm 1.5cm 4cm 1.5cm},clip, width =1.7cm]{ 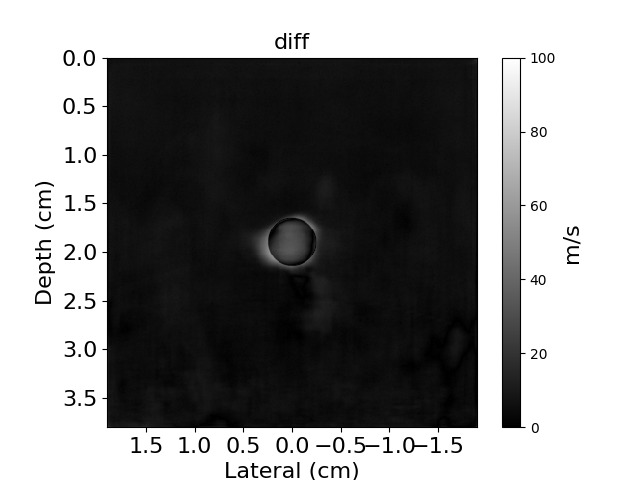}}&
\raisebox{-.5\totalheight}{\includegraphics[trim={3cm 1.5cm 4cm 1.5cm},clip, width =1.7cm]{ 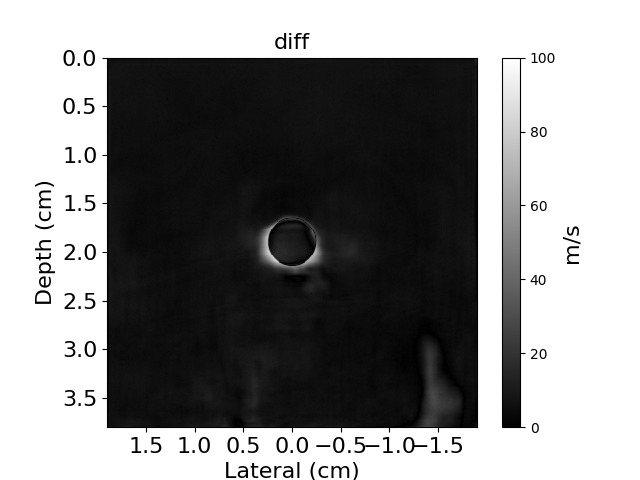}}&
\raisebox{-.5\totalheight}{\includegraphics[height =1.7cm]{img/colormaps/colorbar_diff.pdf}}\\

\raisebox{-.5\totalheight}{\scriptsize \rot{ \makecell{{Predicted} \\ {{SoS}} \\ {Ellipsoids} }}} & 
\raisebox{-.5\totalheight}{\includegraphics[trim={3cm 1.5cm 4cm 1.5cm},clip, width = 1.7cm]{ 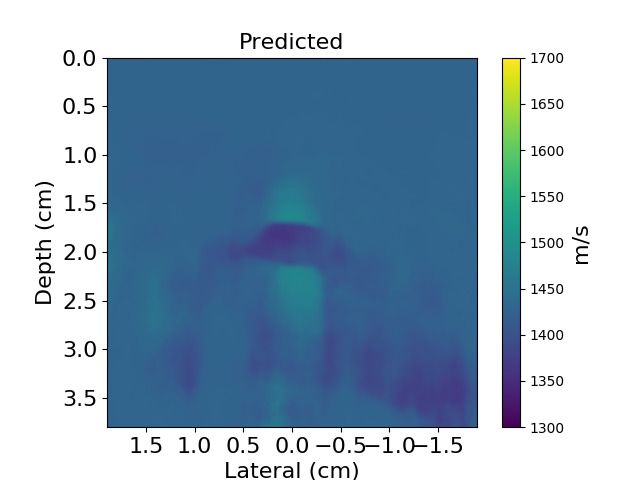}}&
\raisebox{-.5\totalheight}{\includegraphics[trim={3cm 1.5cm 4cm 1.5cm},clip, width = 1.7cm]{ 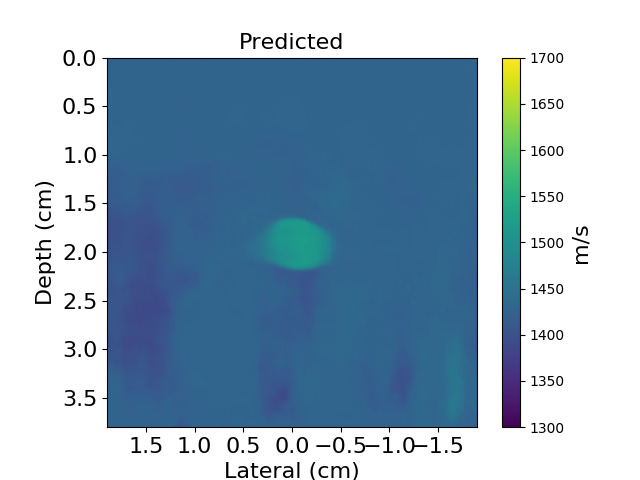}}&
\raisebox{-.5\totalheight}{\includegraphics[trim={3cm 1.5cm 4cm 1.5cm},clip, width = 1.7cm]{ 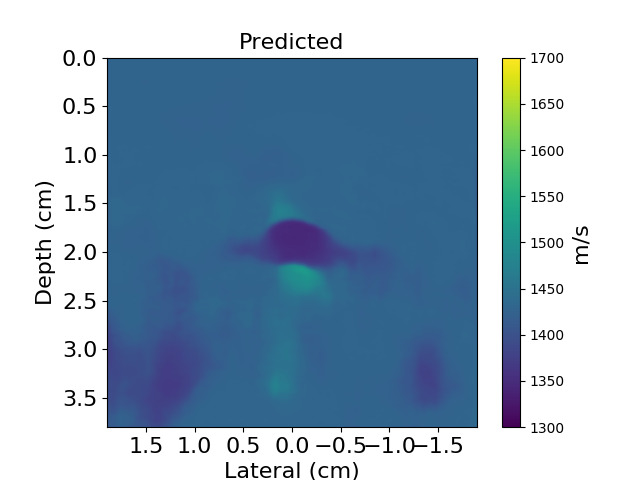}}&
\raisebox{-.5\totalheight}{\includegraphics[trim={3cm 1.5cm 4cm 1.5cm},clip, width = 1.7cm]{ 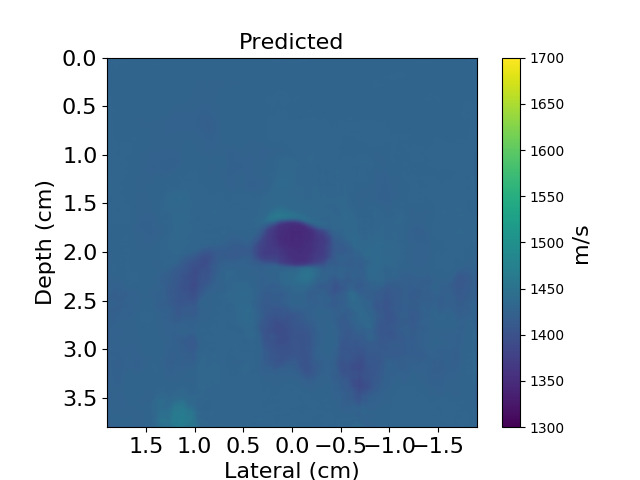}}&
\raisebox{-.5\totalheight}{\includegraphics[trim={3cm 1.5cm 4cm 1.5cm},clip, width = 1.7cm]{ 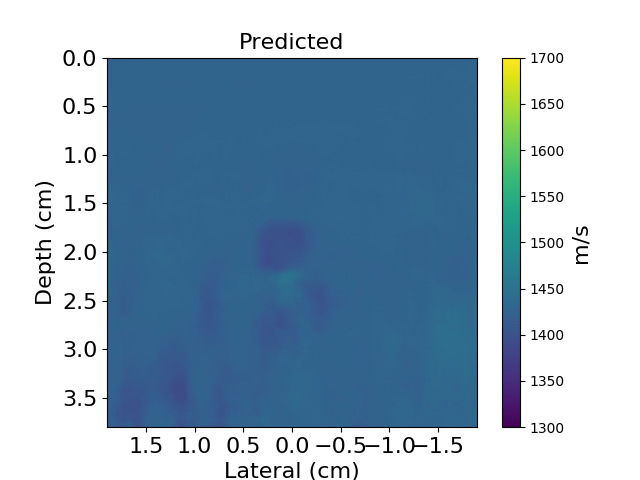}}&
\raisebox{-.5\totalheight}{\includegraphics[trim={3cm 1.5cm 4cm 1.5cm},clip, width = 1.7cm]{ 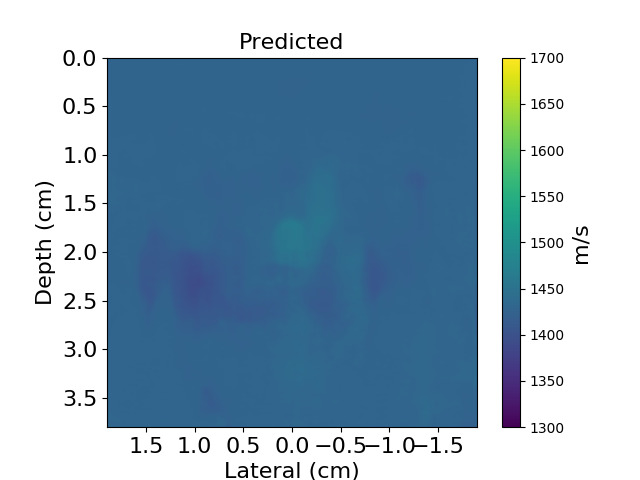}}&
\raisebox{-.5\totalheight}{\includegraphics[trim={3cm 1.5cm 4cm 1.5cm},clip, width =1.7cm]{ 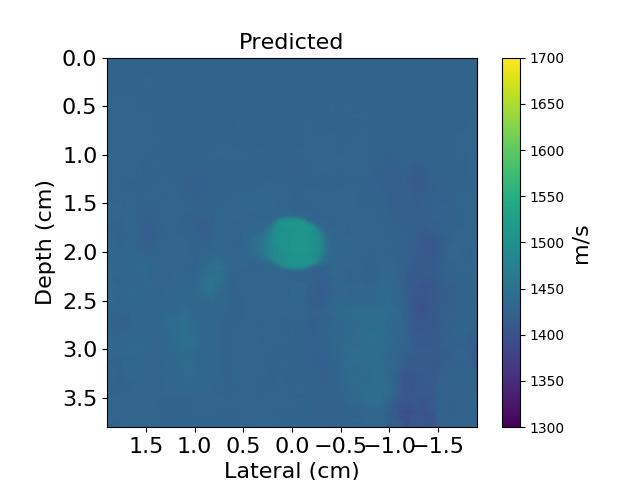}}&
\raisebox{-.5\totalheight}{\includegraphics[trim={3cm 1.5cm 4cm 1.5cm},clip, width =1.7cm]{ 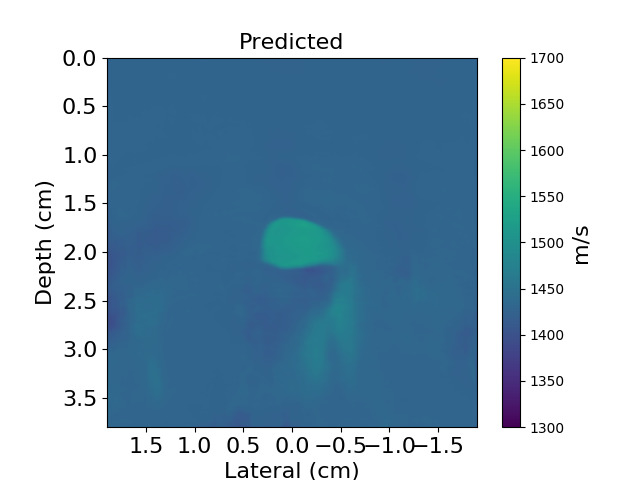}}&
\raisebox{-.5\totalheight}{\includegraphics[height =1.7cm]{img/colormaps/colorbar_simulated.pdf}}\\

\raisebox{-.5\totalheight}{\scriptsize \rot{\makecell{{Absolute} \\  {Difference} }}}& 
\raisebox{-.5\totalheight}{\includegraphics[trim={3cm 1.5cm 4cm 1.5cm},clip, width = 1.7cm]{ 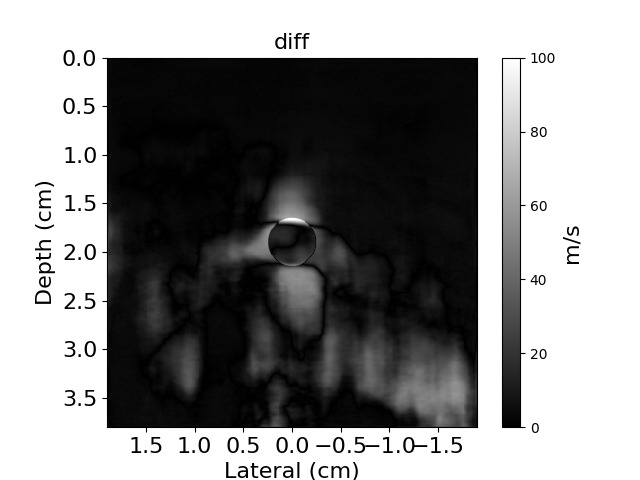}}&
\raisebox{-.5\totalheight}{\includegraphics[trim={3cm 1.5cm 4cm 1.5cm},clip, width = 1.7cm]{ 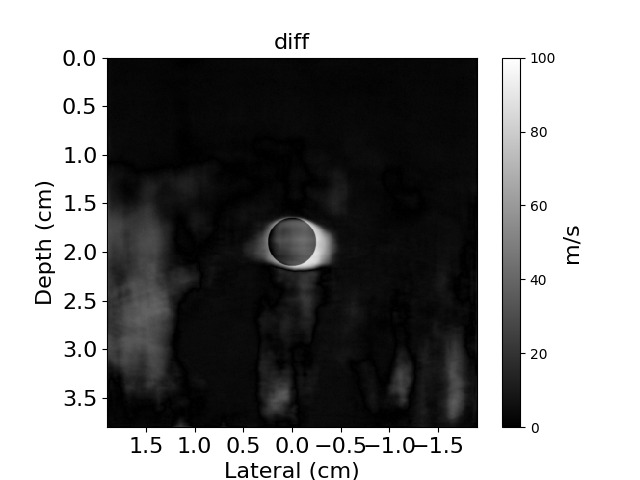}}&
\raisebox{-.5\totalheight}{\includegraphics[trim={3cm 1.5cm 4cm 1.5cm},clip, width = 1.7cm]{ 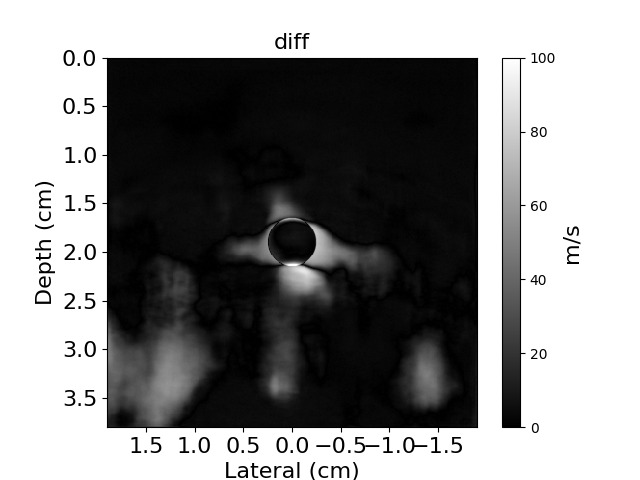}}&
\raisebox{-.5\totalheight}{\includegraphics[trim={3cm 1.5cm 4cm 1.5cm},clip, width = 1.7cm]{ 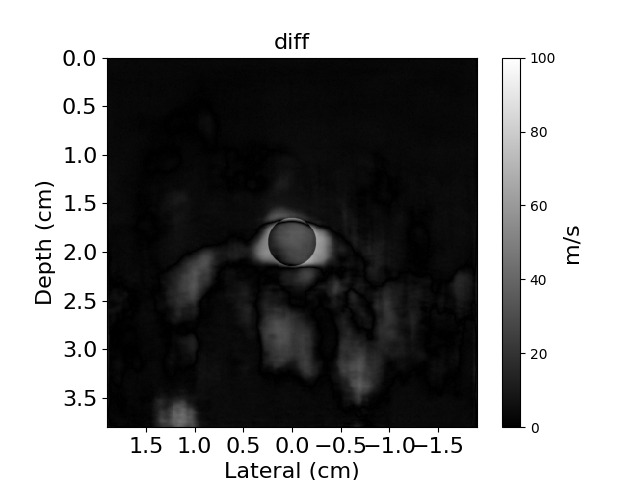}}&
\raisebox{-.5\totalheight}{\includegraphics[trim={3cm 1.5cm 4cm 1.5cm},clip, width = 1.7cm]{ 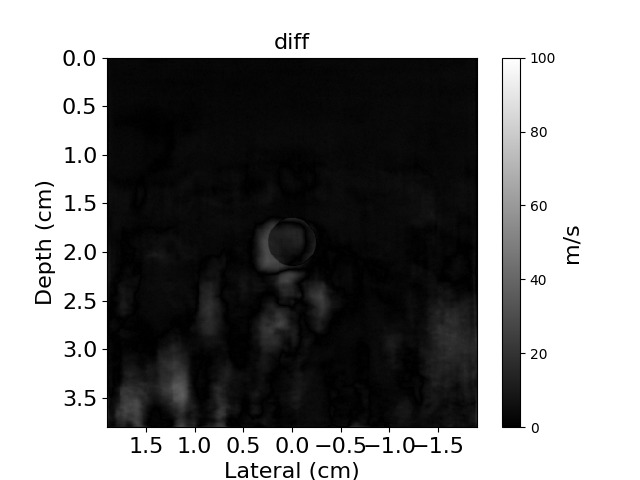}}&
\raisebox{-.5\totalheight}{\includegraphics[trim={3cm 1.5cm 4cm 1.5cm},clip, width = 1.7cm]{ 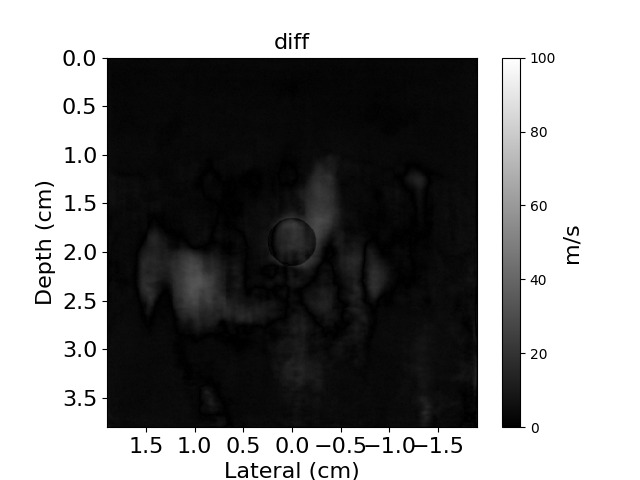}}&
\raisebox{-.5\totalheight}{\includegraphics[trim={3cm 1.5cm 4cm 1.5cm},clip, width =1.7cm]{ 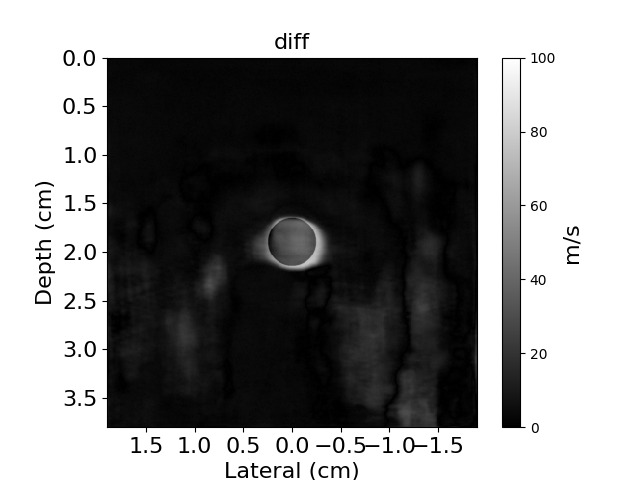}}&
\raisebox{-.5\totalheight}{\includegraphics[trim={3cm 1.5cm 4cm 1.5cm},clip, width =1.7cm]{ 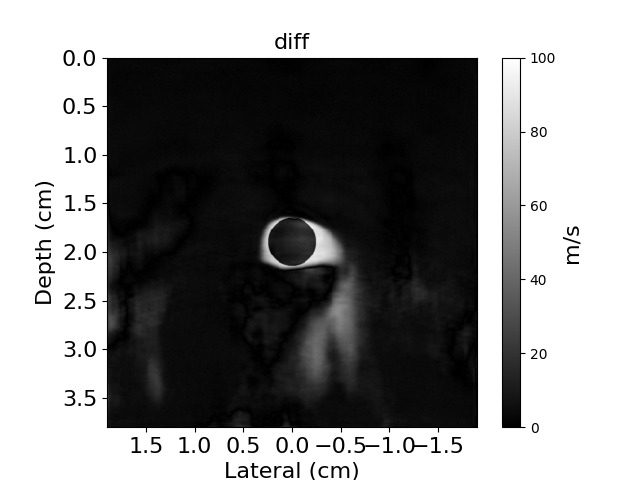}}&
\raisebox{-.5\totalheight}{\includegraphics[height =1.7cm]{img/colormaps/colorbar_diff.pdf}}\\

\end{tabular}
\caption{ \textbf{Echogenicity Contrast with SoS Contrast}: Comparison between the predicted SoS maps by the networks trained with the Combined and the Ellipsoids dataset for homogeneous mediums with anechoic, and isoechoic inclusions \underline{\textit{with SoS contrast}}; Although the SoS contrast in the training data was introduced in combination with only hyperechoic lesions/inclusions both networks can perform with various echogenicity setups. The network trained on the Ellipsoids setup shows more over/underestimations outside the inclusion area.}
\label{fig: digital phantom, SoS contrast}
\end{figure*}

\begin{figure*}[!t]
\centering 

\renewcommand{\arraystretch}{0.05}
\begin{tabular}{@{\hspace{0.5mm}} c @{\hspace{0.5mm}}c @{\hspace{0.5mm}}c @{\hspace{0.5mm}}c @ {\hspace{0.5mm}}c @{\hspace{0.5mm}}c  @{\hspace{0.5mm}}c  @{\hspace{0.5mm}}c  @{\hspace{0.5mm}}c @{\hspace{0.5mm}}l}
& \scriptsize{ Case  1} & \scriptsize{ Case  2} & \scriptsize{Case  3} & \scriptsize{Case  4} & \scriptsize{Case  5}  & \scriptsize{ Case  6} &\scriptsize {Case  7} & \scriptsize{Case 8} & \\

\raisebox{-.5\totalheight}{\scriptsize \rot {B-mode}} & 
\raisebox{-.5\totalheight}{\includegraphics[trim={3cm 1.5cm 4cm 1cm},clip, width = 1.7cm]{ 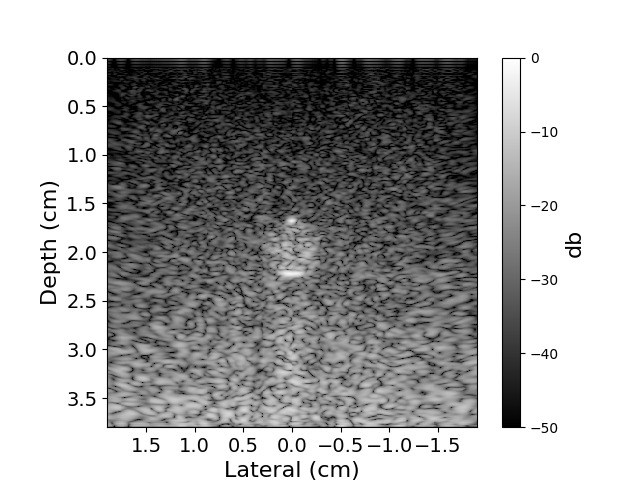}}&
\raisebox{-.5\totalheight}{\includegraphics[trim={3cm 1.5cm 4cm 1cm},clip, width = 1.7cm]{ 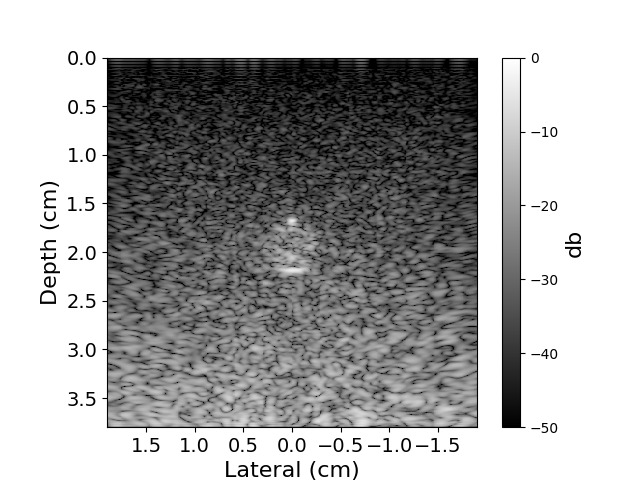}}&
\raisebox{-.5\totalheight}{\includegraphics[trim={3cm 1.5cm 4cm 1cm},clip, width = 1.7cm]{ 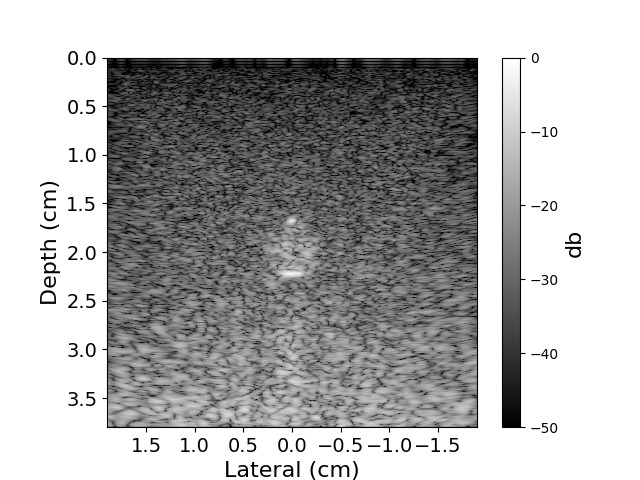}}&
\raisebox{-.5\totalheight}{\includegraphics[trim={3cm 1.5cm 4cm 1cm},clip, width = 1.7cm]{ 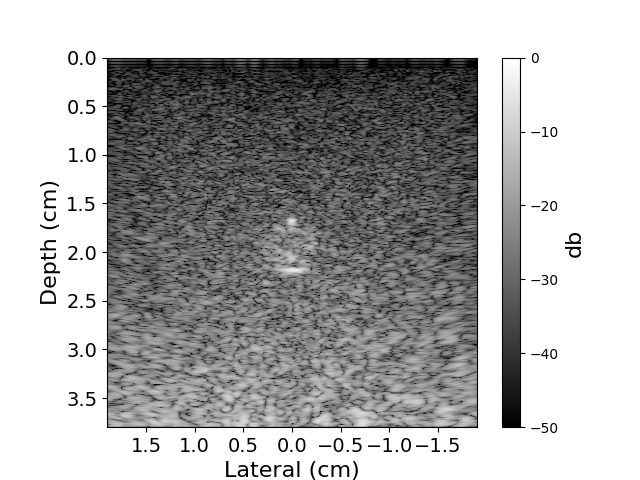}}&
\raisebox{-.5\totalheight}{\includegraphics[trim={3cm 1.5cm 4cm 1cm},clip, width = 1.7cm]{ 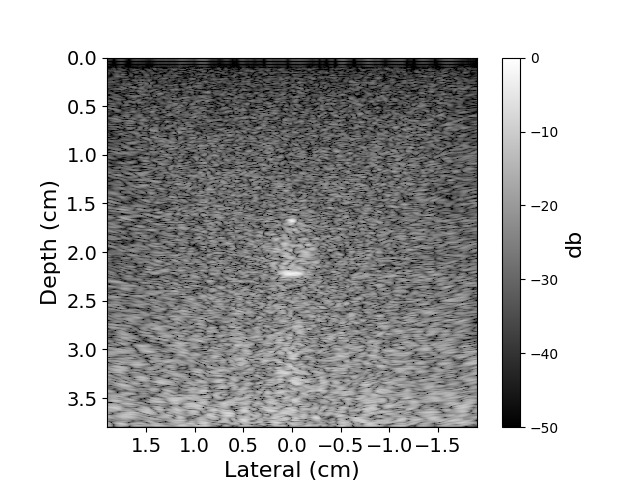}}&
\raisebox{-.5\totalheight}{\includegraphics[trim={3cm 1.5cm 4cm 1cm},clip, width = 1.7cm]{ 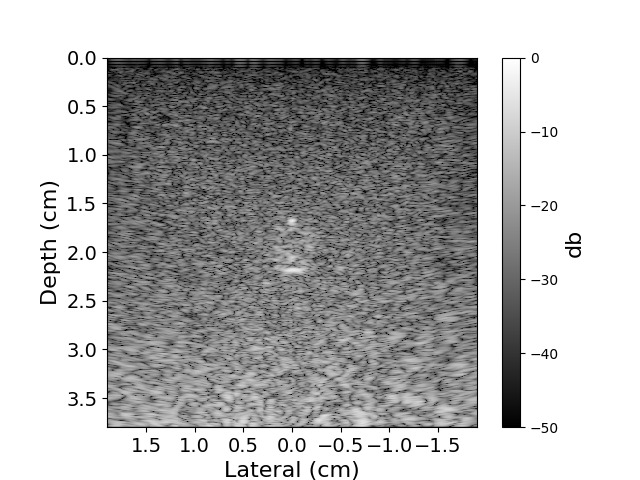}}&
\raisebox{-.5\totalheight}{\includegraphics[trim={3cm 1.5cm 4cm 1cm},clip, width =1.7cm]{ 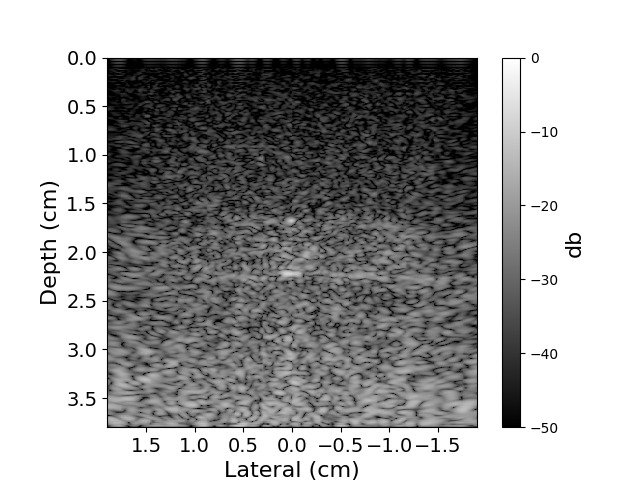}}&
\raisebox{-.5\totalheight}{\includegraphics[trim={3cm 1.5cm 4cm 1cm},clip, width =1.7cm]{ img/Digital_phantom/Noisy/Phase_random/B-mode/1_bmode.jpg}}&
\raisebox{-.5\totalheight}{\includegraphics[height =1.7cm]{img/colormaps/colorbar_bmode.pdf}}\\

\raisebox{-.5\totalheight}{\scriptsize \rot{GT}} & 
\raisebox{-.5\totalheight}{\includegraphics[trim={3cm 1.5cm 4cm 1.5cm},clip, width = 1.7cm]{ 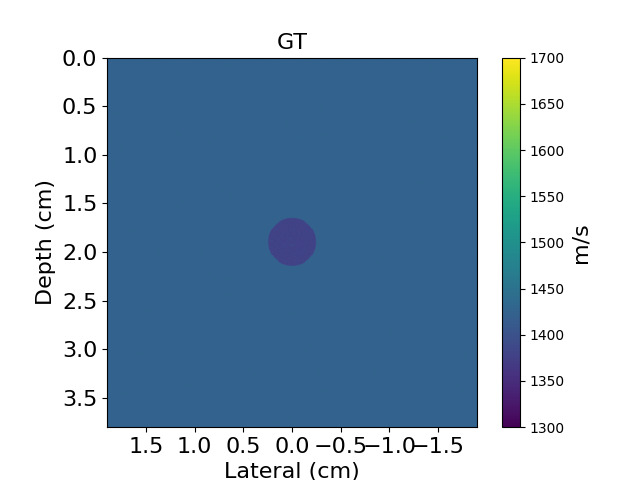}}&
\raisebox{-.5\totalheight}{\includegraphics[trim={3cm 1.5cm 4cm 1.5cm},clip, width = 1.7cm]{ 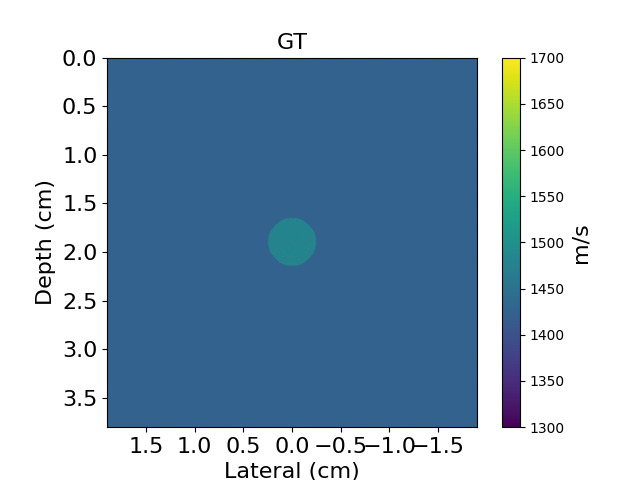}}&
\raisebox{-.5\totalheight}{\includegraphics[trim={3cm 1.5cm 4cm 1.5cm},clip, width = 1.7cm]{ img/Digital_phantom/Noisy/org/GT/GT_1.jpg}}&
\raisebox{-.5\totalheight}{\includegraphics[trim={3cm 1.5cm 4cm 1.5cm},clip, width = 1.7cm]{ img/Digital_phantom/Noisy/org/GT/GT_11.jpg}}&
\raisebox{-.5\totalheight}{\includegraphics[trim={3cm 1.5cm 4cm 1.5cm},clip, width = 1.7cm]{ img/Digital_phantom/Noisy/org/GT/GT_1.jpg}}&
\raisebox{-.5\totalheight}{\includegraphics[trim={3cm 1.5cm 4cm 1.5cm},clip, width = 1.7cm]{ img/Digital_phantom/Noisy/org/GT/GT_11.jpg}}&
\raisebox{-.5\totalheight}{\includegraphics[trim={3cm 1.5cm 4cm 1.5cm},clip, width = 1.7cm]{ img/Digital_phantom/Noisy/org/GT/GT_1.jpg}}&
\raisebox{-.5\totalheight}{\includegraphics[trim={3cm 1.5cm 4cm 1.5cm},clip, width = 1.7cm]{ img/Digital_phantom/Noisy/org/GT/GT_11.jpg}}&
\raisebox{-.5\totalheight}{\includegraphics[height =1.7cm]{img/colormaps/colorbar_simulated.pdf}}\\
 
\raisebox{-.5\totalheight}{\scriptsize \rot {\makecell{{Predicted} \\ {{SoS}} \\ {Combined} }}} & 
\raisebox{-.5\totalheight}{\includegraphics[trim={3cm 1.5cm 4cm 1.5cm},clip, width = 1.7cm]{ 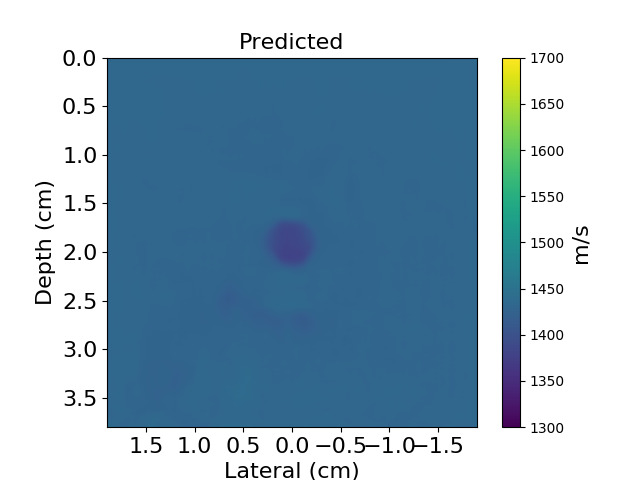}}&
\raisebox{-.5\totalheight}{\includegraphics[trim={3cm 1.5cm 4cm 1.5cm},clip, width = 1.7cm]{ 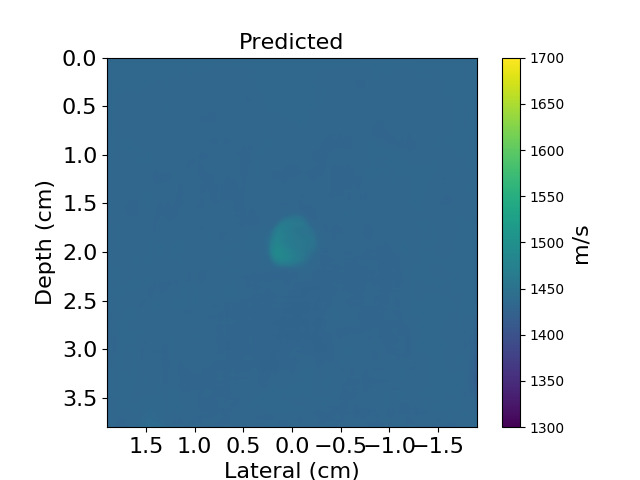}}&
\raisebox{-.5\totalheight}{\includegraphics[trim={3cm 1.5cm 4cm 1.5cm},clip, width = 1.7cm]{ 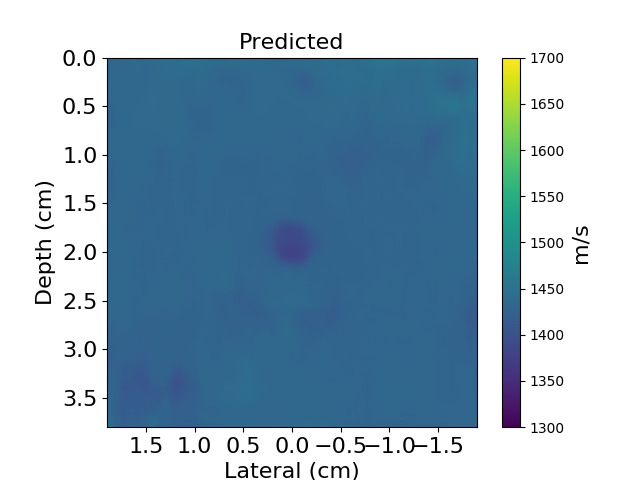}}&
\raisebox{-.5\totalheight}{\includegraphics[trim={3cm 1.5cm 4cm 1.5cm},clip, width = 1.7cm]{ 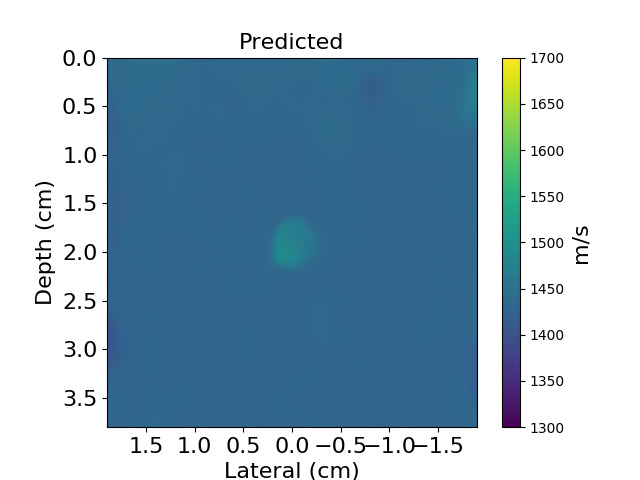}}&
\raisebox{-.5\totalheight}{\includegraphics[trim={3cm 1.5cm 4cm 1.5cm},clip, width = 1.7cm]{ 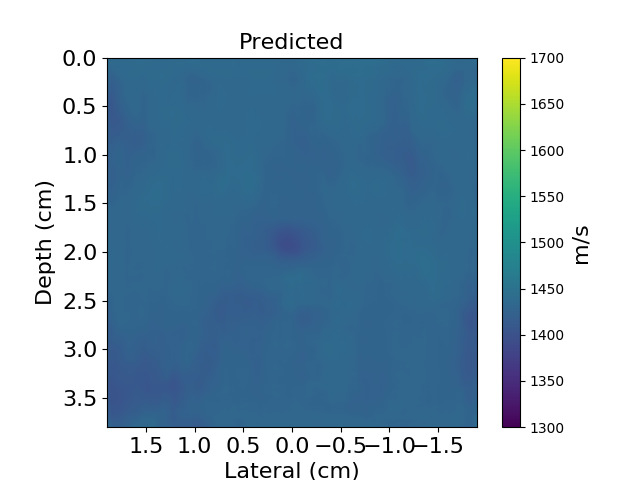}}&
\raisebox{-.5\totalheight}{\includegraphics[trim={3cm 1.5cm 4cm 1.5cm},clip, width = 1.7cm]{ 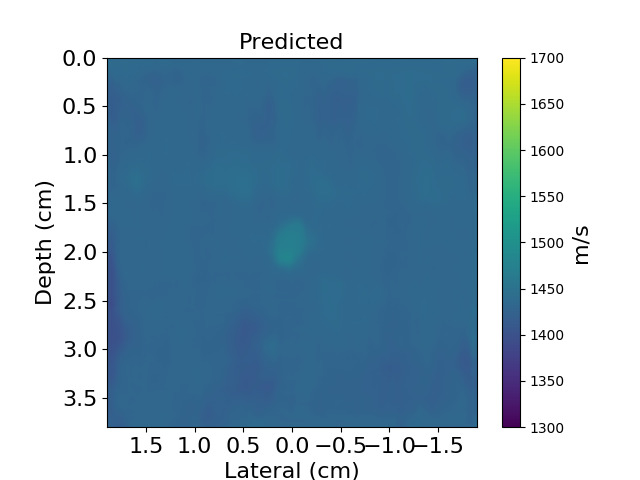}}&
\raisebox{-.5\totalheight}{\includegraphics[trim={3cm 1.5cm 4cm 1.5cm},clip, width =1.7cm]{ 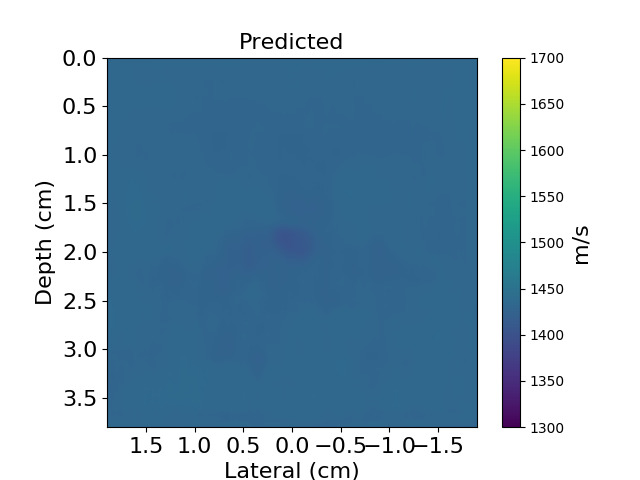}}&
\raisebox{-.5\totalheight}{\includegraphics[trim={3cm 1.5cm 4cm 1.5cm},clip, width =1.7cm]{ 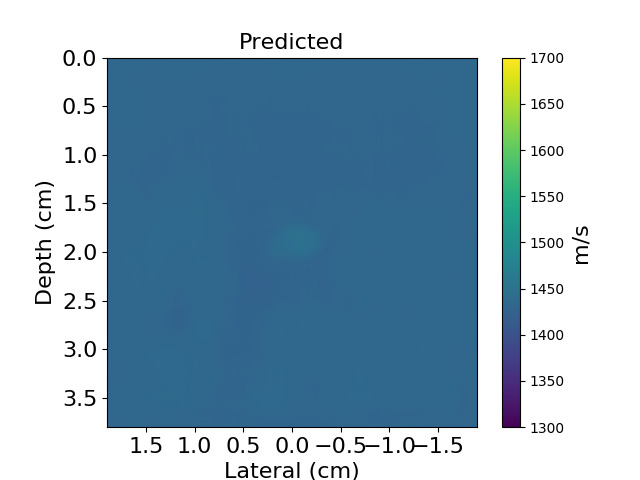}}&
\raisebox{-.5\totalheight}{\includegraphics[height =1.7cm]{img/colormaps/colorbar_simulated.pdf}}\\

\raisebox{-.5\totalheight}{\scriptsize \rot{ \makecell{{Absolute} \\ {{Difference}} }}} & 
\raisebox{-.5\totalheight}{\includegraphics[trim={3cm 1.5cm 4cm 1.5cm},clip, width = 1.7cm]{ 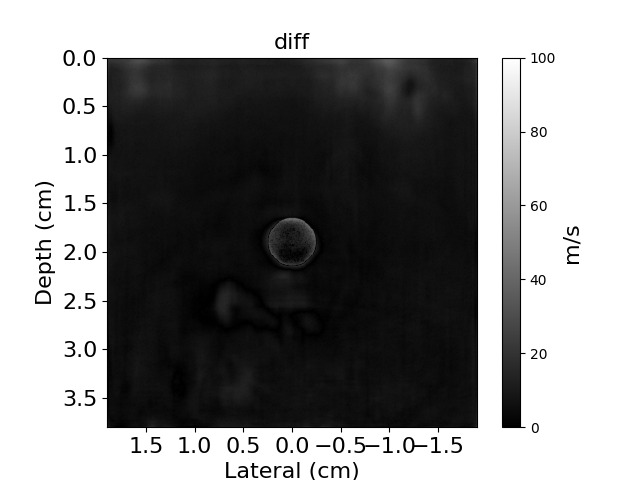}}&
\raisebox{-.5\totalheight}{\includegraphics[trim={3cm 1.5cm 4cm 1.5cm},clip, width = 1.7cm]{ 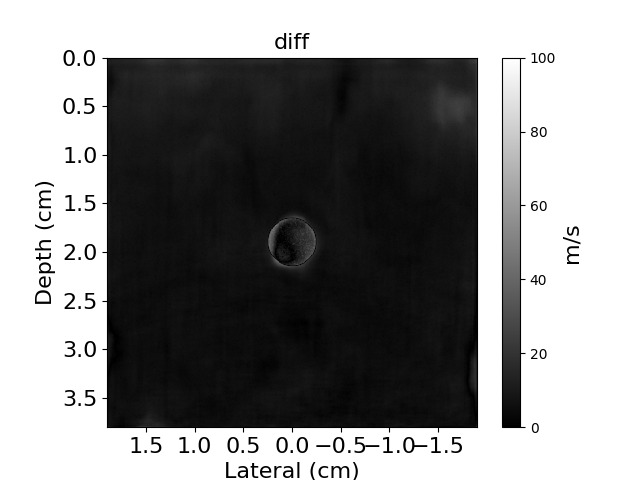}}&
\raisebox{-.5\totalheight}{\includegraphics[trim={3cm 1.5cm 4cm 1.5cm},clip, width = 1.7cm]{ 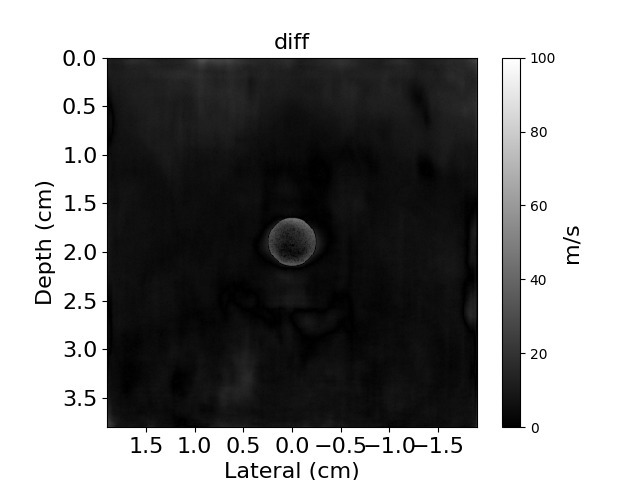}}&
\raisebox{-.5\totalheight}{\includegraphics[trim={3cm 1.5cm 4cm 1.5cm},clip, width = 1.7cm]{ 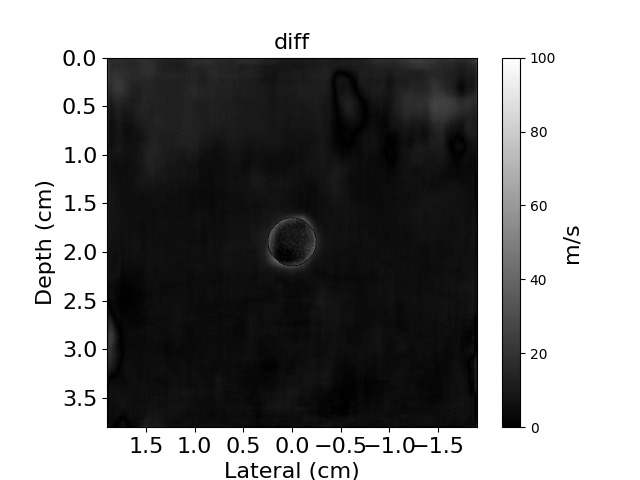}}&
\raisebox{-.5\totalheight}{\includegraphics[trim={3cm 1.5cm 4cm 1.5cm},clip, width = 1.7cm]{ 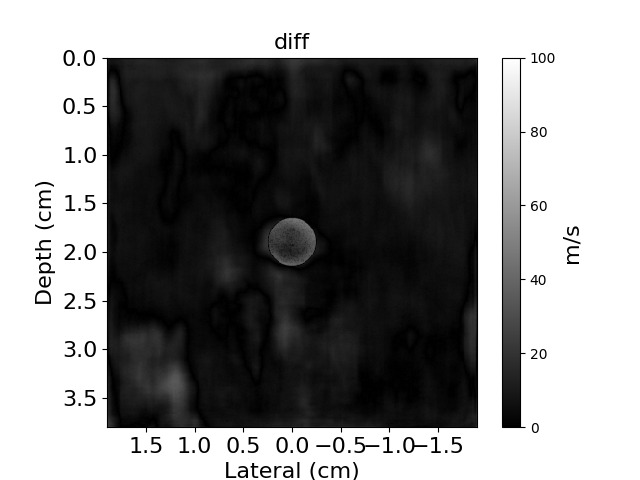}}&
\raisebox{-.5\totalheight}{\includegraphics[trim={3cm 1.5cm 4cm 1.5cm},clip, width = 1.7cm]{ 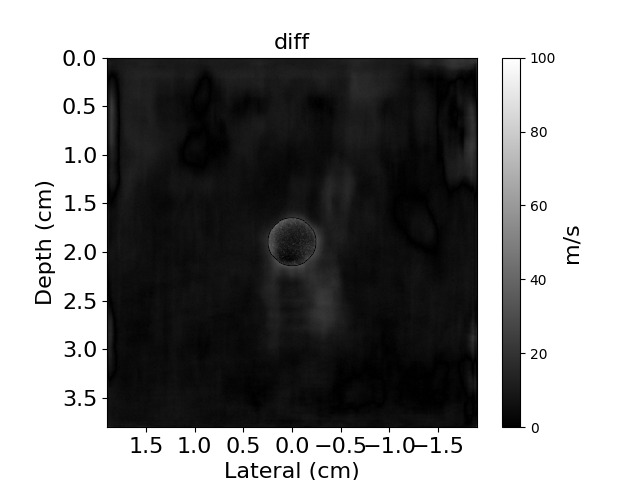}}&
\raisebox{-.5\totalheight}{\includegraphics[trim={3cm 1.5cm 4cm 1.5cm},clip, width =1.7cm]{ 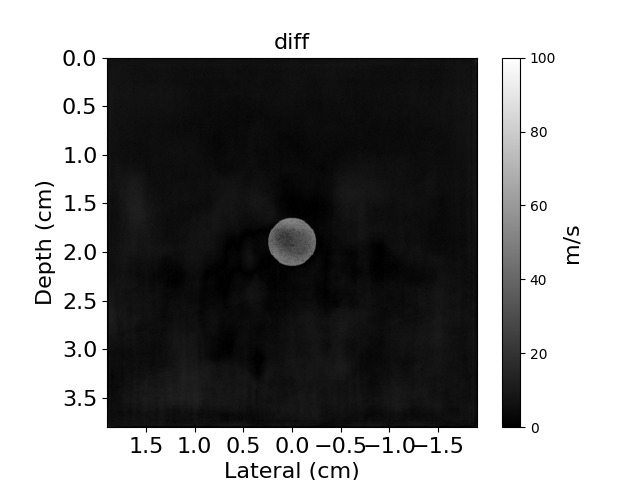}}&
\raisebox{-.5\totalheight}{\includegraphics[trim={3cm 1.5cm 4cm 1.5cm},clip, width =1.7cm]{ 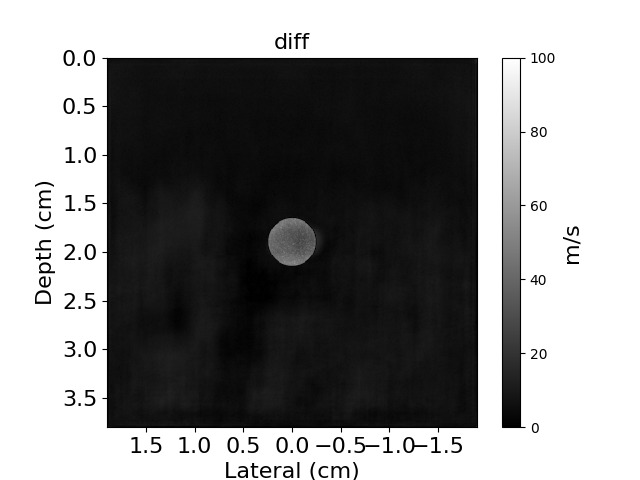}}&
\raisebox{-.5\totalheight}{\includegraphics[height =1.7cm]{img/colormaps/colorbar_diff.pdf}}\\

\raisebox{-.5\totalheight}{\scriptsize \rot{ \makecell{{Predicted} \\ {{SoS}} \\ {Ellipsoids} }}} & 
\raisebox{-.5\totalheight}{\includegraphics[trim={3cm 1.5cm 4cm 1.5cm},clip, width = 1.7cm]{ 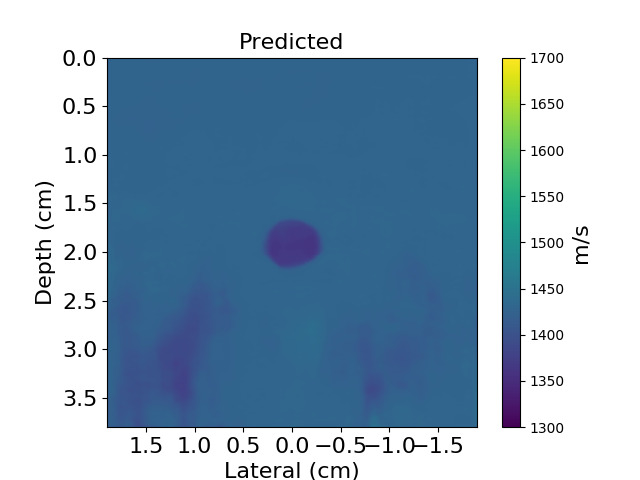}}&
\raisebox{-.5\totalheight}{\includegraphics[trim={3cm 1.5cm 4cm 1.5cm},clip, width = 1.7cm]{ 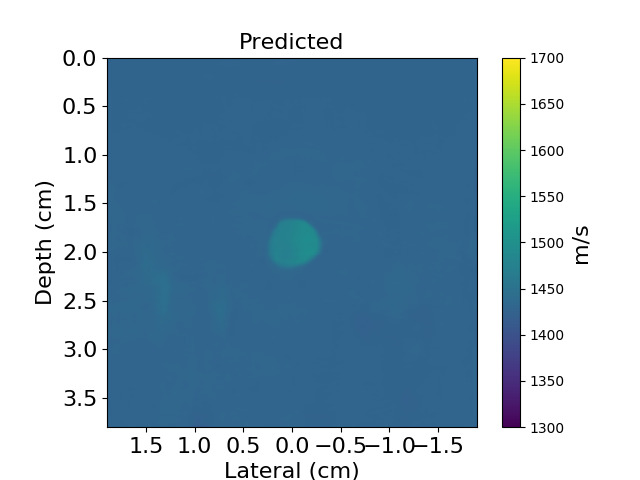}}&
\raisebox{-.5\totalheight}{\includegraphics[trim={3cm 1.5cm 4cm 1.5cm},clip, width = 1.7cm]{ 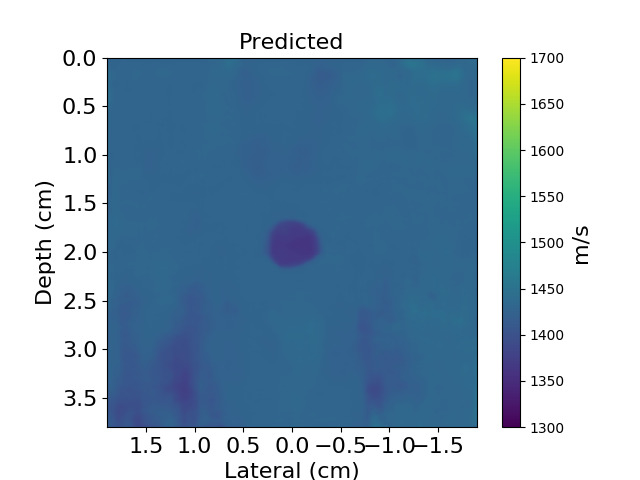}}&
\raisebox{-.5\totalheight}{\includegraphics[trim={3cm 1.5cm 4cm 1.5cm},clip, width = 1.7cm]{ 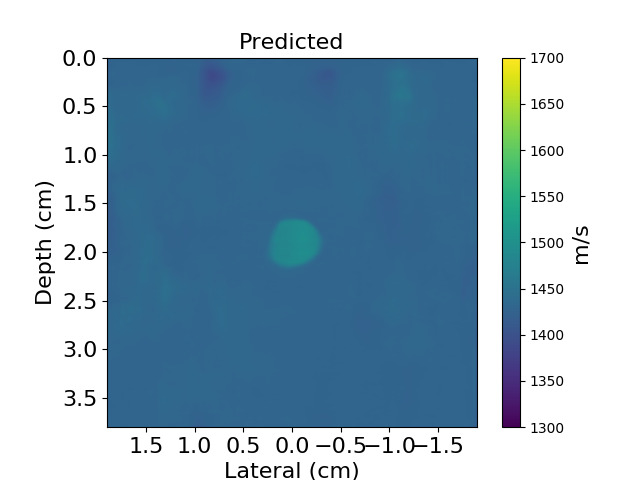}}&
\raisebox{-.5\totalheight}{\includegraphics[trim={3cm 1.5cm 4cm 1.5cm},clip, width = 1.7cm]{ 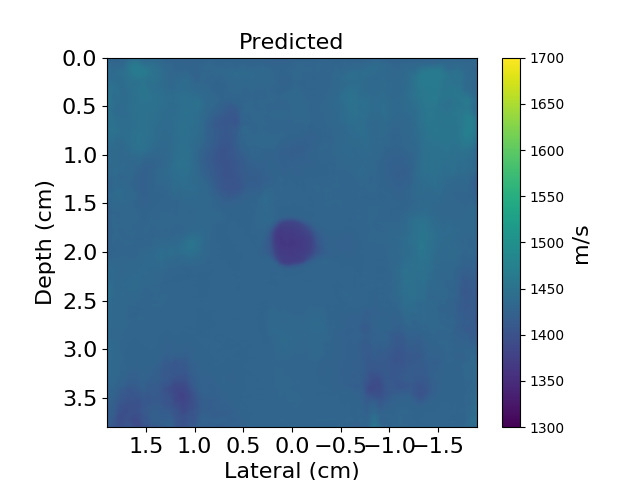}}&
\raisebox{-.5\totalheight}{\includegraphics[trim={3cm 1.5cm 4cm 1.5cm},clip, width = 1.7cm]{ 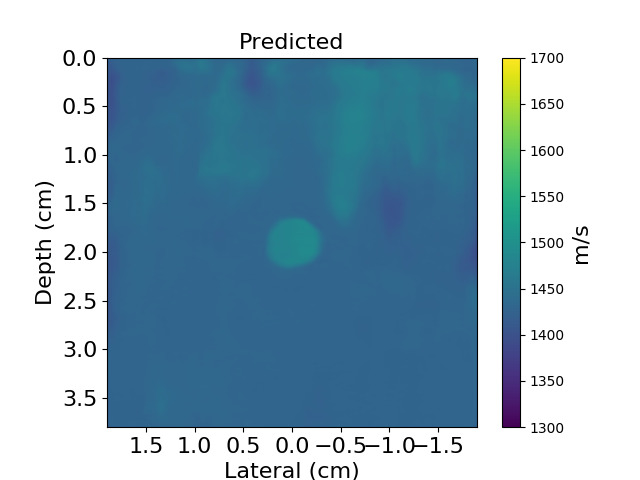}}&
\raisebox{-.5\totalheight}{\includegraphics[trim={3cm 1.5cm 4cm 1.5cm},clip, width =1.7cm]{ 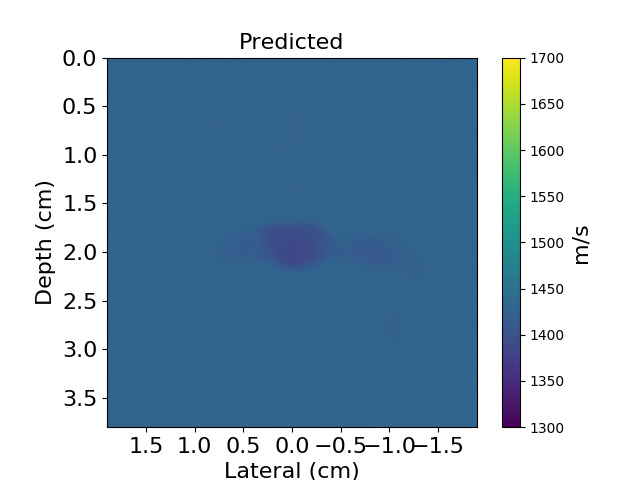}}&
\raisebox{-.5\totalheight}{\includegraphics[trim={3cm 1.5cm 4cm 1.5cm},clip, width =1.7cm]{ 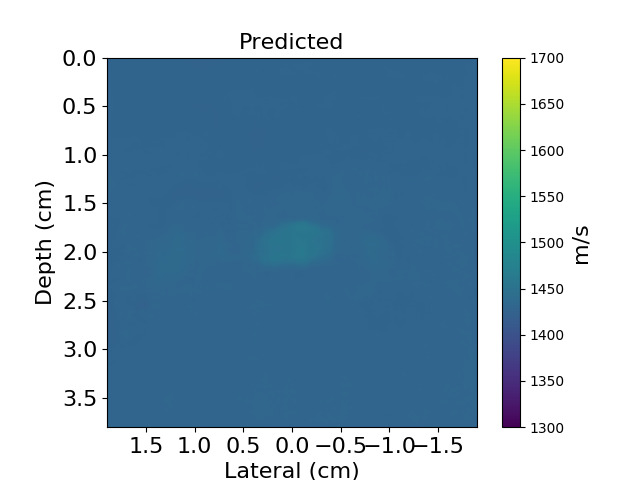}}&
\raisebox{-.5\totalheight}{\includegraphics[height =1.7cm]{img/colormaps/colorbar_simulated.pdf}}\\

\raisebox{-.5\totalheight}{\scriptsize \rot{\makecell{{Absolute} \\  {Difference} }}}& 
\raisebox{-.5\totalheight}{\includegraphics[trim={3cm 1.5cm 4cm 1.5cm},clip, width = 1.7cm]{ 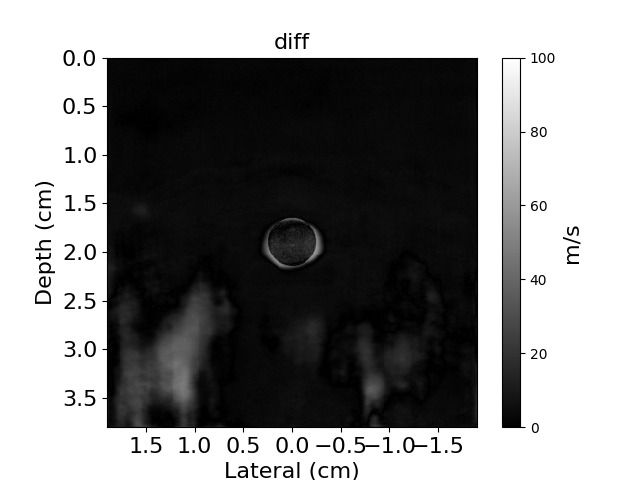}}&
\raisebox{-.5\totalheight}{\includegraphics[trim={3cm 1.5cm 4cm 1.5cm},clip, width = 1.7cm]{ 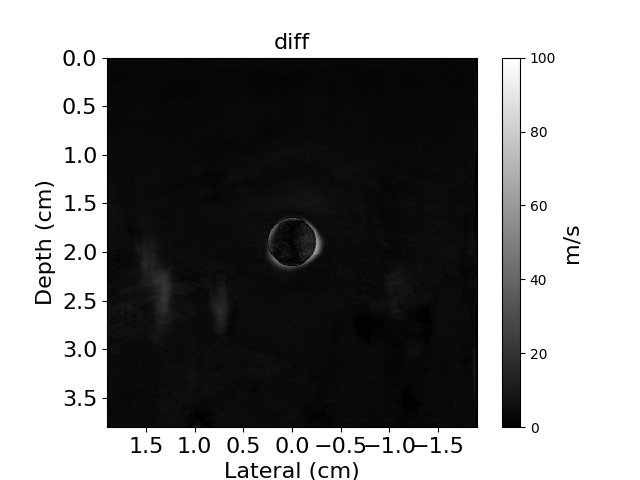}}&
\raisebox{-.5\totalheight}{\includegraphics[trim={3cm 1.5cm 4cm 1.5cm},clip, width = 1.7cm]{ 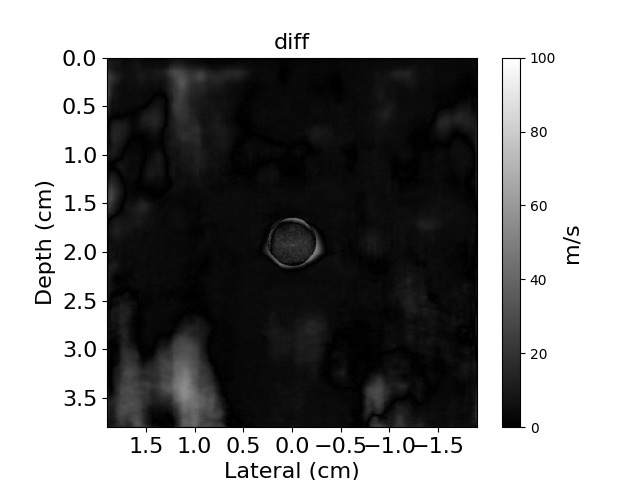}}&
\raisebox{-.5\totalheight}{\includegraphics[trim={3cm 1.5cm 4cm 1.5cm},clip, width = 1.7cm]{ 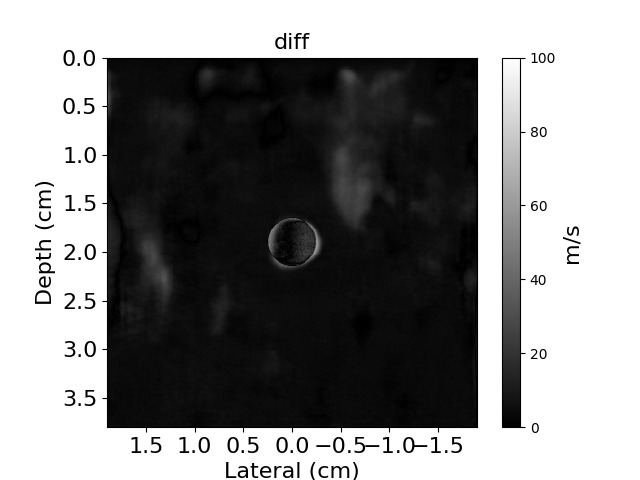}}&
\raisebox{-.5\totalheight}{\includegraphics[trim={3cm 1.5cm 4cm 1.5cm},clip, width = 1.7cm]{ 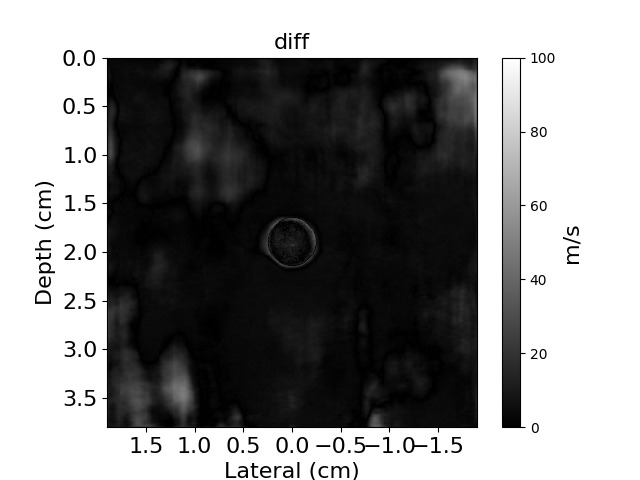}}&
\raisebox{-.5\totalheight}{\includegraphics[trim={3cm 1.5cm 4cm 1.5cm},clip, width = 1.7cm]{ 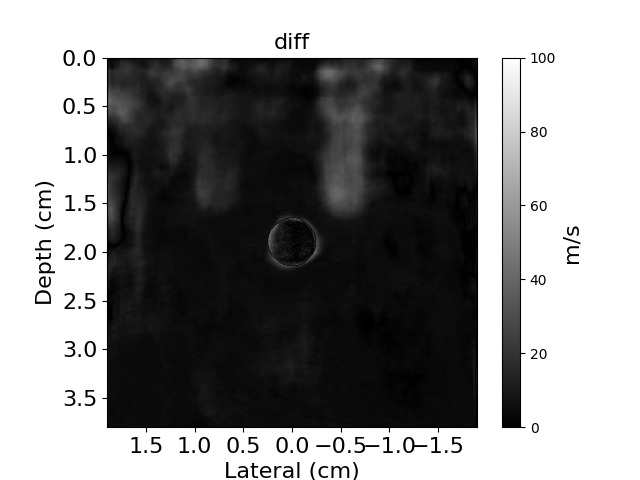}}&
\raisebox{-.5\totalheight}{\includegraphics[trim={3cm 1.5cm 4cm 1.5cm},clip, width =1.7cm]{ 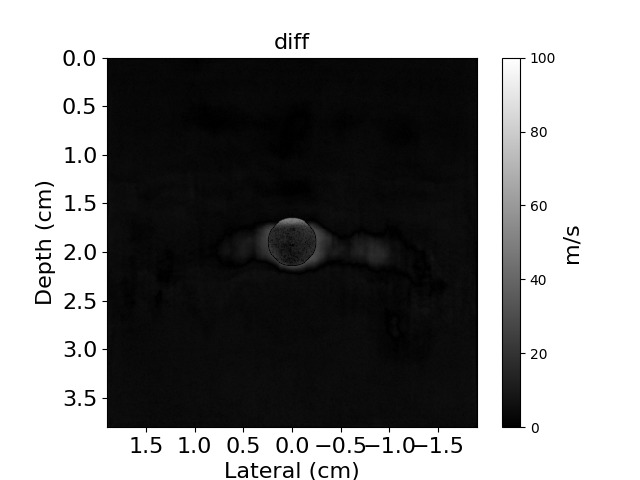}}&
\raisebox{-.5\totalheight}{\includegraphics[trim={3cm 1.5cm 4cm 1.5cm},clip, width =1.7cm]{ 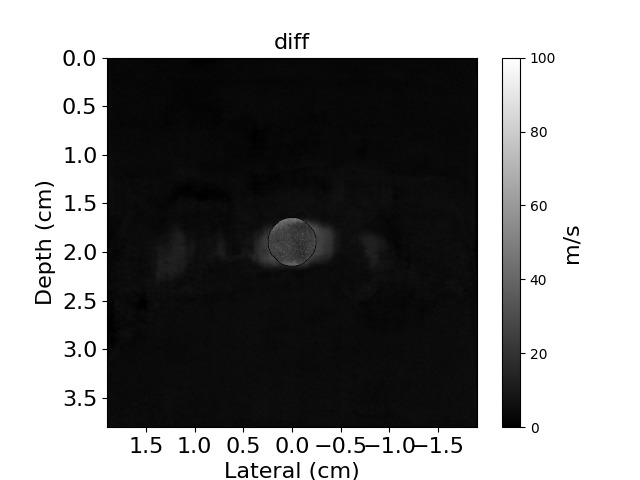}}&
\raisebox{-.5\totalheight}{\includegraphics[height =1.7cm]{img/colormaps/colorbar_diff.pdf}}\\

\end{tabular}
\caption{\textbf{Noise}: Comparison between the predicted SoS maps of noisy data by the networks trained with the Combined and the Ellipsoids dataset for homogeneous mediums with hyperechoic inclusions with SoS contrast; \textbf{(Cases 1 and 2)}: original training data with hyperechoic inclusions with SoS contrast of \(\pm50\)~\(m/s\), \textbf{(Cases 3 and 4)}: Cases 1 and 2 plus AWGN with target SNR of 15 db, \textbf{(Cases 5 and 6)}: Cases 1 and 2 plus AWGN with target SNR of 10 db, \textbf{(Cases 7 and 8)}: Cases 1 and 2 with random uncorrelated channel-wise phase distortions.}
\label{fig: digital phantom, noise}
\end{figure*}

\subparagraph{Echogenicity Contrast with SoS Contrast} Since in the training set we only considered hyperechoic regions, the aim of this investigation is to show how the networks behave in cases with SoS contrast with varied echogenicity characteristics. 

Figure~\ref{fig: rmse_in} \say{Echogenicity contrast (with SoS contrast)}, compares the RMSE values inside the inclusion and in the background for 20 cases with SoS contrast in the range \(SoS_{back}+[-50,50]\)~\(m/s\) with \(7.5\)~\(m/s\) step, where $SoS_{back}$ is the SoS in the background. 
For these sets, the Combined setup outperforms the Ellipsoids setup both inside the inclusion and in the background region. 

Figure~\ref{fig: digital phantom, SoS contrast} shows anechoic and isoechoic inclusions with SoS contrast from the background. 
Cases 1 and 2, the anechoic inclusions, have \(\pm50\)~\(m/s\) SoS contrast from the background. 
Cases 3-8 show SoS contrast in range \(SoS_{in}+[-50,50]\)~\(m/s\) with \(15\)~\(m/s\) steps. 

Generally, the RMSE values are higher compared to the previous cases shown in Figure~\ref{fig: rmse_in} \say{Echogenicity contrast (No SoS contrast)} and \say{\#Reflective speckles}. This can be seen in Figure \ref{fig: digital phantom, SoS contrast} absolute difference images. 
Anechoic inclusions are more challenging due to the absence of scatterers inside the inclusions. 
For isoechoic cases, the predicted SoS maps by the Combined setup have clearer margins and fewer under/overestimations outside the inclusion. 
Additionally, a shadowing effect can be seen directly below the inclusion which increases the overall RMSE. 

The interesting finding from this investigation is that the networks can indicate the presence of inclusion even when the inclusion is not visible in the b-mode images (isoechoic cases). Although isoechoic lesions are often benign, they are known to be challenging to find in conventional ultrasound imaging, and often a complementary tool is required to find those lesions \citep{kim2011find}. SoS reconstruction can be a potential solution to find these kinds of lesions and prevent false-negative interpretations and a delayed diagnosis of breast cancers.

\paragraph{Noise} The aim of the following sections is to show how additive and phase noise affects the predictions of the networks:

\subparagraph{Additive Noise} Additive White Gaussian Noise (AWGN) is added with a target signal-to-noise ratio (SNR) of 10, 15, and 20 db, where $SNR = P_s / P_n$. $\text P_s$  is the power of the signal and $\text P_n$ is the power of the noise. 
Figure \ref{fig: rmse_in}, Noise boxes, show the RMSE comparison inside the inclusion and in the background regions of 20 cases, for each target SNR, with SoS contrast in the range $SoS_{\text{back}}+[-75,75]$~\(m/s\) with \(7.5\)~\(m/s\) steps. $SoS_{\text{back}}$ is the SoS value in the background. 
For this investigation, only AWGN is added and echogenicity is considered to be hyperechoic (same as the training set).

Based on Figure \ref{fig: rmse_in} \say{Noise (20db)}, \say{Noise (15db)}, and \say{Noise (10db)} boxes, as the additive noise increases, the RMSE values inside the inclusions are increased for both Ellipsoids and Combined setup. 
However, the variation of the RMSE in the background regions is lower for the Combined setup, and as the noise increases the gap between the two setups increases.
This shows that the Combined setup is more robust to the noise. 
In Figure \ref{fig: digital phantom, noise}, Cases 3 and 4, and Cases 5 and 6 examples of the data with target SNR of 15 db and 10 db are shown, respectively. 
 Figure \ref{fig: digital phantom, noise}, Cases 1 and 2 show the original data (without additive noise).
The SoS contrast for each pares is \(\pm50\)~\(m/s\) of SoS in the background.  

\subparagraph{Phase Noise} Studies showed that phase distortions can affect SoS predictions \citep{stahli2020bayesian,jush2021data}.
Therefore, we  applied uncorrelated phase distortions by shifting the phase of each channel by a random value in the range \([-0.7,0.7]\) radian.
Figure~\ref{fig: rmse_in} \say{Phase noise} boxes show the corresponding RMSE values inside the inclusion and for the background in the presence of such a noise for 20 cases. 
The highest RMSE values and the highest variation between cases for the Combined setup belong to the RMSE inside the inclusions.
Figure~\ref{fig: digital phantom, noise} Cases 7 and 8 show cases with described phase distortions. Although there is an indication of inclusion by both networks, the margins are distorted. Based on Figure \ref{fig: rmse_in}, the absolute value of the SoS predictions has a higher offset for the Combined setup. 
This indicates that the Combined setup is highly sensitive to phase distortions. 

The predicted SoS values in the background regions for the Combined setup show a similar pattern as \say{Echogenicity contrast (No SoS contrast)} and \say{\#Reflective speckles} where the RMSE is slightly lower than the Ellipsoids setup.

\begin{figure*}[!t]
\centering 

\renewcommand{\arraystretch}{0.05}
\begin{tabular}{@{\hspace{0.5mm}} c @{\hspace{0.5mm}}c @{\hspace{0.5mm}}c @{\hspace{0.5mm}}c @ {\hspace{0.5mm}}c @{\hspace{0.5mm}}c  @{\hspace{0.5mm}}c  @{\hspace{0.5mm}}c  @{\hspace{0.5mm}}c @{\hspace{0.5mm}}l }
& \scriptsize{ Case  1} & \scriptsize{ Case  2} & \scriptsize{Case  3} & \scriptsize{Case  4} & \scriptsize{Case  5}  & \scriptsize{ Case  6} & \scriptsize{Case  7} & \scriptsize{Case 8 } &  \\

\raisebox{-.5\totalheight}{\scriptsize \rot {B-mode}} & 
\raisebox{-.5\totalheight}{\includegraphics[trim={3cm 1.5cm 4cm 1cm},clip, width = 1.7cm]{ 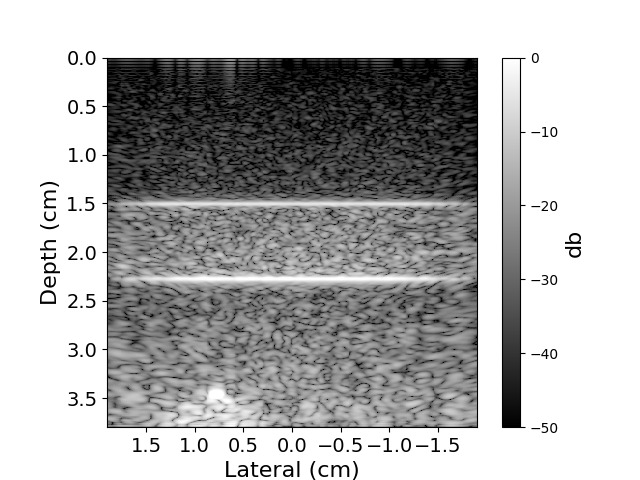}}&
\raisebox{-.5\totalheight}{\includegraphics[trim={3cm 1.5cm 4cm 1cm},clip, width = 1.7cm]{ 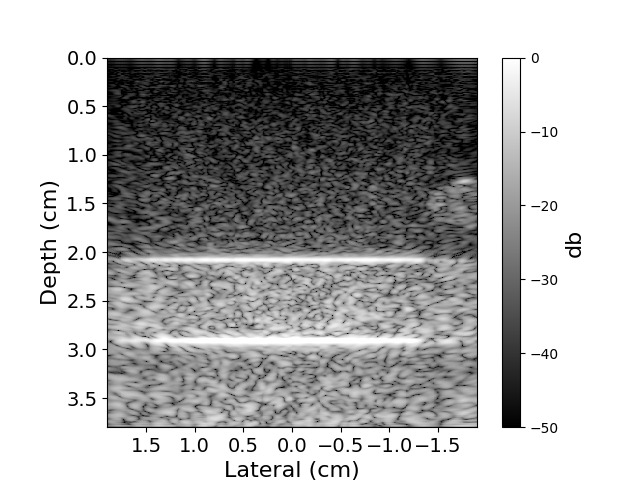}}&
\raisebox{-.5\totalheight}{\includegraphics[trim={3cm 1.5cm 4cm 1cm},clip, width = 1.7cm]{ 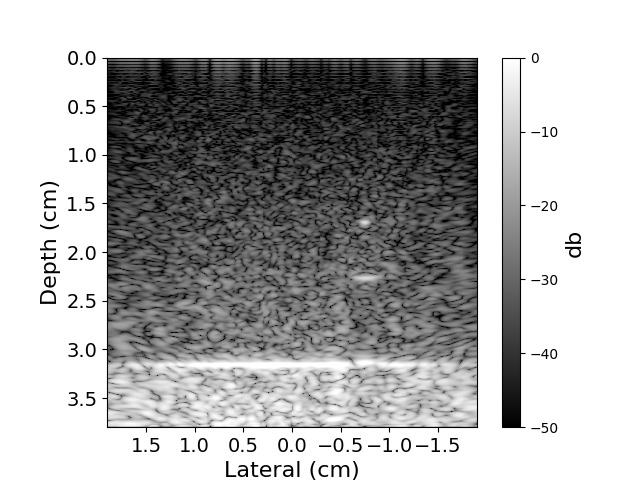}}&
\raisebox{-.5\totalheight}{\includegraphics[trim={3cm 1.5cm 4cm 1cm},clip, width = 1.7cm]{ 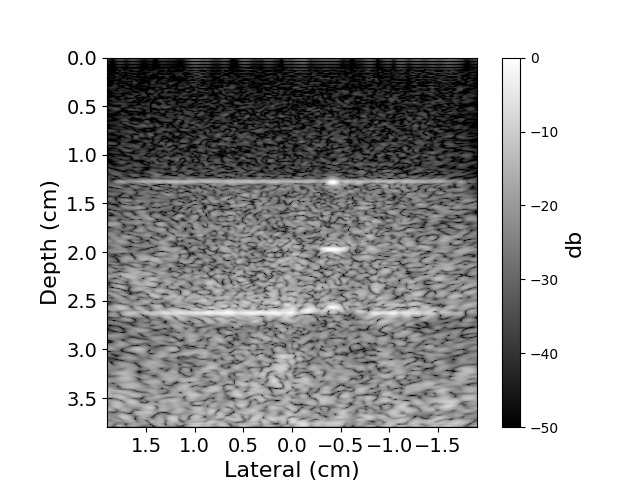}}&
\raisebox{-.5\totalheight}{\includegraphics[trim={3cm 1.5cm 4cm 1cm},clip, width = 1.7cm]{ 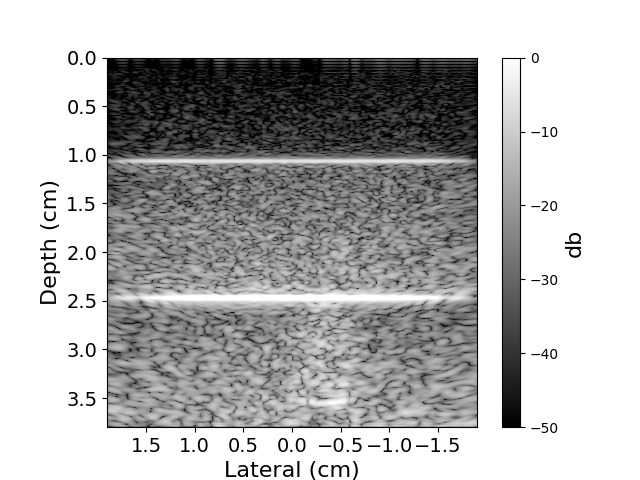}}&
\raisebox{-.5\totalheight}{\includegraphics[trim={3cm 1.5cm 4cm 1cm},clip, width =1.7cm]{ 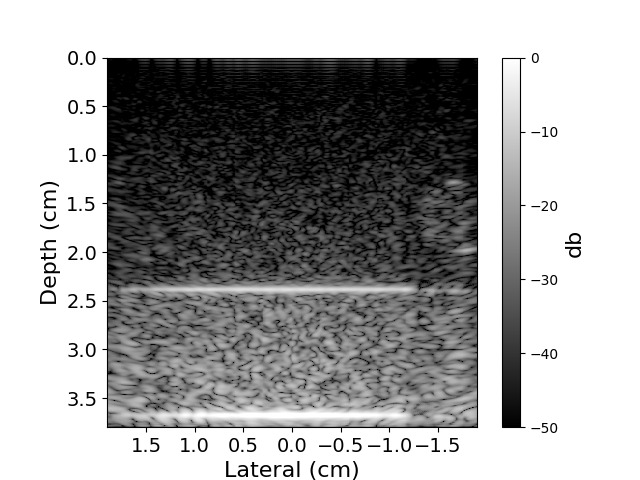}}&
\raisebox{-.5\totalheight}{\includegraphics[trim={3cm 1.5cm 4cm 1cm},clip, width =1.7cm]{ 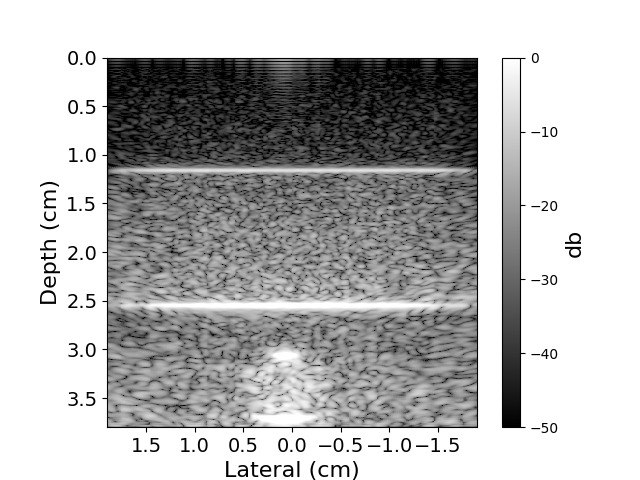}}&
\raisebox{-.5\totalheight}{\includegraphics[trim={3cm 1.5cm 4cm 1cm},clip, width = 1.7cm]{ 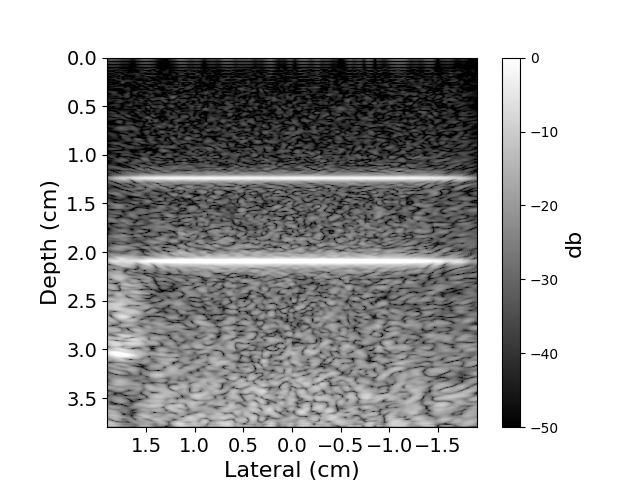}}&
\raisebox{-.5\totalheight}{\includegraphics[height =1.7cm]{img/colormaps/colorbar_bmode.pdf}}\\

\raisebox{-.5\totalheight}{\scriptsize \rot{GT}} & 
\raisebox{-.5\totalheight}{\includegraphics[trim={3cm 1.5cm 4cm 1.5cm},clip, width = 1.7cm]{ 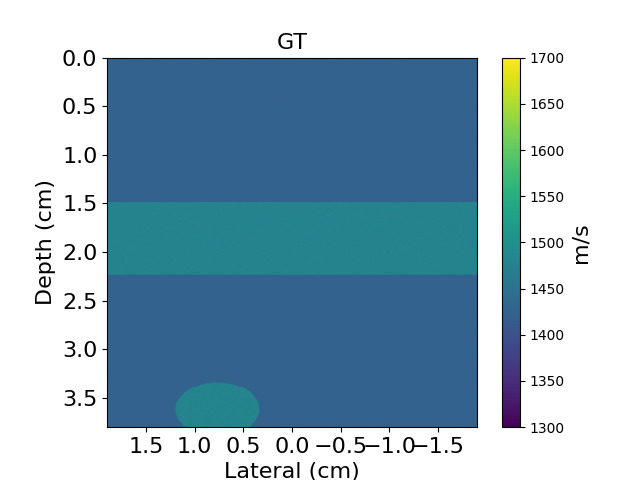}}&
\raisebox{-.5\totalheight}{\includegraphics[trim={3cm 1.5cm 4cm 1.5cm},clip, width = 1.7cm]{ 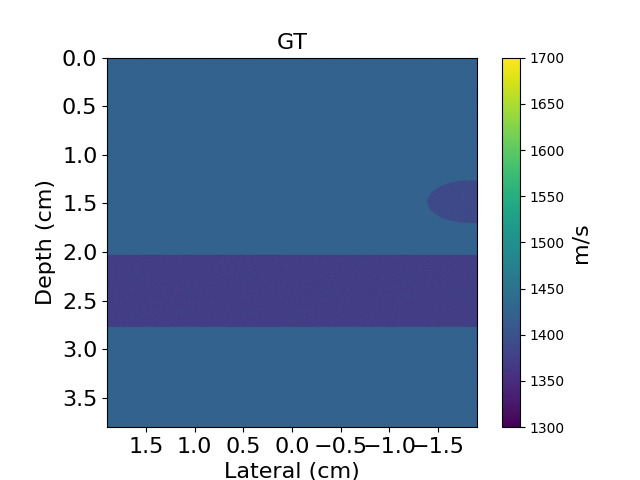}}&
\raisebox{-.5\totalheight}{\includegraphics[trim={3cm 1.5cm 4cm 1.5cm},clip, width = 1.7cm]{ 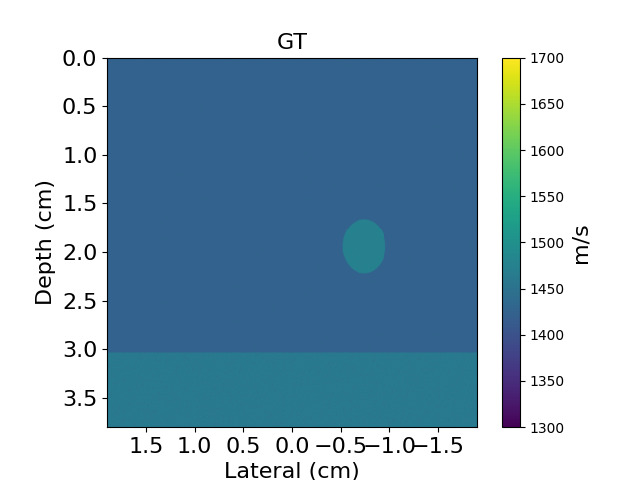}}&
\raisebox{-.5\totalheight}{\includegraphics[trim={3cm 1.5cm 4cm 1.5cm},clip, width = 1.7cm]{ 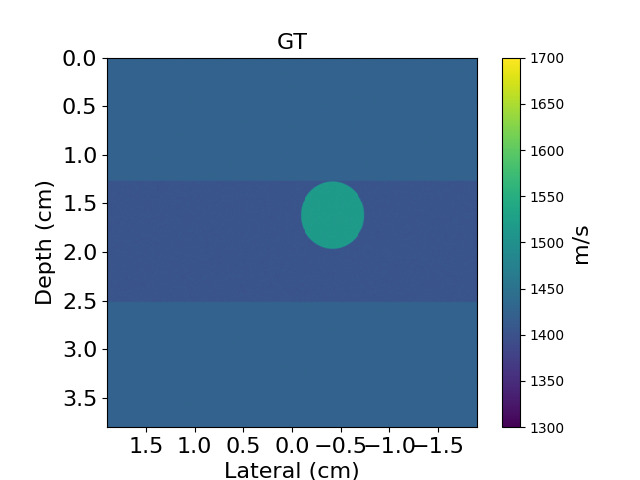}}&
\raisebox{-.5\totalheight}{\includegraphics[trim={3cm 1.5cm 4cm 1.5cm},clip, width = 1.7cm]{ 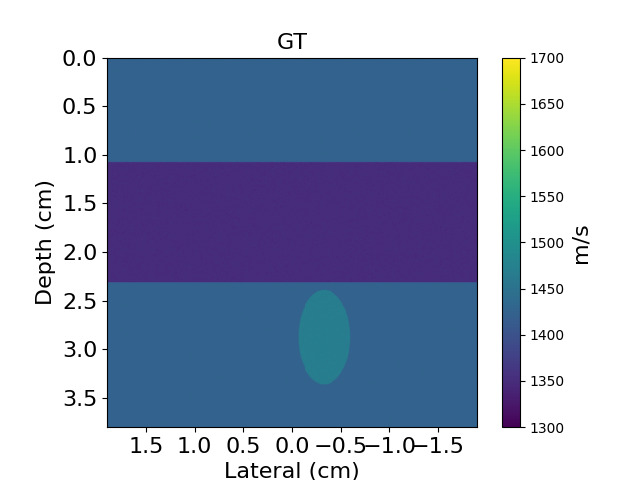}}&
\raisebox{-.5\totalheight}{\includegraphics[trim={3cm 1.5cm 4cm 1.5cm},clip, width =1.7cm]{ 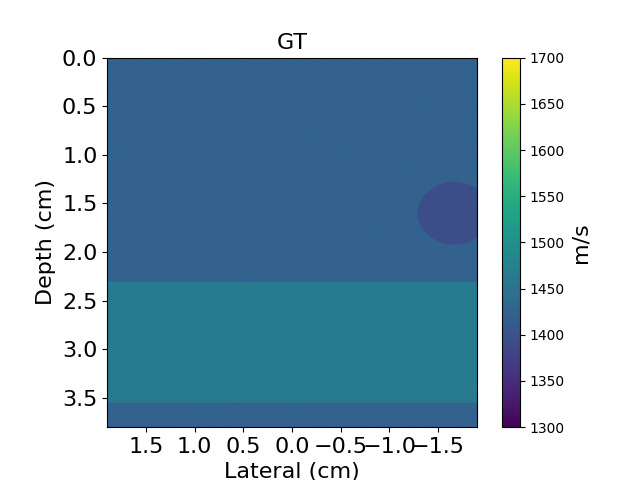}}&
\raisebox{-.5\totalheight}{\includegraphics[trim={3cm 1.5cm 4cm 1.5cm},clip, width =1.7cm]{ 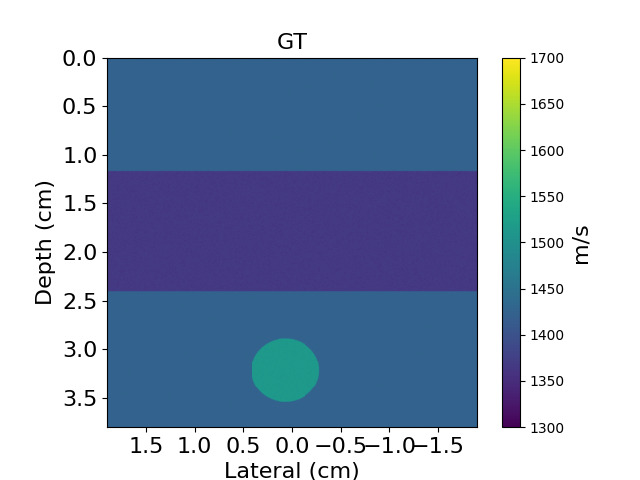}}&
\raisebox{-.5\totalheight}{\includegraphics[trim={3cm 1.5cm 4cm 1.5cm},clip, width = 1.7cm]{ 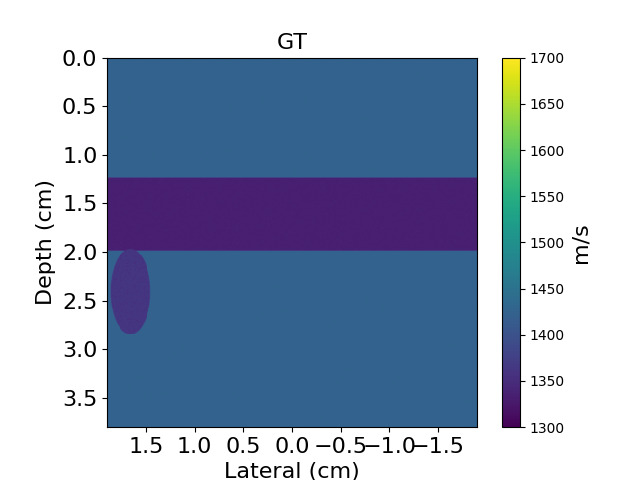}}&
\raisebox{-.5\totalheight}{\includegraphics[height =1.7cm]{img/colormaps/colorbar_simulated.pdf}}\\
 
\raisebox{-.5\totalheight}{\scriptsize \rot {\makecell{{Predicted} \\ {{SoS}} \\ {Combined} }}} & 
\raisebox{-.5\totalheight}{\includegraphics[trim={3cm 1.5cm 4cm 1.5cm},clip, width = 1.7cm]{ 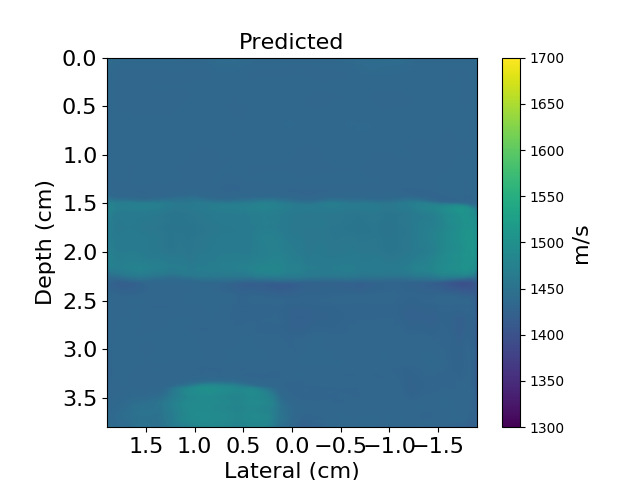}}&
\raisebox{-.5\totalheight}{\includegraphics[trim={3cm 1.5cm 4cm 1.5cm},clip, width = 1.7cm]{ 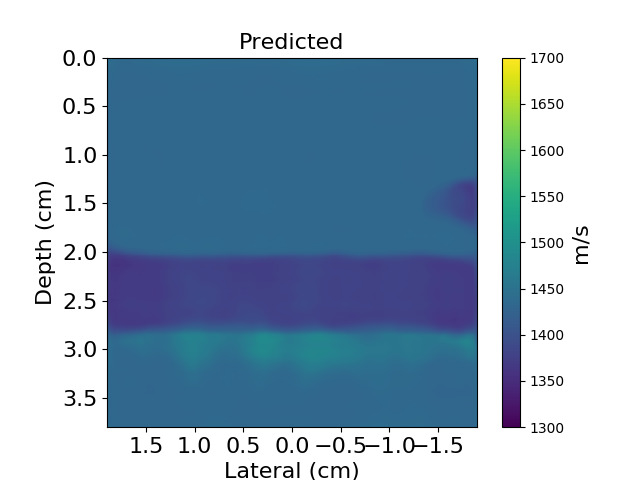}}&
\raisebox{-.5\totalheight}{\includegraphics[trim={3cm 1.5cm 4cm 1.5cm},clip, width = 1.7cm]{ 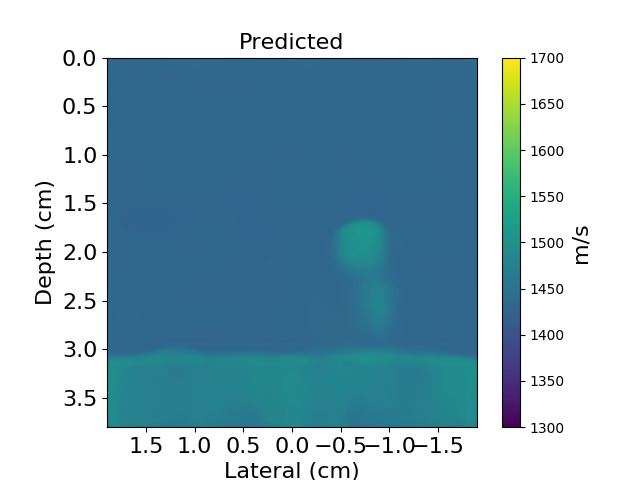}}&
\raisebox{-.5\totalheight}{\includegraphics[trim={3cm 1.5cm 4cm 1.5cm},clip, width = 1.7cm]{ 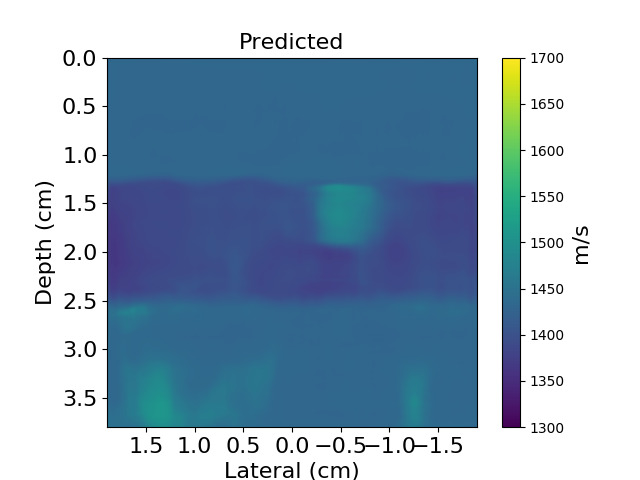}}&
\raisebox{-.5\totalheight}{\includegraphics[trim={3cm 1.5cm 4cm 1.5cm},clip, width = 1.7cm]{ 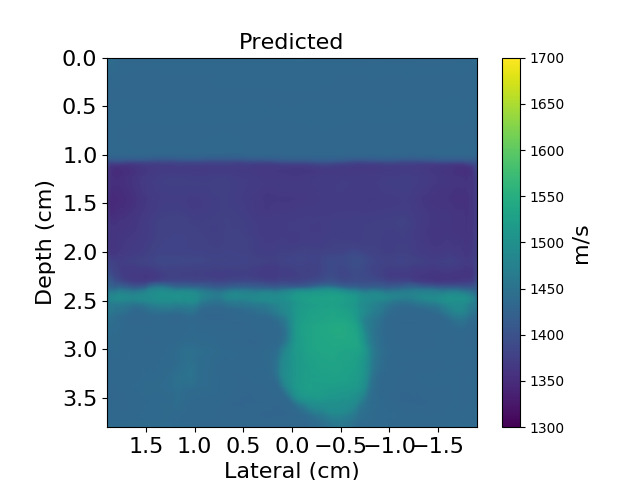}}&
\raisebox{-.5\totalheight}{\includegraphics[trim={3cm 1.5cm 4cm 1.5cm},clip, width =1.7cm]{ 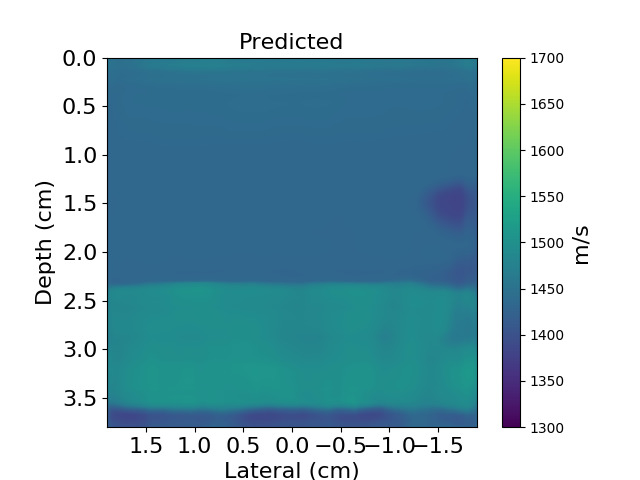}}&
\raisebox{-.5\totalheight}{\includegraphics[trim={3cm 1.5cm 4cm 1.5cm},clip, width =1.7cm]{ 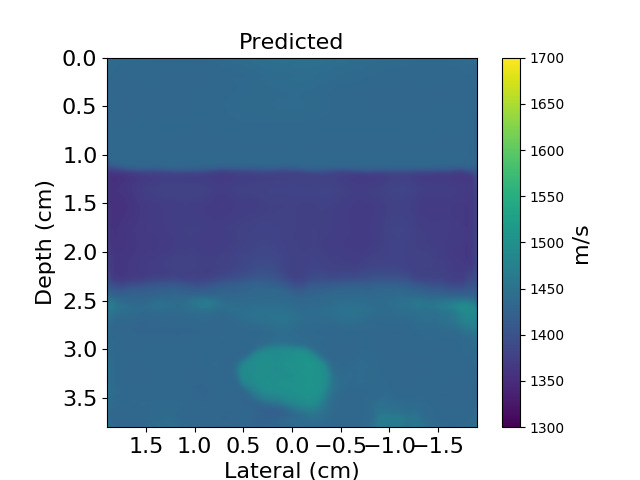}}&
\raisebox{-.5\totalheight}{\includegraphics[trim={3cm 1.5cm 4cm 1.5cm},clip, width = 1.7cm]{ 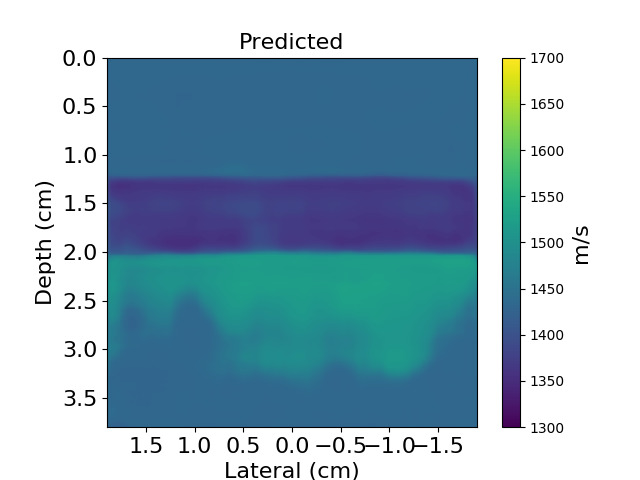}}&
\raisebox{-.5\totalheight}{\includegraphics[height =1.7cm]{img/colormaps/colorbar_simulated.pdf}}\\

\raisebox{-.5\totalheight}{\scriptsize \rot{ \makecell{{Absolute} \\ {{Difference}} }}} & 
\raisebox{-.5\totalheight}{\includegraphics[trim={3cm 1.5cm 4cm 1.5cm},clip, width = 1.7cm]{ 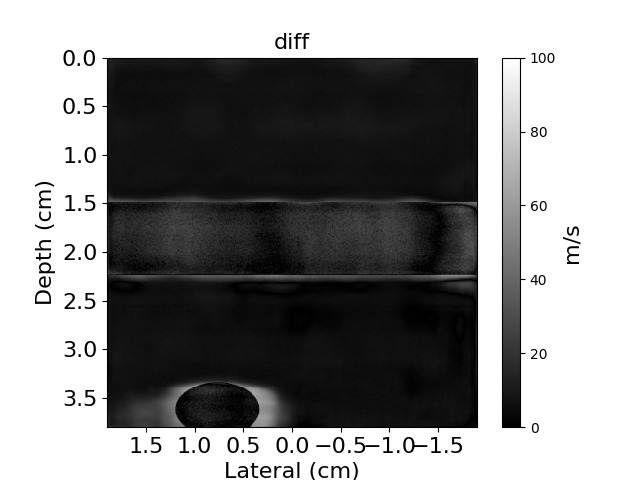}}&
\raisebox{-.5\totalheight}{\includegraphics[trim={3cm 1.5cm 4cm 1.5cm},clip, width = 1.7cm]{ 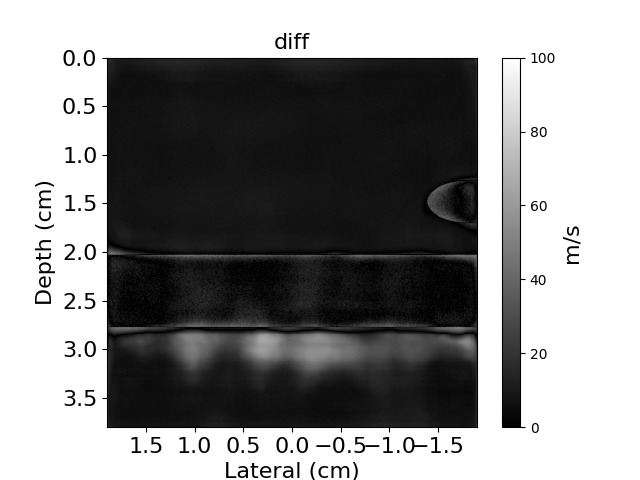}}&
\raisebox{-.5\totalheight}{\includegraphics[trim={3cm 1.5cm 4cm 1.5cm},clip, width = 1.7cm]{ 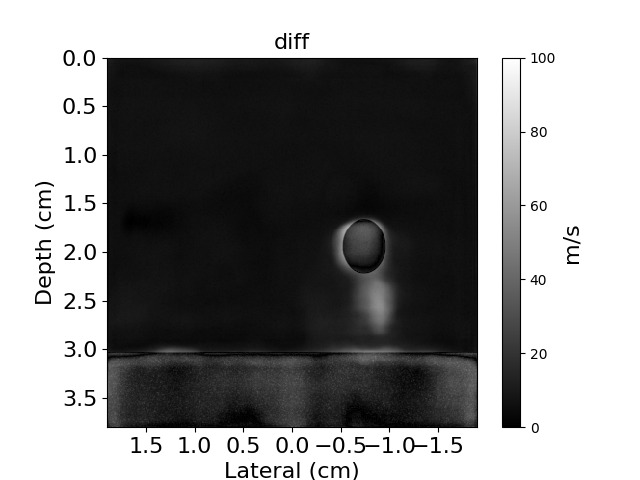}}&
\raisebox{-.5\totalheight}{\includegraphics[trim={3cm 1.5cm 4cm 1.5cm},clip, width = 1.7cm]{ 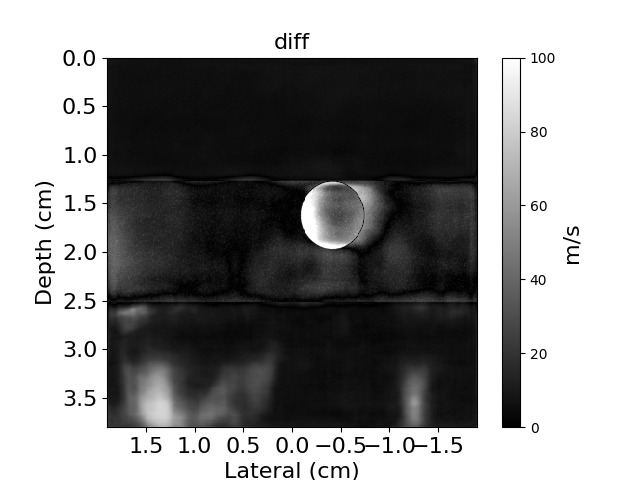}}&
\raisebox{-.5\totalheight}{\includegraphics[trim={3cm 1.5cm 4cm 1.5cm},clip, width = 1.7cm]{ 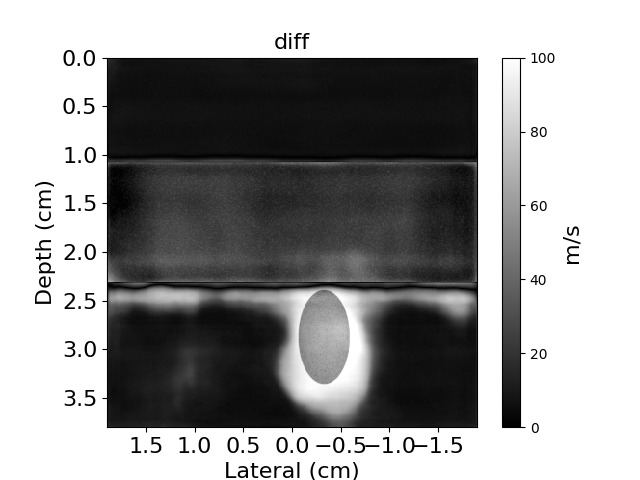}}&
\raisebox{-.5\totalheight}{\includegraphics[trim={3cm 1.5cm 4cm 1.5cm},clip, width =1.7cm]{ 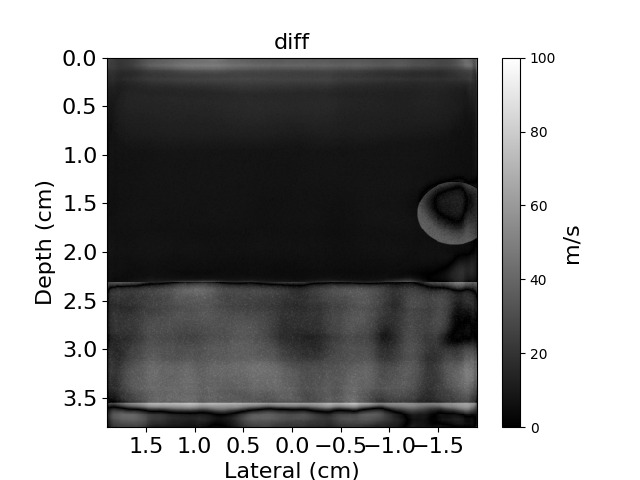}}&
\raisebox{-.5\totalheight}{\includegraphics[trim={3cm 1.5cm 4cm 1.5cm},clip, width =1.7cm]{ 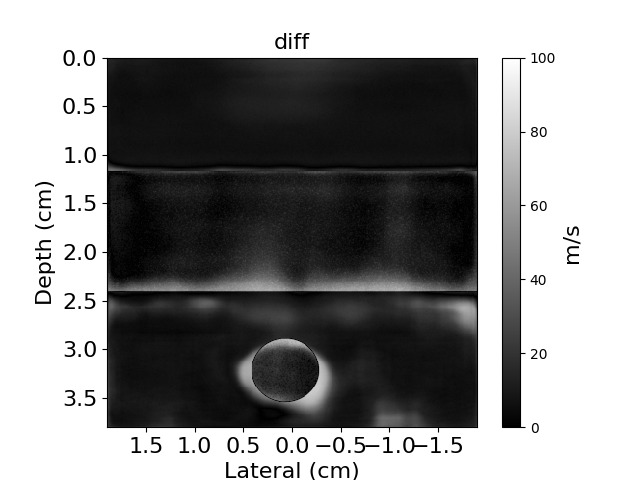}}&
\raisebox{-.5\totalheight}{\includegraphics[trim={3cm 1.5cm 4cm 1.5cm},clip, width = 1.7cm]{ 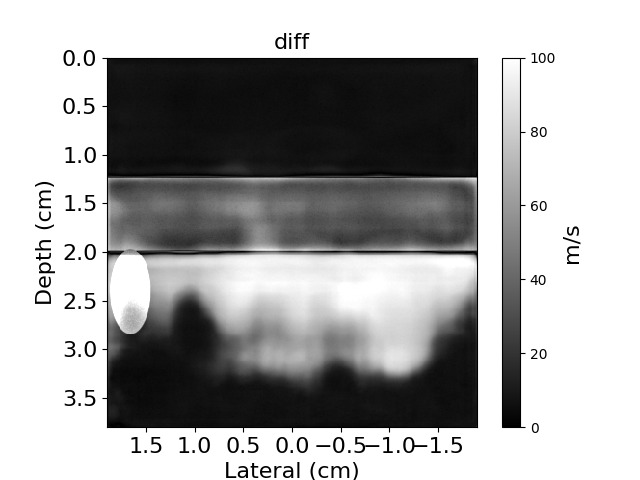}}&
\raisebox{-.5\totalheight}{\includegraphics[height =1.7cm]{img/colormaps/colorbar_diff.pdf}}\\

\rule{0pt}{15pt}{\scriptsize {RMSE} }& 
\scriptsize {9.17}& 
\scriptsize {11.17}& 
\scriptsize {9.72}&
\scriptsize {16.00}&
\scriptsize{25.19}& 
\scriptsize{19.30}&
\scriptsize{13.41}&
\scriptsize{43.34} & \scriptsize [m/s] \\  [10pt]

\raisebox{-.5\totalheight}{\scriptsize \rot{ \makecell{{Predicted} \\ {{SoS}} \\ {Ellipsoids} }}} & 
\raisebox{-.5\totalheight}{\includegraphics[trim={3cm 1.5cm 4cm 1.5cm},clip, width = 1.7cm]{ 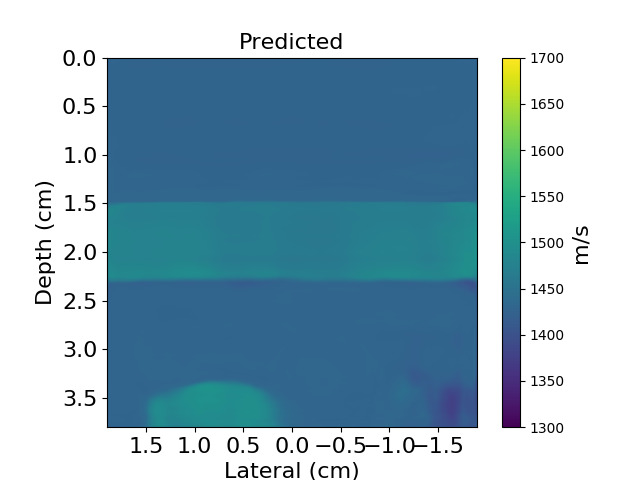}}&
\raisebox{-.5\totalheight}{\includegraphics[trim={3cm 1.5cm 4cm 1.5cm},clip, width = 1.7cm]{ 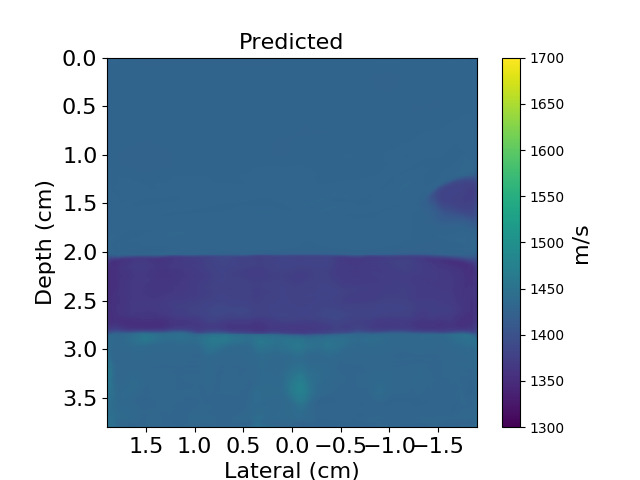}}&
\raisebox{-.5\totalheight}{\includegraphics[trim={3cm 1.5cm 4cm 1.5cm},clip, width = 1.7cm]{ 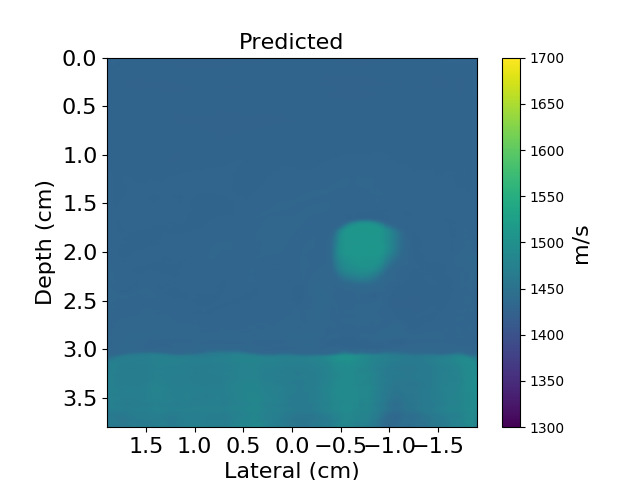}}&
\raisebox{-.5\totalheight}{\includegraphics[trim={3cm 1.5cm 4cm 1.5cm},clip, width = 1.7cm]{ 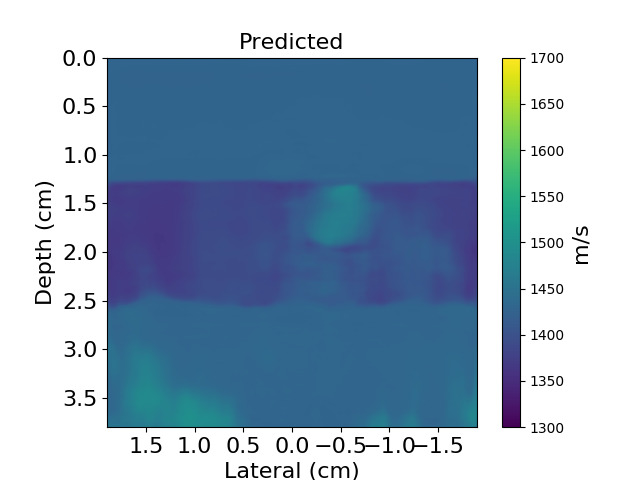}}&
\raisebox{-.5\totalheight}{\includegraphics[trim={3cm 1.5cm 4cm 1.5cm},clip, width = 1.7cm]{ 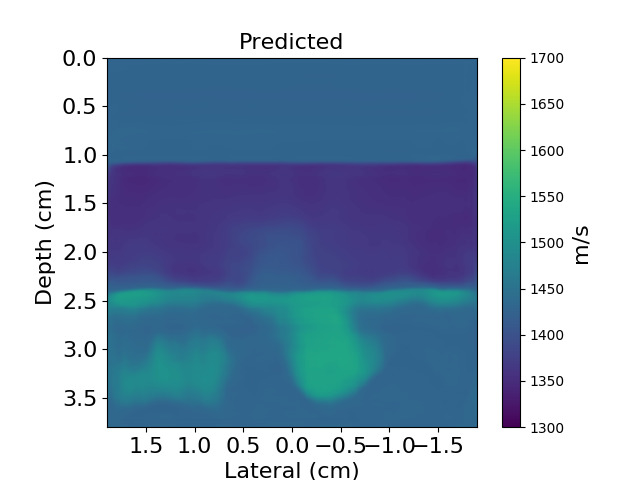}}&
\raisebox{-.5\totalheight}{\includegraphics[trim={3cm 1.5cm 4cm 1.5cm},clip, width =1.7cm]{ 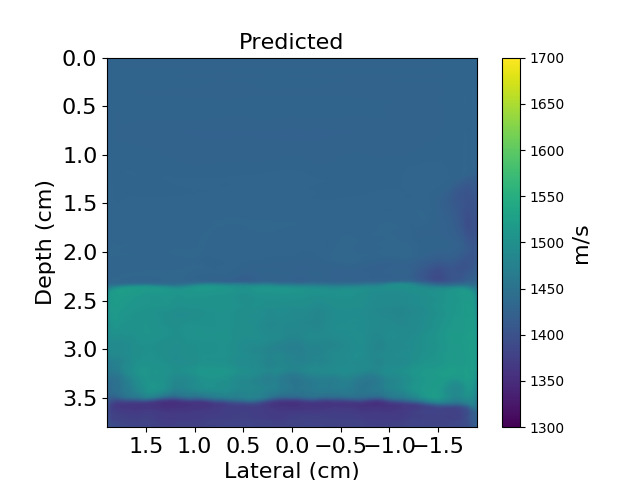}}&
\raisebox{-.5\totalheight}{\includegraphics[trim={3cm 1.5cm 4cm 1.5cm},clip, width =1.7cm]{ 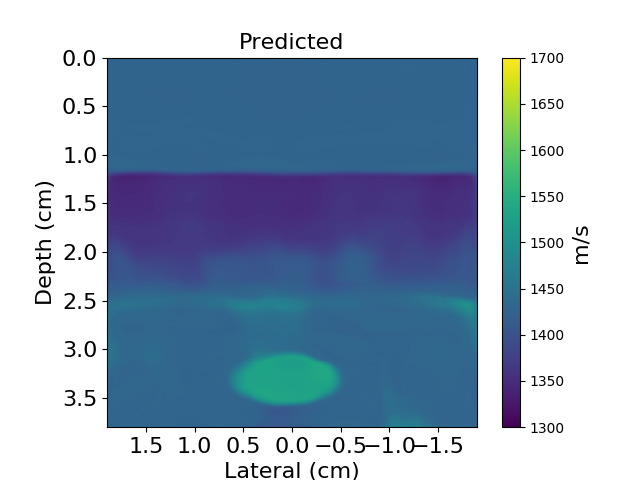}}&
\raisebox{-.5\totalheight}{\includegraphics[trim={3cm 1.5cm 4cm 1.5cm},clip, width = 1.7cm]{ 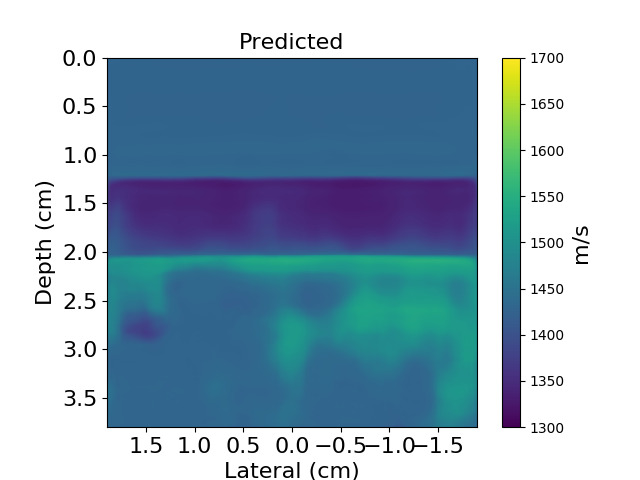}}&
\raisebox{-.5\totalheight}{\includegraphics[height =1.7cm]{img/colormaps/colorbar_simulated.pdf}}\\

\raisebox{-.5\totalheight}{\scriptsize \rot{\makecell{{Absolute} \\  {Difference} }}}& 
\raisebox{-.5\totalheight}{\includegraphics[trim={3cm 1.5cm 4cm 1.5cm},clip, width = 1.7cm]{ 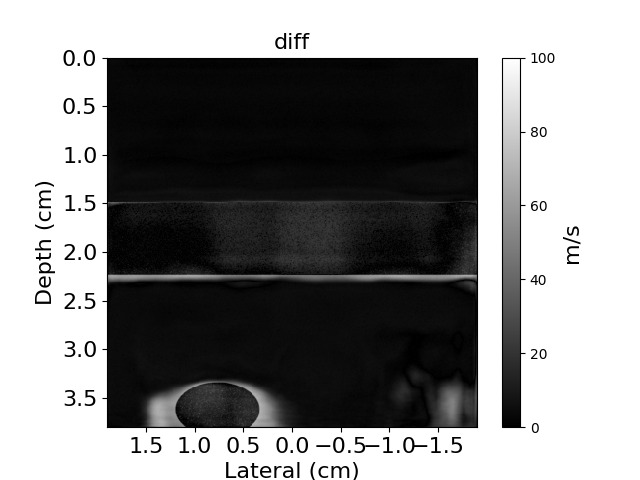}}&
\raisebox{-.5\totalheight}{\includegraphics[trim={3cm 1.5cm 4cm 1.5cm},clip, width = 1.7cm]{ 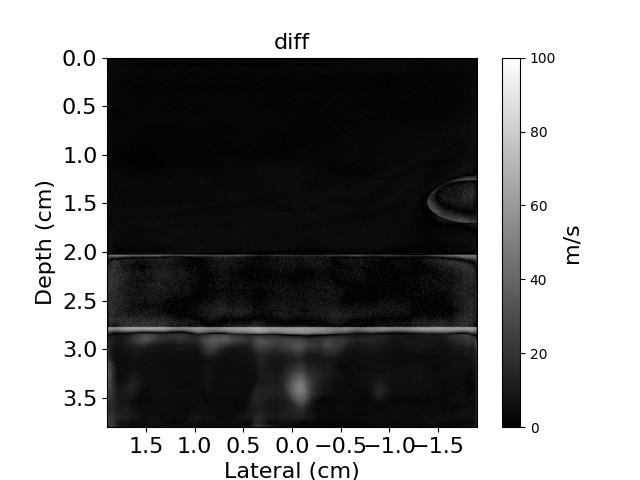}}&
\raisebox{-.5\totalheight}{\includegraphics[trim={3cm 1.5cm 4cm 1.5cm},clip, width = 1.7cm]{ 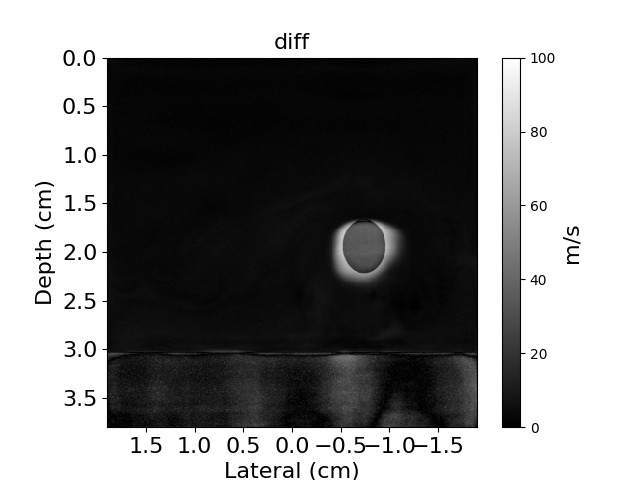}}&
\raisebox{-.5\totalheight}{\includegraphics[trim={3cm 1.5cm 4cm 1.5cm},clip, width = 1.7cm]{ 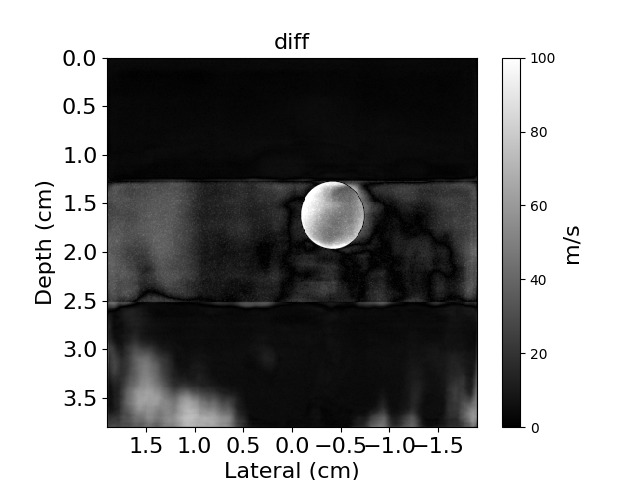}}&
\raisebox{-.5\totalheight}{\includegraphics[trim={3cm 1.5cm 4cm 1.5cm},clip, width = 1.7cm]{ 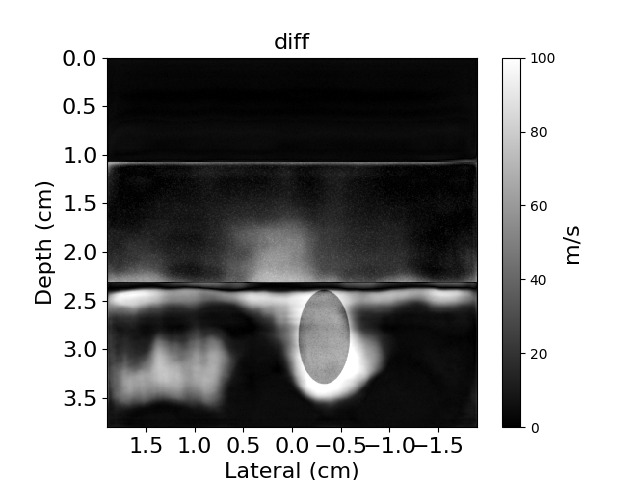}}&
\raisebox{-.5\totalheight}{\includegraphics[trim={3cm 1.5cm 4cm 1.5cm},clip, width =1.7cm]{ 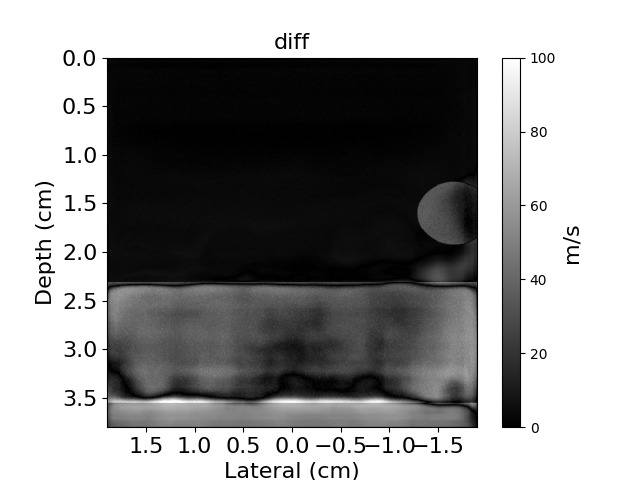}}&
\raisebox{-.5\totalheight}{\includegraphics[trim={3cm 1.5cm 4cm 1.5cm},clip, width =1.7cm]{ 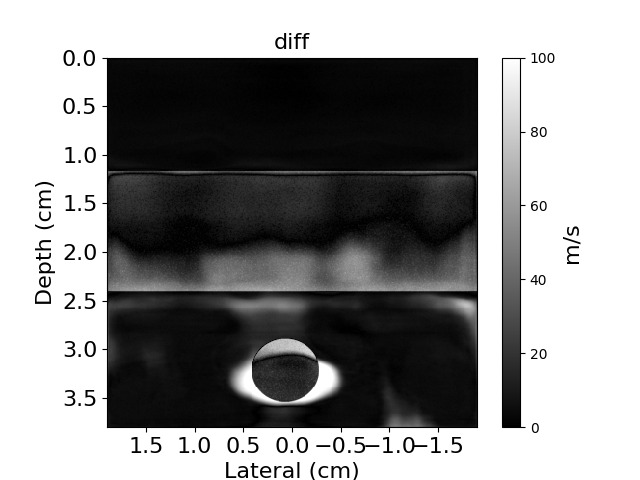}}&
\raisebox{-.5\totalheight}{\includegraphics[trim={3cm 1.5cm 4cm 1.5cm},clip, width = 1.7cm]{ 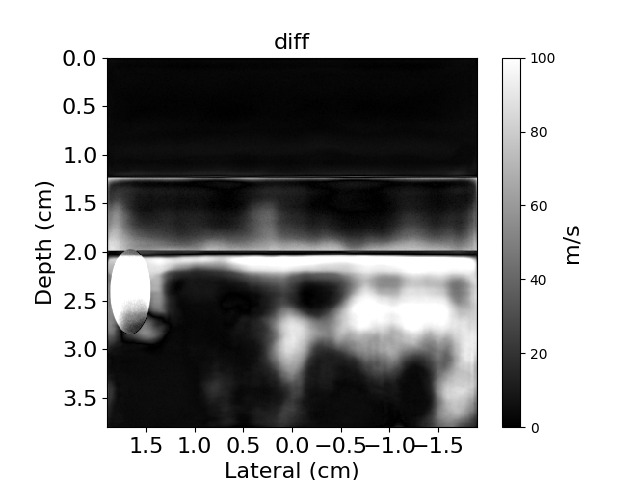}}&
\raisebox{-.5\totalheight}{\includegraphics[height =1.7cm]{img/colormaps/colorbar_diff.pdf}}\\

\rule{0pt}{15pt}{\scriptsize {RMSE} }& 
\scriptsize{12.04}& 
\scriptsize{10.74}& 
\scriptsize{10.87}&
\scriptsize{18.31}&
\scriptsize{19.90}& 
\scriptsize{25.11}&
\scriptsize{20.82}&
\scriptsize{40.20}& \scriptsize [m/s] \\  [10pt]

\end{tabular}
\caption{ \textbf{Layered Phantoms}: Comparison between the predicted SoS maps by the networks trained with the Combined and the Ellipsoids dataset for mediums with layered structure and an embedded inclusion with various echogenicity and SoS contrasts. }
\label{fig: experiments, Layers}
\end{figure*}

\paragraph{Layered Phantoms} The aim of this investigation is to demonstrate how the networks behave in the presence of out-of-domain geometry with various echogenicity and SoS contrasts. 

We simulated digital phantoms with layered structures and an elliptical inclusion randomly placed inside one of the layers.
A layer with a random thickness in the range \([0.5-2]\)~cm is placed in the medium. 
The SoS value inside the layer and the inclusions are assigned randomly in the range \([1300-1700]\)~\(m/s\). 
The echogenicity of each layer and the inclusion is assigned randomly as well, hence, the layers or/and the inclusions can have echogenicity contrast compared to the training sets. 

Figure \ref{fig: experiments, Layers} shows 8 example phantoms. 
The figure shows that in all cases, the layers could be separated by both networks. 
Additionally, inclusions independent of their echogenicity are detectable in SoS maps, except for Case 8 for both networks. 
This case has an inclusion that is only partially in the field of view. Since a single PW is used, it is expected that with increasing the depth, the performance drops in the sides of the field of view.  
Case 8 is more challenging because the inclusion is directly below the layer and both networks often under/overestimate the SoS values directly below the layer. 
Similar observations can be found in Figure\ref{fig: experiments, Layers} Cases 2, 5, and 7. 

We created 500 cases of layered phantoms, the average RMSE of 500 cases for the Combined network is \(18.89\)~\(m/s\) with a standard deviation of \(11.59\)~\(m/s\) and for the Ellipsoids network, the average RMSE is \(19.03\)~\(m/s\) with a standard deviation of \(12.47\)~\(m/s\). The average RMSE  for both networks is in a similar range as the RMSE of T2US and Combined setup but the standard deviation is higher. This indicates that it is probable that cases similar to e.g., Figure \ref{fig: experiments, Layers} Case 8 appear where an over/underestimation artifact is present directly below the layer structures. Nevertheless, in most cases, the margins and inclusions are detected correctly. Thus, although the training data does not include layer-like structures the network can handle out-of-domain geometry with any given echogenicity.

\subsection{Experimental Data} 

\subsubsection{Single Frame}

In this section, we test the trained networks on a dataset acquired from the CIRS multi-modality breast phantom. 
The phantom consists of an anechoic skin layer on top (directly below the transducer). 
Directly below the skin layer is the breast tissue mimicking layer with hyperechoic characteristics with randomly mixed embedded fibers. 
Hereafter, we will refer to this layer as background tissue. 
Inside the tissue-mimicking layer, there is an embedded hyperechoic inclusion with SoS contrast from the background layer. 
We do not have the SoS GT map for this phantom but based on the information provided by the manufacturer the SoS values in this phantom are in the range \([1440-1610]\pm 10\)~\(m/s\), where the skin layer has the lower
SoS and the inclusion has a higher SoS contrast compared to the background. 

\begin{figure}[!t]
\centering
\begin{tabular}{l @{\hspace{0.8mm}\vspace{0.8mm}} c @{\hspace{0.8mm}}c @{\hspace{0.8mm}}c @{\hspace{0.8mm}}c}
& & \scriptsize {Frame 0} &  & \\ 
& & 
\raisebox{-.5\totalheight}{ \includegraphics[width = 2.2cm]{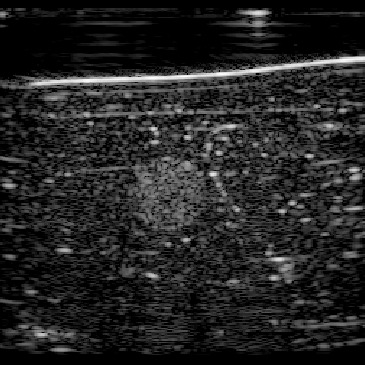}}&
\\
& \scriptsize {Ellipsoids } & \scriptsize {T2US} & \scriptsize{Combined} &  \scriptsize[m/s]
\\
\raisebox{-.5\totalheight}{\scriptsize \rot {\makecell{{Predicted} \\ {{SoS}} }}} & 
\raisebox{-.5\totalheight}{\includegraphics[width = 2.2cm]{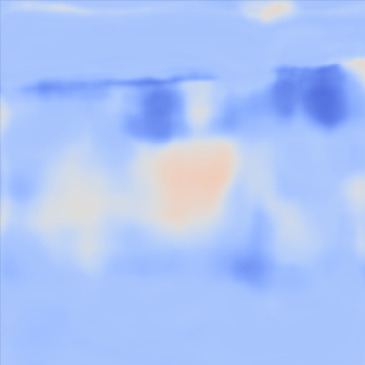}}&
\raisebox{-.5\totalheight}{\includegraphics[width = 2.2cm]{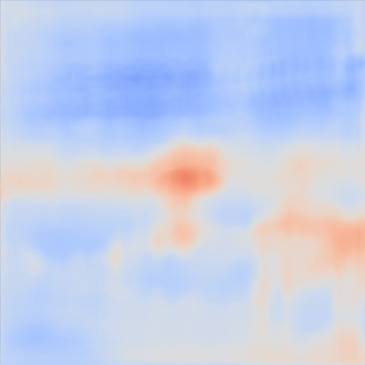}}&
\raisebox{-.5\totalheight}{\includegraphics[width = 2.2cm]{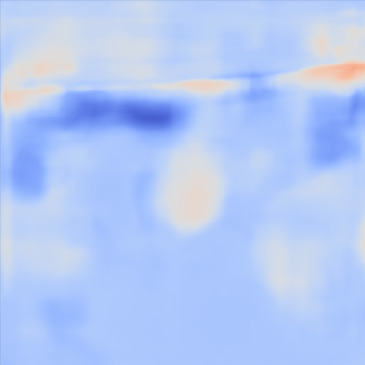}}& 
\raisebox{-.5\totalheight}{\includegraphics[height =2.2cm]{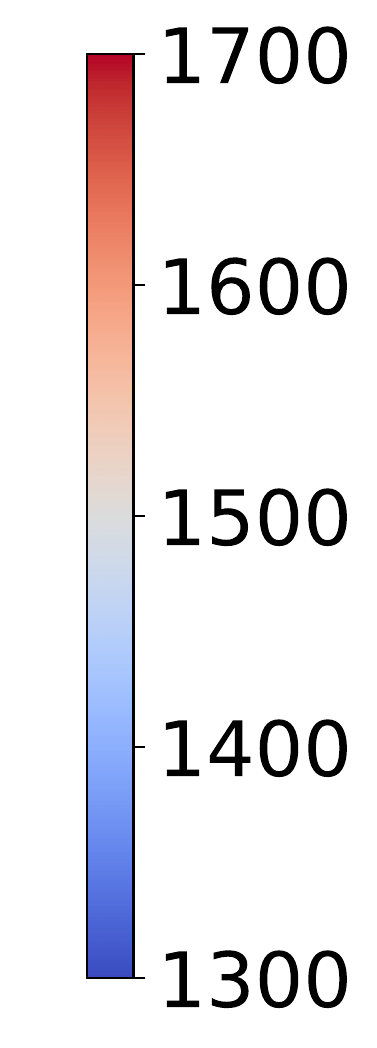}}\\
\raisebox{-.5\totalheight}{\scriptsize \rot {\makecell{{Overlay} \\ {{image} }}}} & 
\raisebox{-.5\totalheight}{\includegraphics[width = 2.2cm]{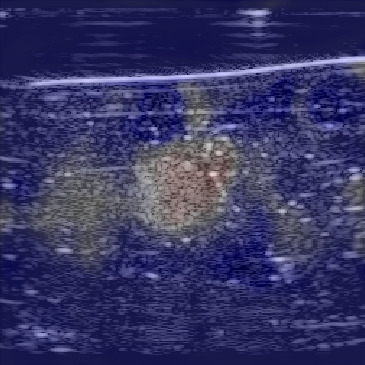}}&
\raisebox{-.5\totalheight}{\includegraphics[width = 2.2cm]{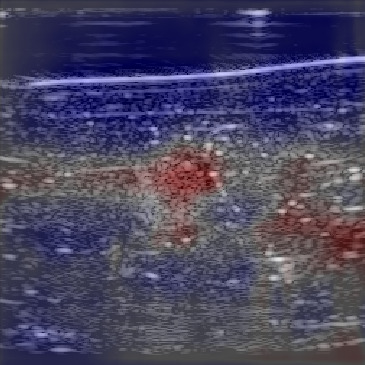}}&
\raisebox{-.5\totalheight}{\includegraphics[width = 2.2cm]{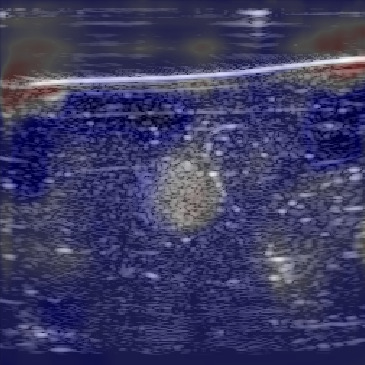}}&
\raisebox{-.5\totalheight}{\includegraphics[height =2.3cm]{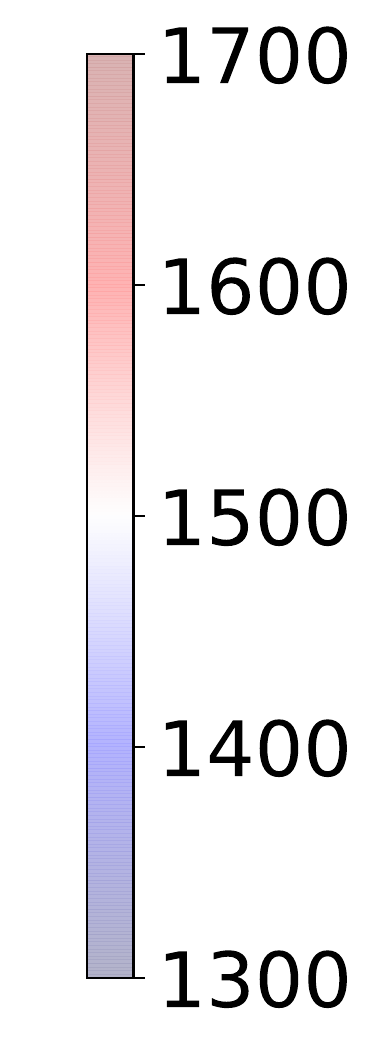}}
\\
\end{tabular}

\caption{1st row: the reconstructed b-mode image of CIRS multi-modality biopsy phantom (Model 073), the image is reconstructed with delay-and-sum beamforming from
single PW acquisition (45 db is shown). The size of the field of view is $3.8 \times 3.8$~$cm$., 2nd row: the predicted SoS map for measured data when the networks are trained with the Ellipsoids, the T2US, and the Combined setup, respectively, 3rd row: the overlay images of the b-mode and the predicted SoS maps.}
\label{fig: Sos Predictions, measured data single}
\end{figure}

Figure~\ref{fig: Sos Predictions, measured data single} shows the predicted SoS values for all three setups.
Since the exact SoS GT is not available, from this point on  we are going to use the relative SoS values to compare the results:

\begin{enumerate}[label=(\roman*)]
\item The predicted SoS values are in the expected range (except for the artifact directly below the skin layer, this effect was observed in layered structured digital phantoms as well). 

\item The Ellipsoids and the Combined setups were capable of separating the skin margin from the background tissue, whereas the T2US setup fails to find the skin layer correctly. 

\item The inclusions are separated with higher SoS values compared to the background for all three setups.
The margins of the inclusion are best detected by the Combined setup and second best by the Ellipsoids setup but there are several artifacts present in the predicted maps using T2US setup.
Although this setup worked well on the simulated data, it is too sensitive to the highly reflective scatterers embedded in the phantom when tested on measured data.
Due to the high sensitivity, we removed this setup from the robustness investigation for consecutive frames presented in the next section.   
\end{enumerate}

\subsubsection{Consecutive Frames}

\begin{figure}[!t]
\centering 

\renewcommand{\arraystretch}{0.1}
\begin{tabular}{@{\hspace{0.5mm}} c  @{\hspace{0.5mm}} c @{\hspace{0.5mm}}c @{\hspace{0.5mm}}c @{\hspace{0.5mm}}c @ {\hspace{0.5mm}}c @{\hspace{0.5mm}}c }
& \scriptsize{ Frame  1} & \scriptsize{ Frame  2} & \scriptsize{Frame  3} & \scriptsize{Frame  4} & \scriptsize{Frame  5} &  \\
& & & & & & \\ 
 \raisebox{-.5\totalheight}{\scriptsize \rot {B-mode}} & 
 \raisebox{-.5\totalheight}{\includegraphics[width = 2.2cm]{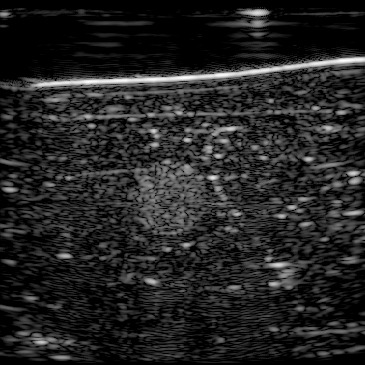}}&
 \raisebox{-.5\totalheight}{\includegraphics[width = 2.2cm]{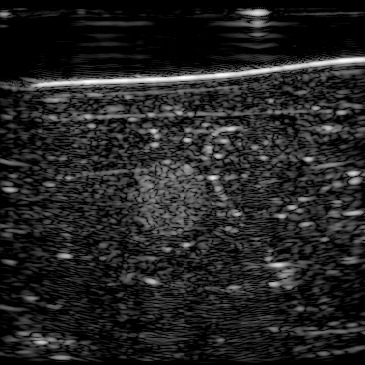}}&
 \raisebox{-.5\totalheight}{\includegraphics[width = 2.2cm]{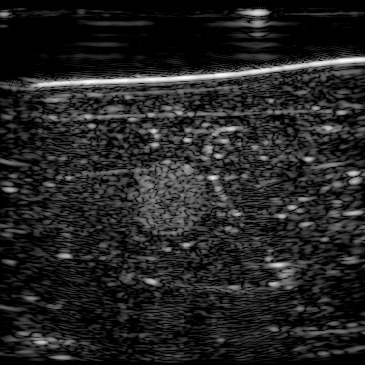}}&
 \raisebox{-.5\totalheight}{\includegraphics[width = 2.2cm]{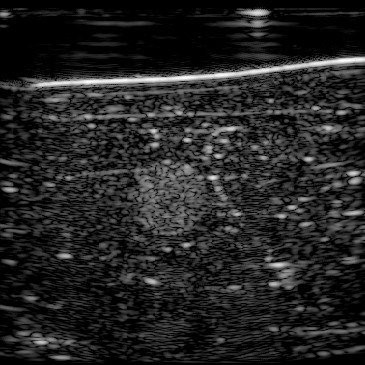}}&
 \raisebox{-.5\totalheight}{\includegraphics[width = 2.2cm]{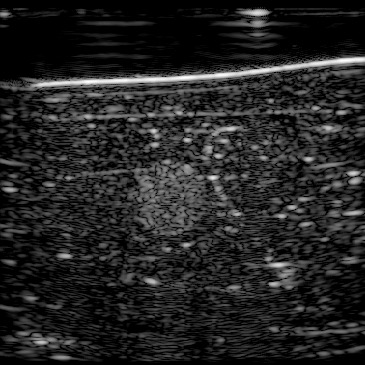}}&
\raisebox{-.5\totalheight}{\includegraphics[height =2.1cm]{img/colormaps/colorbar_bmode.pdf}}\\
\\

\raisebox{-.5\totalheight}{\scriptsize \rot{\makecell{{Absolute} \\  {Difference} \\ {B-mode}}}}& 
 \raisebox{-.5\totalheight}{\includegraphics[width = 2.2cm]{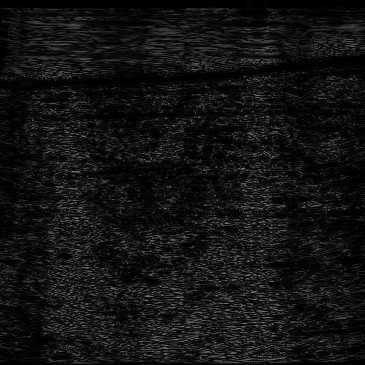}}&
 \raisebox{-.5\totalheight}{\includegraphics[width = 2.2cm]{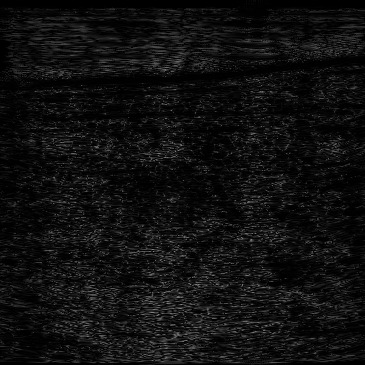}}&
 \raisebox{-.5\totalheight}{\includegraphics[width = 2.2cm]{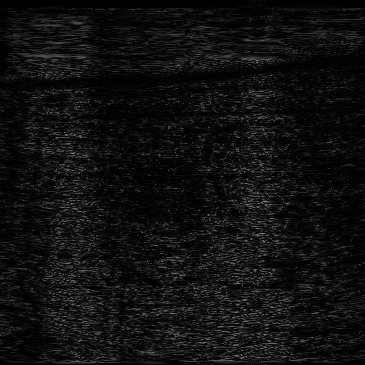}}&
 \raisebox{-.5\totalheight}{\includegraphics[width = 2.2cm]{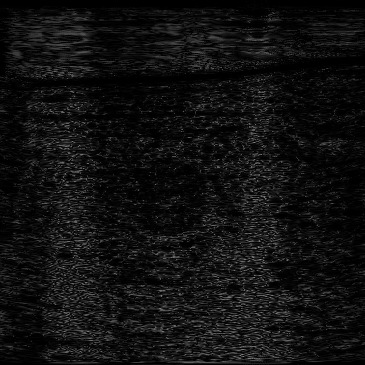}}&
 \raisebox{-.5\totalheight}{\includegraphics[width = 2.2cm]{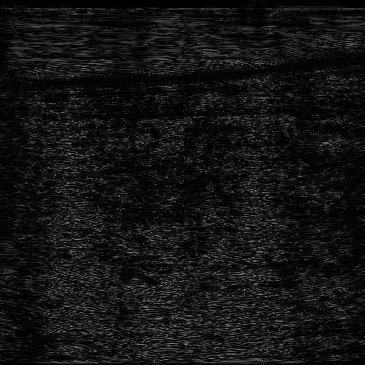}}&
\raisebox{-.5\totalheight}{\includegraphics[height =2.1cm]{img/colormaps/colorbar_bmode.pdf}}\\
\\

 \raisebox{-.5\totalheight}{\scriptsize \rot{\makecell{{Predicted} \\  {SoS} \\ {Combined} }}}& 
 \raisebox{-.5\totalheight}{\includegraphics[width = 2.2cm]{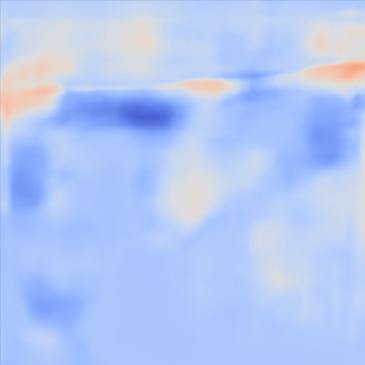}}&
 \raisebox{-.5\totalheight}{\includegraphics[width = 2.2cm]{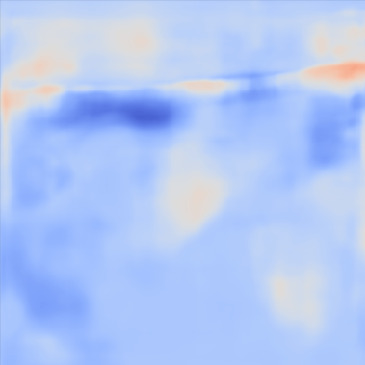}}&
 \raisebox{-.5\totalheight}{\includegraphics[width = 2.2cm]{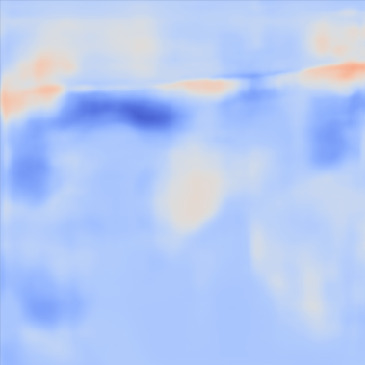}}&
 \raisebox{-.5\totalheight}{\includegraphics[width = 2.2cm]{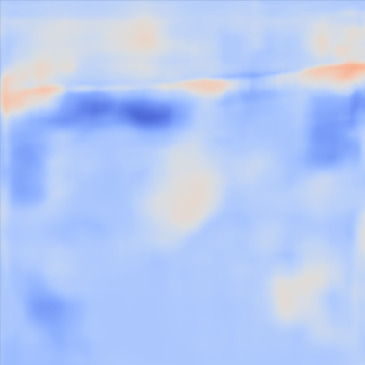}}&
 \raisebox{-.5\totalheight}{\includegraphics[width = 2.2cm]{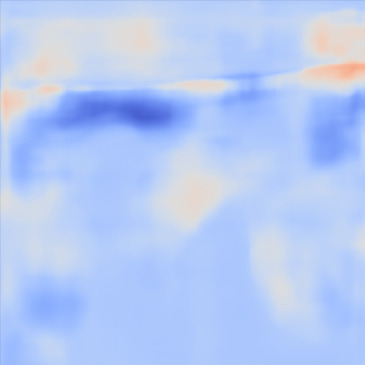}}&
\raisebox{-.5\totalheight}{\includegraphics[height =2.1cm]{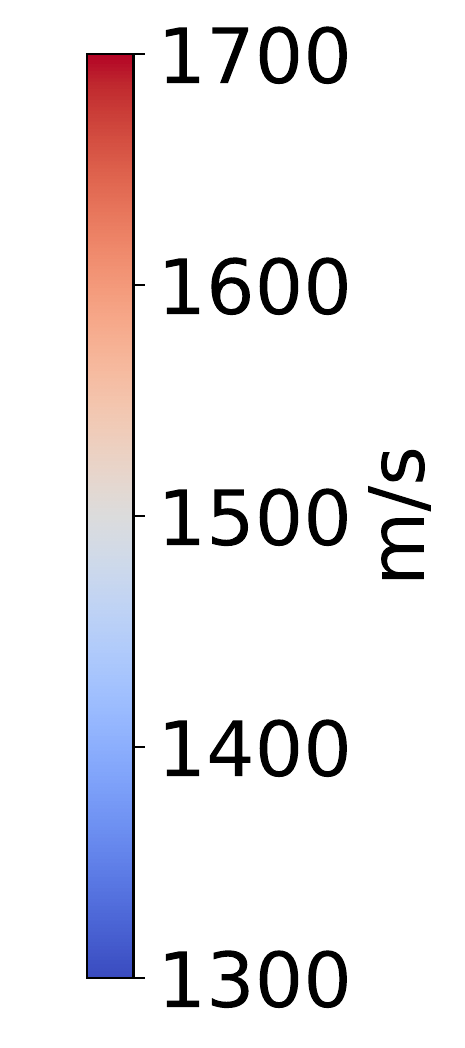}}\\
\\

 \raisebox{-.5\totalheight}{\scriptsize \rot{\makecell{{Predicted} \\  {SoS} \\ {Ellipsoids} }}}& 
 \raisebox{-.5\totalheight}{\includegraphics[width = 2.2cm]{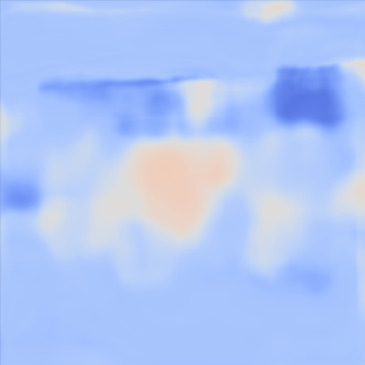}}&
 \raisebox{-.5\totalheight}{\includegraphics[width = 2.2cm]{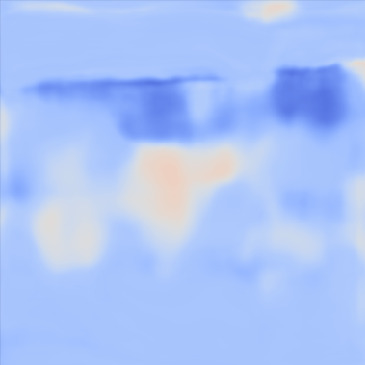}}&
 \raisebox{-.5\totalheight}{\includegraphics[width = 2.2cm]{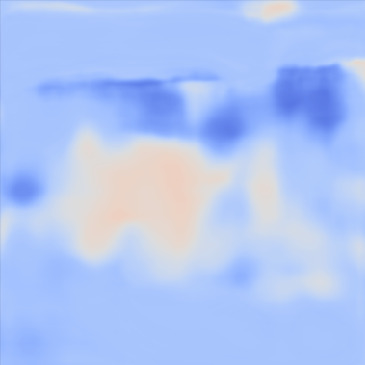}}&
 \raisebox{-.5\totalheight}{\includegraphics[width = 2.2cm]{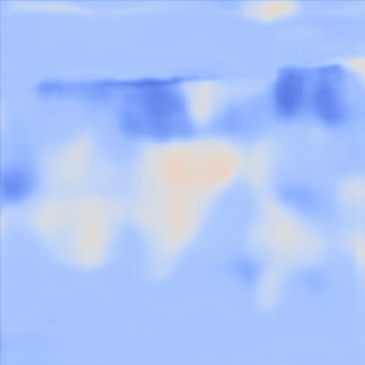}}&
 \raisebox{-.5\totalheight}{\includegraphics[width = 2.2cm]{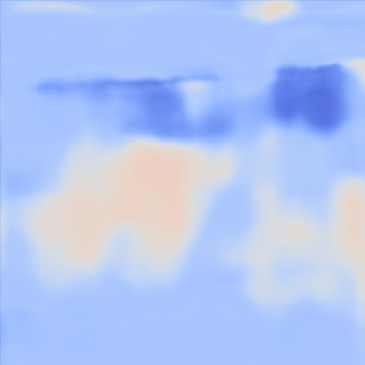}}&
\raisebox{-.5\totalheight}{\includegraphics[height =2.1cm]{img/colormaps/colorbar_real.pdf}}\\
\\

\end{tabular}
\caption{1st row: reconstructed b-mode images of 5 consecutive frames, 2nd row: the absolute difference matrices of 6 consecutive frames (starting from frame 0, shown in Figure~\ref{fig: Sos Predictions, measured data single}), 3rd row: the predicted SoS for the consecutive frames when the network is trained on the Combined dataset, 4th row: the predicted SoS for the consecutive frames when the network is trained on the Combined dataset.}
\label{fig: Sos Predictions, measured data multiple}
\end{figure}

\begin{figure}[!t]
\centering 
\renewcommand{\arraystretch}{0.05}
\begin{tabular}{ @{\hspace{0.5mm}}c | @{\hspace{0.5mm}}c   }

Inside Inclusion & Background\\

\raisebox{-.5\totalheight}{\includegraphics[width=6cm]{ 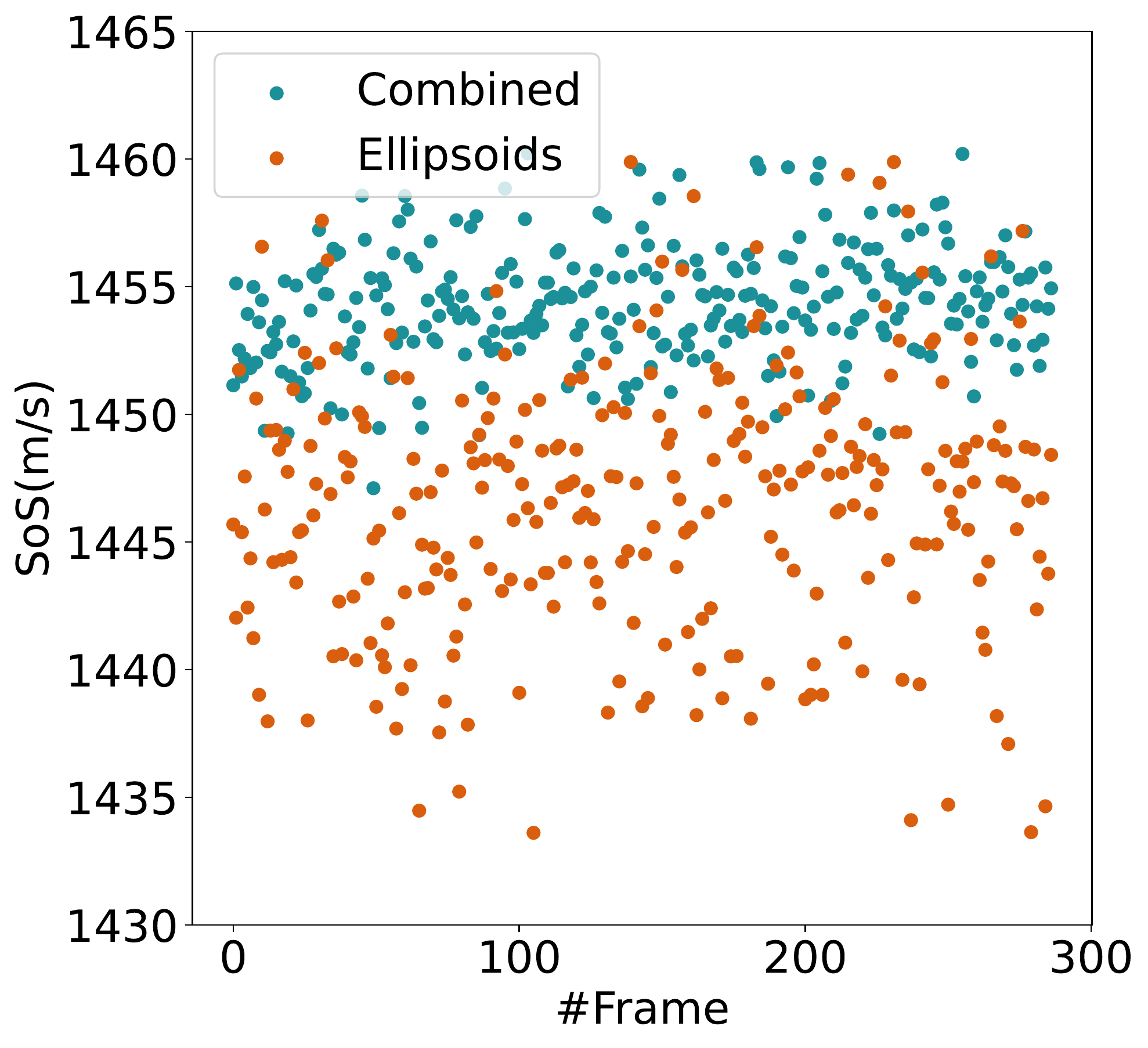}}&
\raisebox{-.5\totalheight}{\includegraphics[width=6cm]{ 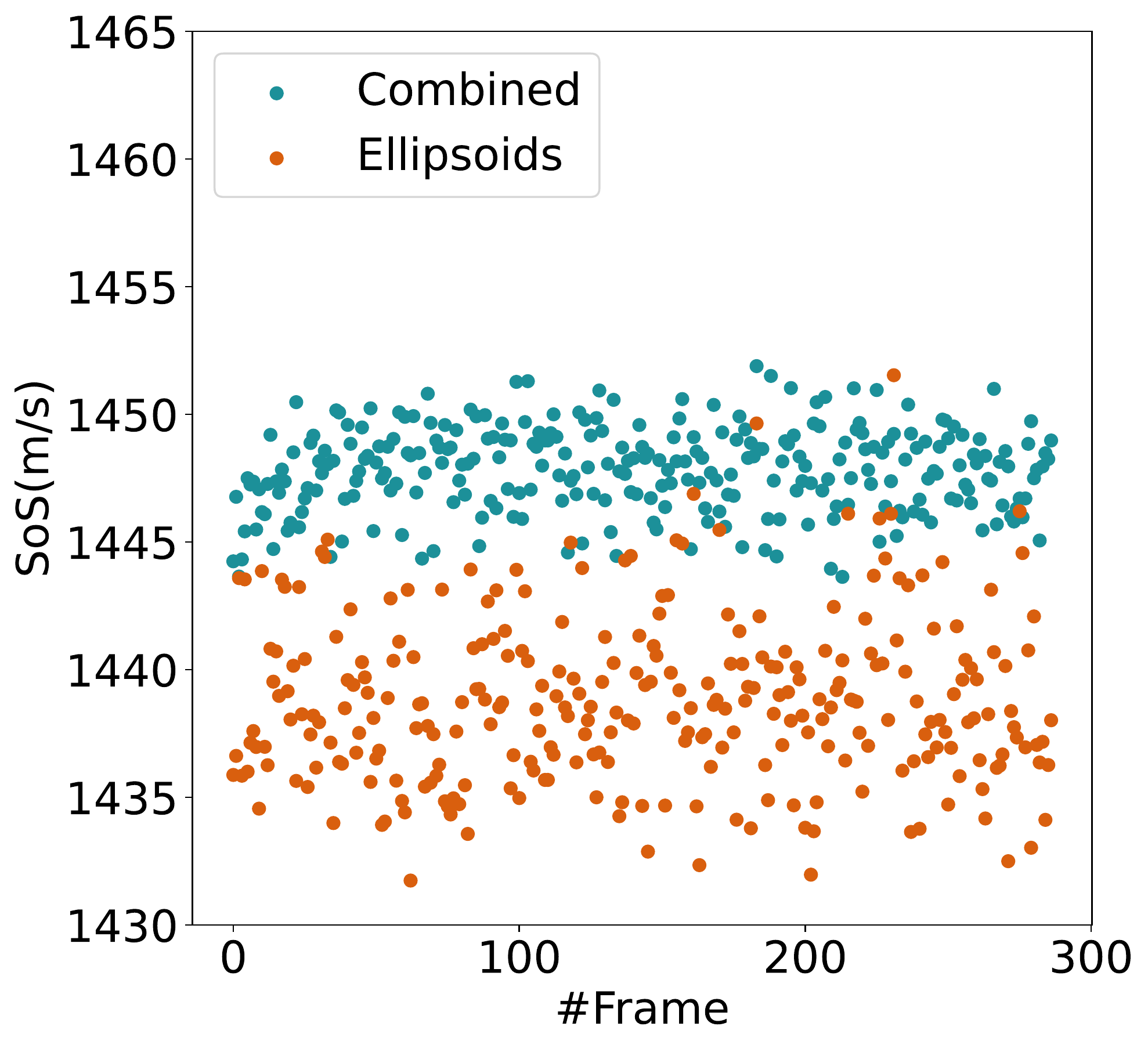}}\\
(a) & (b) \\ 
\raisebox{-.5\totalheight}{\includegraphics[width=6cm]{ 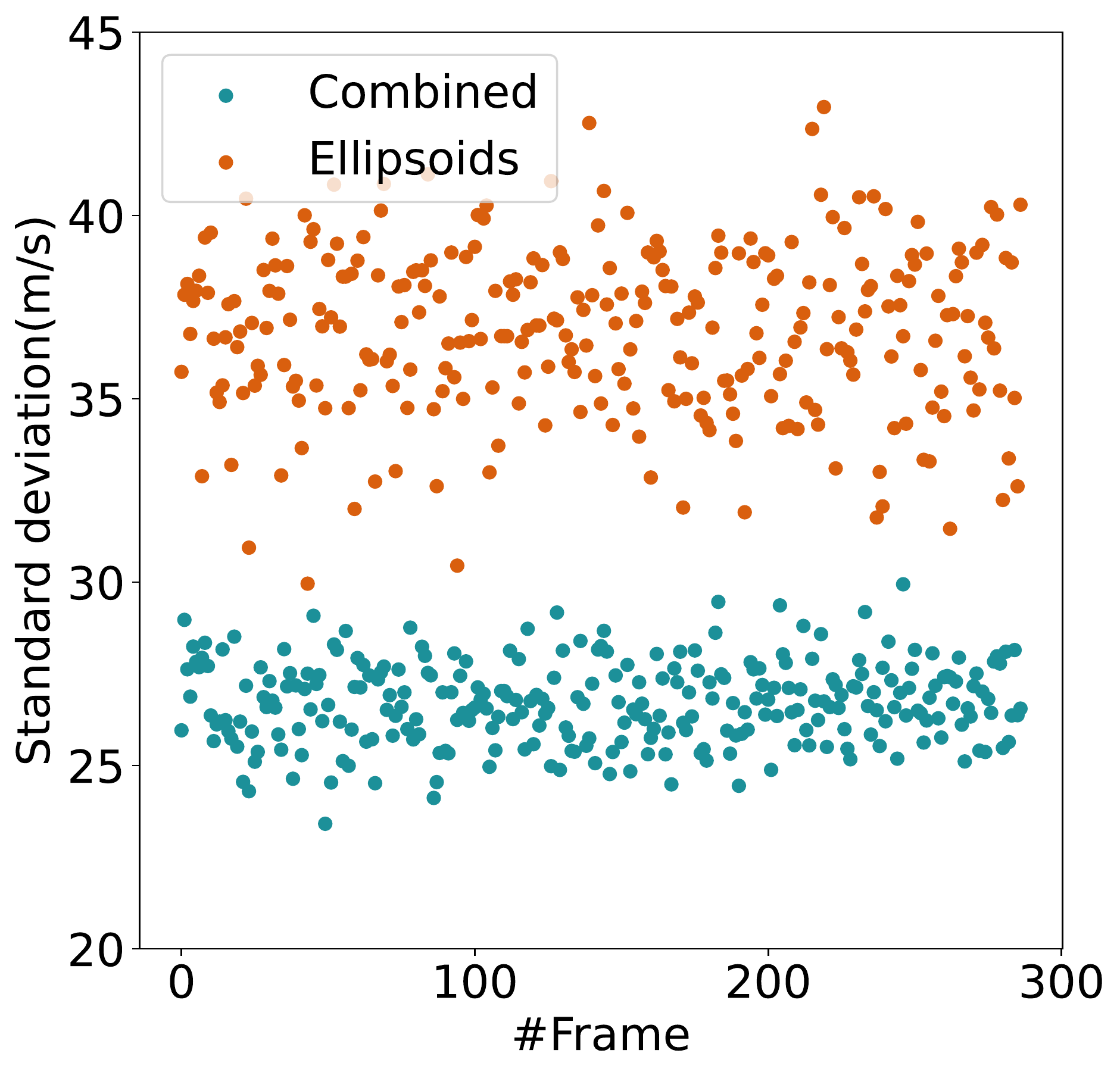}}&
\raisebox{-.5\totalheight}{\includegraphics[width=6cm]{ 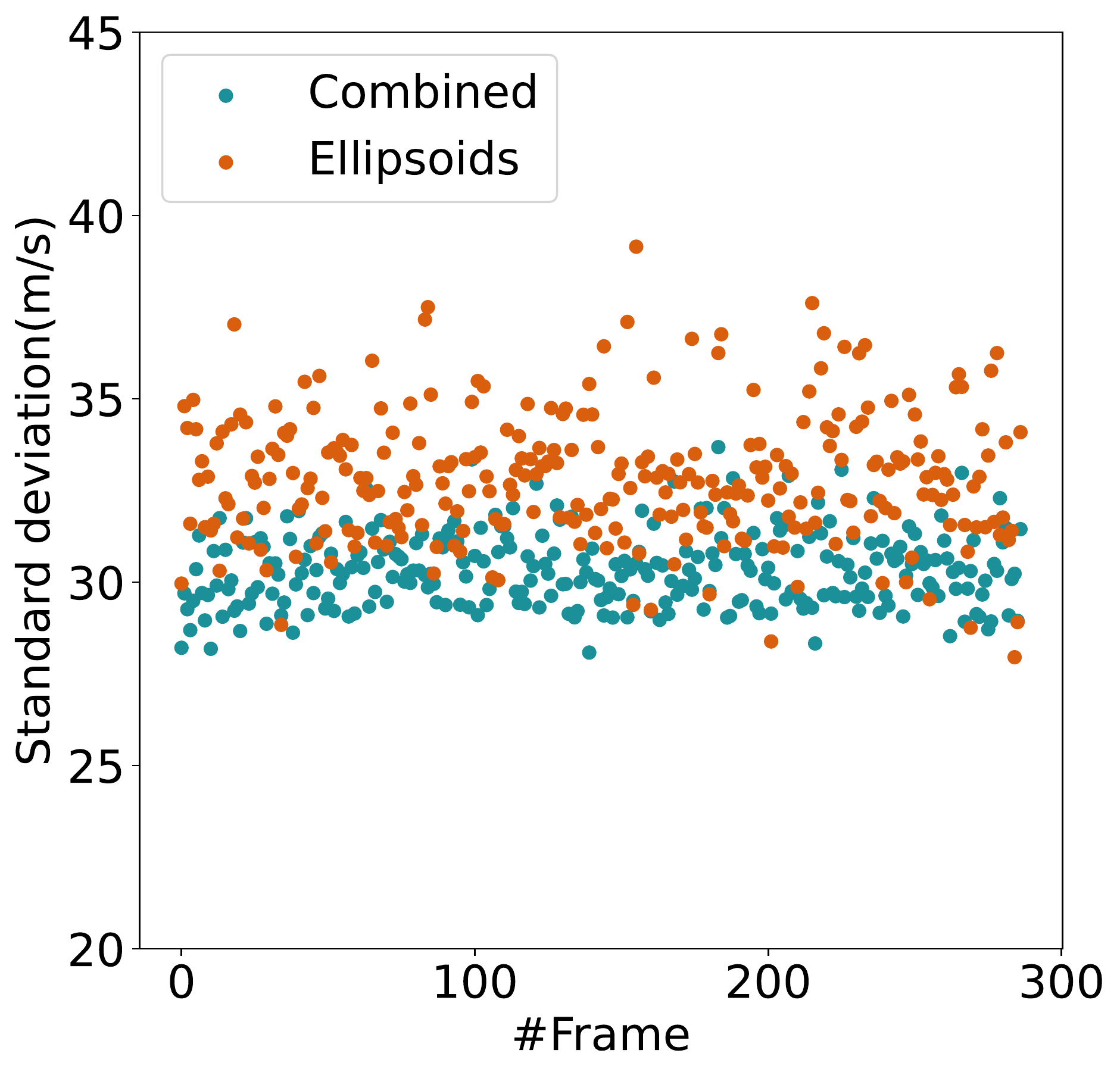}}\\
(c) & (d) \\ 
\end{tabular}
 \caption{the Combined vs. the Ellipsoids setup, over 287 consecutive frames from the same field of view: \textbf{(a)}: Comparison of average predicted SoS values \underline{inside the inclusion}, \textbf{(b)}: Comparison of average predicted SoS values \underline{in the background},  \textbf{(c)}: Comparison of the standard deviation of the predictions \underline{inside the inclusion}, \textbf{(d)}:  Comparison of the standard deviation of the predictions \underline{in the background}; average predicted SoS inside the inclusion for the Ellipsoids setup varies in a larger range for each frame compared to the Combined setup. Predicted SoS in the background region has an offset between the two setups but still, the Ellipsoids setup has a higher variation from frame to frame.}
\label{fig: measured, frames, scatter}
\end{figure}

In our previous work \citep{jush2020dnn}, when tested on the measured data, we realized that the network is sensitive to slight movements during acquisition and electronic noise in the raw data. 
We mechanically fixed the probe head and phantom to reduce unwanted movements and vibrations during data acquisition and recorded multiple consecutive acquisitions, five of which are shown in Figure~\ref{fig: Sos Predictions, measured data multiple}. 
Given that the setup is fixed we expect consistent and reproducible predictions for the acquired frames.

Figure~\ref{fig: Sos Predictions, measured data multiple}, 1st row demonstrates the reconstructed b-mode images for 5 frames. 
We computed the absolute difference of each frame compared to its previous frame which resulted in the image shown in Figure~\ref{fig: Sos Predictions, measured data multiple}, 2nd row. Note that frame 0 is shown in Figure~\ref{fig: Sos Predictions, measured data single}, and the difference matrices for frame 1 are computed using frame 0. 
There is no apparent difference between these frames suggesting phantom movement, thus, the difference is only caused by electrical noise.

In Figure~\ref{fig: Sos Predictions, measured data multiple}, 4th row, the corresponding SoS map of each frame is shown for the network trained on the Ellipsoids dataset alone. 
Although there are minimum movements between the acquired frames, the inclusion localization margins vary for each frame, thus, the network trained with the Ellipsoids dataset is highly sensitive to the electronic noise.

Figure~\ref{fig: Sos Predictions, measured data multiple}, 3rd row demonstrates the SoS map of each frame for the network trained on the Combined dataset. 
The variations between predictions from frame to frame are significantly lower than the cases shown in the 3rd row. 
The variations between frames are mostly present on sides that can possibly be related to the single PW transmission setup. 
Nevertheless, in comparison, this setup shows lower sensitivity.

For quantitative comparison, we computed the mean and standard deviation of the predicted SoS maps of 287 consecutive frames of the same view shown in Figure~\ref{fig: Sos Predictions, measured data multiple}. 
The corresponding scatter plots are shown in Figure \ref{fig: measured, frames, scatter}.
The variation between average SoS values inside the inclusion as well as the background shows that for the same field of view, the network trained with the Combined dataset has more consistent predictions. 
The mean standard deviation for the Combined setup over 287 frames is \(26.70\)~\(m/s\) inside the inclusion and \(30.32\)~\(m/s\) in the background. 
In comparison, the mean standard deviation for the Ellipsoids setup inside the inclusion is  \(36.84\)~\(m/s\), and in the background is \(32.85\)~\(m/s\).
The overall mean standard deviation of the Combined setup is \(6.33\)~\(m/s\) lower than the Ellipsoids setup which is equal to \(18\)\% improvement. Thereby, mixing the Ellipsoids dataset with the T2US dataset improved the stability of the setup and increased the robustness and reproducibility of the network for consecutive frames. 

\section{Summary and Conclusion}

Deep-learning-based algorithms have predictable run times and therefore have great potential for real-time deployment. However, since they are data-hungry, they pose challenges regarding training data requirements. Particularly, due to the lack of a gold standard method, for SoS reconstruction acquiring real-world data alongside their exact GT sufficient to train these networks is even more challenging.
Employing simulated data for training can be advantageous because firstly, it resolves the problem of creating a huge amount of data alongside their exact GT. 
Secondly, it allows for designing diverse mediums with arbitrary heterogeneous structures and acoustic properties that otherwise would have been time-consuming and costly. In prior studies, a simplified model for tissue modeling was used to create training datasets \citep{feigin2019deep, feigin2020detecting, vishnevskiy2019deep, oh2021neural}. 

Transferring deep-learning-based models that are trained on the simulated data to perform robustly on real-world setups is a non-trivial task and models trained on simulated data often only perform well on the same distributions on which they are trained and perform poorly on real setups. 
In this study, we proposed a new simulation setup for modeling the tissues to reduce the sensitivity and increase the stability of an existing deep-learning-based SoS reconstruction method. 
Our proposed medium structure is extracted from Tomosynthesis images of breast tissues that are diagnosed with malignant or benign lesions. This setup includes fine, complicated, tissue-like structures.
We compared our model with a method that uses artificial ellipsoid structures for modeling organs and tissues introduced by \cite{feigin2019deep}. 
Although our proposed model consists of fine and complicated structures, the deep neural network is still able to reconstruct the underlying map and perform well on the simulated data. 

Testing each trained network on its dissimilar test dataset demonstrated that neither the training dataset based on artificial ellipsoids structures nor the training dataset derived from Tomosynthesis data contains sufficient diversified information to generalize the learning of the network. 
Hence, we mixed two datasets and investigated the possibility of training the network jointly with both data representations. 
The network is capable of learning both sets at once. 

Additionally, in the Out-of-Domain section, we addressed the echogenicity characteristics of the simulated mediums and compared the performance of the networks with out-of-domain echogenicity, scatterers, medium structures, and noise presence for simulated data. 
Both networks can handle various echogenicity and different densities of scatterers. 
However, for the Ellipsoids setup varying these properties results in over/underestimations, especially in the background regions which can lead to false positives. 
We showed that the network can detect SoS contrast even when no inclusion is visible in b-mode images. 
In the presence of noise, the network trained with the Combined dataset outperforms the Ellipsoids setup. The gap increases by increasing the noise. 
In the presence of phase noise, both networks showed high error rates and inconsistent predictions that show the sensitivity of the networks to phase distortions. 

In the setup with measured data, all models could localize the embedded inclusion inside the tissue-mimicking phantom but the estimated SoS maps of the network trained with the Combined dataset, have cleaner margins and fewer misestimations. 

Furthermore, we tested the performance of the networks in terms of reproducibility and stability on consecutive frames of the same field of view over 287 frames and demonstrated that by mixing the training datasets we could improve the stability of the network by 18\%.

We conclude two points; firstly, compared to the baseline training data generation method proposed by \cite{feigin2019deep}, our proposed method improves the overall performance of the network both on simulated and measured data setup (especially in the presence of noise) by using diverse simulated dataset. Secondly, the investigations on the out-of-domain simulated data showed that the number of the scatterers and the echogenicity of the simulation has the least impact on the stability of the network, whereas, the network is highly sensitive to phase noises. 
This should be taken into account in the postprocessing and filtering steps of the ultrasound RF data. Avoiding phase distortions is the key to creating a robust setup.

The investigated method has great potential and is advantageous because first of all it is based on a pulse-echo setup, thus, does not require special hardware and can be used with handheld probes. Secondly, it only requires a single PW acquisition which makes it easy to create simulated data to integrate deep neural networks. The biggest advantage of employing deep neural networks is that they can be used in real-time during inference. Plus, a diverse dataset that covers a wide range of shapes, SoS values, and acoustic properties eliminates the burden of hyperparameter tuning and prior assumptions often required by the analytical and optimization methods.

\section{Limitations and Future Research}

Despite encouraging initial results, the proposed method has the following limitations that can be considered for future works. Firstly, there still remains the challenge of quantifying the error on measured data for these setups.
Currently, the exact GT for our measured data is not available. 
Nevertheless, the accuracy of the predicted SoS values can be evaluated by using ultrasound tomography devices. In ultrasound tomography, due to multi-sided access to the tissues, it is possible to calculate the SoS values of the tissues under examination using Time-of-Flight (ToF). Thus, using this approach the expected range of the SoS values can be determined and consequently the accuracy of the predicted values in the pulse-echo setup can be evaluated in comparison. 
It is noteworthy that the effectiveness of SoS methods (using ultrasound tomography or pulse-echo setup) for clinical integration is still in the research phase. 

Secondly, although the new proposed setup improved the performance of the deep neural network under investigation, further research can focus on designing more generalized simulation setups: e.g., the presented setup is a 2D model, and therefore off-axis reflections from the z-direction are not modeled due to high computational effort and memory requirements. Thus, one idea would be to investigate 3D modeling and its effect on the stability of the network. Another approach can include prior knowledge about the tissue type present in Tomosynthesis images by accurately segmenting and annotating the images and mapping the intensity values to the expected SoS values of the corresponding tissue types. 
In general, despite extensive research, developing models that represent realistic tissue structures poses many challenges due to the complex structure and mechanical response of soft tissues. 
For deep learning approaches the characteristics of the measurement devices affect the network generalization to a great extent and therefore, should be taken into account. This further complicates the problem, instead, employing domain adaptation techniques can be a possible solution to bridge the gap between simulated and measured data, an example of such a solution for beamforming is examined by \cite{tierney2021training}.

\acks{We thank Prof. Michael Golatta and the university hospital of Heidelberg for providing the Tomosynthesis dataset.}

\ethics{The work follows appropriate ethical standards in conducting research and writing the manuscript, following all applicable laws and regulations regarding the treatment of animals or human subjects.}

\section*{Disclaimer}
The information in this paper is based on research results that
are not commercially available.

\bibliography{sample}

\newpage
\appendix 

\appendix 
\section{Simulation Setup}

In this appendix, we included details regarding the transducer and transmit setup and baseline simulation setup.  

\noindent


\subsection{Transducer and Transmit Setup} 

\subsubsection{Transducer} 
Based on our available setup \citep{hager2019lightabvs,jush2020dnn}, a linear transducer with 192 active channels, \(200~\mu m\) pitch, and \(3.8~cm\) active area is modeled. 
The transducer is modeled by considering $7$ grid points per channel with $1$ point as kerf. 
The center frequency of the transducer is set to \(5~MHz\). 
A tone-burst pulse with a rectangular envelope, two cycles, and a center frequency of \(5~MHz\) is chosen to fit the pulse shape of the measured data.
The pulse shape and center frequency are chosen in a way to match our hardware setup. 

\subsubsection{Transmit Strategy} 
In conventional focused ultrasound imaging, the medium is insonified for each focusing depth and a line-by-line scan is performed \citep{szabo2004diagnostic}.

In PW imaging a large field of view is insonified by a single transmission.
Deep learning algorithms require a huge amount of diverse training datasets for better generalization.
Therefore, to provide a comprehensive dataset, we need to create a huge number of heterogeneous mediums.
The conventional line-by-line imaging strategy adds a significant computational burden.
Moreover, a huge amount of data will be created for each medium which increases storage and memory requirements as well as training time.
    
Studies showed that PW imaging with multiple steering angles can be used as an alternative to conventional focusing scans \citep{garcia2013stolt,jensen2016optimized}.
Nevertheless, often in b-mode imaging to achieve comparable quality, various steering angles are required \citep{garcia2013stolt,jensen2016optimized}. 
Employing deep learning techniques, it is possible to reduce the number of steering angles to three acquisitions while preserving the image quality \citep{gasse2017high}.
    
For SoS reconstruction, \cite{feigin2019deep} first proposed a network with three PWs, and then \cite{feigin2020detecting} modified the network to a single PW setup.
As such, in this study, we chose the single PW transmission setup as well.
Since we have 192 channels, due to hardware limitations, memory requirements, and simulation time, a single PW is the best choice for our current setup. 
    
Although using a single PW can be a challenging task, deep  neural  networks  are  powerful  tools  that can  extract or/and interpolate relevant information from known data.
As an analogy, take single-shot depth estimation (SIDE) in the field of computer vision, where deep neural networks are employed to learn an implicit representation between color pixels and depth from a monocular image \citep{laina2016deeper,chen2016single,eigen2015predicting,xu2015empirical}.

\subsection{Geometry and Acoustic Properties}

Mediums are simulated on a 2D grid of size \(1536\times3072\) (time step: $dt=5.7692$~$ns$ and Courant-Friedrichs-Lewy value: $CFL=0.3$). 
The mediums translate to \(3.8~cm\) in depth and \(7.6~cm\) in the lateral direction and the probe head is placed above the central section of the medium. 
The area directly under the transducer is recovered and reconstructed, resulting in a \(3.8\times3.8\)~\(cm\) field of view. 

The acoustic properties of the medium are set based on the properties of the breast tissue, with acoustic attenuation of \(0.75\)~\(dB/MHz.cm\) and the mass density of \(1020\)~\(kg/m^3\) (\cite{szabo2004diagnostic}, Table B.1).
The power law absorption exponent (alpha power) is set to $1.5$ and the non-linearity parameter (BonA) is equal to $9.63$ (\cite{szabo2004diagnostic}, Table B.1). 
Speckle modeling in the density domain is based on \cite{feigin2019deep}, uniformly distributed random speckles are added with a mean distribution of 2 reflectors per \(\lambda^2\) and \(\pm3\%\) variation from background density.
Speckle modeling in the SoS domain is also uniformly distributed random scatterers based on k-wave b-mode example for linear transducers \citep{bmodeexample}. 

\subsection{Simulation Setup, Prior Work: Ellipsoids Setup} 

This setup is based on \cite{feigin2019deep}. The medium consists of a background with homogeneous SoS and elliptical-shaped inclusions randomly placed inside the medium. The SoS values are chosen randomly in a range of \([1300-1700]\)~\(m/s\). 

 We introduced echogenicity contrast inside inclusions  by increasing the standard deviation of \(10\%\) of the speckles by \([4.4-5.5]\%\) in the SoS domain. This means that \(10\%\) of the speckles inside lesions have slightly higher SoS values which result in hyperechoic ultrasound characteristics which create hyperechoic inclusions. 
 Hyperechoic lesions are often benign \citep{gokhale2009ultrasound} but there are cases of malignant lesions among them \citep{linda2011hyperechoic}. 
 Although generally for malignant lesions other factors, i.e., ill-defined borders, spiculated margins, posterior acoustic shadowing, and microcalcifications are important factors to take into account \citep{gokhale2009ultrasound,ramani2021hyperechoic}.

One to five inclusions with elliptical shapes are randomly placed inside the medium. 
Since the lateral direction is twice the size of the probe head, inclusions can be outside of the recovered region as well. 
This is done to handle off-plane reflections.
Ellipsoids are randomly rotated between \([-60,60\)] degrees. 
The radius of ellipsoids in lateral direction varies between \([2-20]\)~\(mm\) and in axial direction between \([2-10]\)~\(mm\). 

\section{Data Processing and Training} 

In this appendix, we included data pre/post-processing for the network.
Before training, the channel data is pre-processed in the respective order:
\subsection{Time-gain Compensation} The simulation setup is configured with acoustic attenuation in the medium, thus, a time-gain compensation with \(0.75~dB/MHz.cm\) at SoS of \(1540~m/s\) is applied. 
    
\subsection{Input Pulse} The first samples of RF channel data contain unwanted signals due to electrical crosstalk from the input pulse. 
These samples are removed by setting the first 100 samples to zero.

\subsection{Normalization} Each dataset is normalized on a per-channel basis with a sample range from -1 to 1 and a mean value of zero.  
    
\subsection{Quantization Noise} In the measured data setup due to the properties of ADCs, data quantization is inevitable. To take measured data characterization into account, a uniformly distributed quantization noise is added to the simulated data. 
    
\subsection{GT Map} The medium size is \(1536 \times 3072\) grid points. The central section of the grid is recovered which results in the size of \(1536\times1536\) grid points. This is still a large matrix. Therefore, we resized the medium with a factor of \(4\) using bilinear interpolation. This results in a matrix of size \(384 \times 384\) with \(0.1\)~\(mm\) resolution.

The network is implemented with the GPU-accelerated Tensorflow framework and the following setup is used during training: 

\subsection{Regularization} During training, Gaussian noise with a standard deviation of 1 is added to the input data.
Dropout layers are activated in layers \(4\), \(5\), and \(7\). 

\subsection{Convolution kernels} Non-square convolution kernels are used based on \citep{feigin2020detecting}.
In the contracting path, kernel size starts from size \(3\times15\), in each convolution steps the width of the kernel size is decreased by 2.
For example, the second layer has a kernel size of \(3\times13\), and the third layer has a kernel size of \(3\times11\).
The kernel size is decreased up to the deepest layer, resulting in \(3\times3\) squared kernel.
In the expanding path excluding the last two layers, the kernel sizes are increased in a reverse manner. 
In layer \(13\), a \(3\times3\) kernel is used and the last layer is a \(1\times1\) convolution layer. 
    
\subsection{Weight Initialization} The Xavier uniform initializer (Glorot uniform initializer) is used for weight initialization \citep{glorot2010understanding}.
This initialization draws samples from a uniform distribution within \([-limit,+limit]\), where the \(limit\) value is defined based on the number of input and output units in the weight tensor \citep{glorot2010understanding}.

\subsection{Loss Function} The mean squared error (MSE), a.k.a. \(L_2\) loss, is used as the loss function.
    
\subsection{Optimizer} Mini-batch gradient descent with a batch size of \(10\), \(0.9\) momentum with a decaying learning rate of \(0.00001\) is used. 

\end{document}